\documentclass[aps,prd,tightenlines,nofootinbib,reprint,floatfix]{revtex4-1}

\usepackage{graphicx}
\usepackage{dcolumn}
\usepackage{bm}
\usepackage{amsmath}
\usepackage{multirow}
\usepackage{longtable}
\usepackage{dcolumn}

\begin{document}

\newcommand{\mevcc}{\!\mathrm{MeV}\!/c^2}
\newcommand{\mevc}{\!\mathrm{MeV}/\!c}
\newcommand{\mev}{\!\mathrm{MeV}}
\newcommand{\gevcc}{\!\mathrm{GeV}/\!c^2}
\newcommand{\gevc}{\!\mathrm{GeV}/\!c}
\newcommand{\gev}{\!\mathrm{GeV}}

\title{First Measurements of Exclusive Hadronic Decays of $\Upsilon(1S)$ and $\Upsilon(2S)$}

\author{S.~Dobbs}
\author{Z.~Metreveli}
\author{A.~Tomaradze}
\author{T.~Xiao}
\author{Kamal~K.~Seth}
\affiliation{Northwestern University, Evanston, Illinois 60208, USA}

\date{May 22, 2012}

\begin{abstract}
Using data taken with the CLEO~III detector, 1.09~fb$^{-1}$ at $\Upsilon(1S)$, and 1.28~fb$^{-1}$ at $\Upsilon(2S)$, branching fractions have been measured for the first time for exclusive decays of each resonance into different final states consisting of 4 to 10 light hadrons, pions (including up to $2\pi^0$), kaons, and protons.  Significant strength is found in 73 decay modes of $\Upsilon(1S)$ and 17 decay modes of $\Upsilon(2S)$, with branching fractions ranging from $0.3\times10^{-5}$ to $110\times10^{-5}$.
Upper limits at 90\% confidence level are presented for the other decay modes.

\end{abstract}

\maketitle

\section{Introduction}

The hadronic decays of charmonium S--wave states have been extensively studied.
For example, for $J/\psi$ ($\psi(1S)$) and $\psi(2S)$ 97 and 70 exclusive hadronic decays, respectively, have been measured~\cite{pdg}.  In sharp contrast, branching fractions for not even a single exclusive hadronic decay of $\Upsilon(1S)$ or $\Upsilon(2S)$ have been measured, and only five upper limits have been established for $\Upsilon(1S)$~\cite{pdg}.  
In this paper we report on the first measurements of exclusive decays of both $\Upsilon(1S)$ and $\Upsilon(2S)$ into $4-10$ light hadrons (including up to $2\pi^0$) using the CLEO~III detector.  
For $\Upsilon(1S)$, branching fractions for 73 decays have been measured and upper limits have been established for 27 decays.  For $\Upsilon(2S)$, branching fractions for 17 decays have been measured and upper limits have been established for 83 decays.

\section{The CLEO~III Detector}

The CLEO III detector, which has been described before \cite{cleodetector}, consists of a CsI electromagnetic calorimeter, an inner silicon vertex detector, a central drift chamber, and a ring-imaging Cherenkov (RICH) detector, all inside a superconducting solenoid magnet with a 1.5~T magnetic field.  Layers of proportional counters embedded in the flux--return iron are used to identify muons.
The detector has a total acceptance of 93$\%$ of $4\pi$ for charged and neutral particles.  The photon energy resolution in the central ($81\%$ of $4\pi$) part of the calorimeter is about $2\%$ at $E_{\gamma}=1$~GeV and about $5\%$ at $100\;\,\mathrm{MeV}$.  The charged particle momentum resolution is about 0.6$\%$ at $1~\mathrm{GeV}/c$.

\section{Data Samples \& Decay Modes}

The on--resonance $\Upsilon(1S)$ data consist of $e^+e^-$ annihilations with an integrated luminosity of 1.09~fb$^{-1}$, with 21.5~million $\Upsilon(1S)$ produced. The on--resonance $\Upsilon(2S)$ data consist of $e^+e^-$ annihilations with an integrated luminosity of 1.28~fb$^{-1}$, with 9.3~million~$\Upsilon(2S)$ produced.  We also use 0.20~fb$^{-1}$ of data off--$\Upsilon(1S)$ resonance (36~MeV below $\Upsilon(1S)$) and 0.43~fb$^{-1}$ of data off--$\Upsilon(2S)$ resonance (28~MeV below $\Upsilon(2S)$) to determine non--resonance contributions.

We reconstruct final states containing 4 to 10 hadrons, pions, kaons, and protons,  including 0, 1, or 2 $\pi^0$'s.  

Monte Carlo (MC) simulations were used to evaluate efficiencies, and $10^5$ events were simulated for each decay mode.  The hadronic decays were generated with phase space distributions.

\section{Event Selections}

We require that there be either 4, 6, 8, or 10 charged particle tracks in the event and that the total charge of these tracks be zero.
Charged particles are required to have $|\cos\theta|<0.93$, and to be consistent with originating from the $e^+e^-$ interaction point.
 Photon candidates are calorimeter showers which do not contain any of the few known noisy calorimeter cells, are not consistent with the projection of any charged track to the calorimeter, and whose transverse energy distributions are consistent with an electromagnetic shower. 
No restriction on the number of photon candidates in the event is made.

To identify the charged hadrons, we use the energy loss in the drift chamber ($dE/dx$) and information from the RICH subdetector.
To utilize $dE/dx$ information, for each particle hypothesis, $h=\pi,~K,~p$ or $\bar{p}$, we calculate $\chi_{h}^{dE/dx}=[(dE/dx)_\mathrm{measured}-(dE/dx)_\mathrm{predicted}]/\sigma_{X}$, where $\sigma_{h}$ is the standard deviation of the measured $dE/dx$ for hypothesis $h$.
For low momentum particles (defined as $p<0.6$ GeV/$c$ for $\pi$ and $K$, and $p<1.5$ GeV/$c$ for protons), we use only $dE/dx$ information for particle identification (ID) and require
\begin{itemize}
\item $\pi$ ID: \quad $|\chi_\pi^{dE/dx}|<3$, \qquad $|\chi_\pi^{dE/dx}| < |\chi_K^{dE/dx}|$
\item $K$ ID:   \quad $|\chi_K^{dE/dx}|<3$, \qquad $|\chi_K^{dE/dx}| < |\chi_\pi^{dE/dx}|$
\item proton ID:   \quad $|\chi_p^{dE/dx}|<3$, \qquad $|\chi_p^{dE/dx}| < |\chi_K^{dE/dx}|$
\end{itemize}
For higher momentum particles, we also use the log-likelihood, $L^\mathrm{RICH}=-2\log(L_h)$, where $L_h$ is the likelihood that a particle corresponds with a given hypothesis ($h=\pi,K,p$ or $\bar{p}$) based on Cherenkov photons detected in the RICH subdetector.  We use this information if the charged particle passes through the RICH detector ($|\cos\theta|<0.8$) and at least three photons observed in the RICH are associated with the particle hypothesis.  We distinguish between different charged particle types using the combined variable
$$\Delta \mathcal{L}_{i,j} = L^\mathrm{RICH}_i - L^\mathrm{RICH}_j + (\chi^{dE/dx}_i)^2 - (\chi^{dE/dx}_j)^2,$$
and require
\begin{itemize}
\item $\pi$ ID: \quad  $\Delta \mathcal{L}_{\pi,K} < 0$
\item $K$ ID: \quad   $\Delta \mathcal{L}_{K,\pi} < 0$
\item proton ID: \quad   $\Delta \mathcal{L}_{p,K} < 0$
\end{itemize}
Single particle efficiencies are found to be $>80\%$ with a fake rate of $\lesssim10\%$ over the momentum ranges considered.

\begin{figure*}[!p]
\begin{center}
\includegraphics[width=2.in]{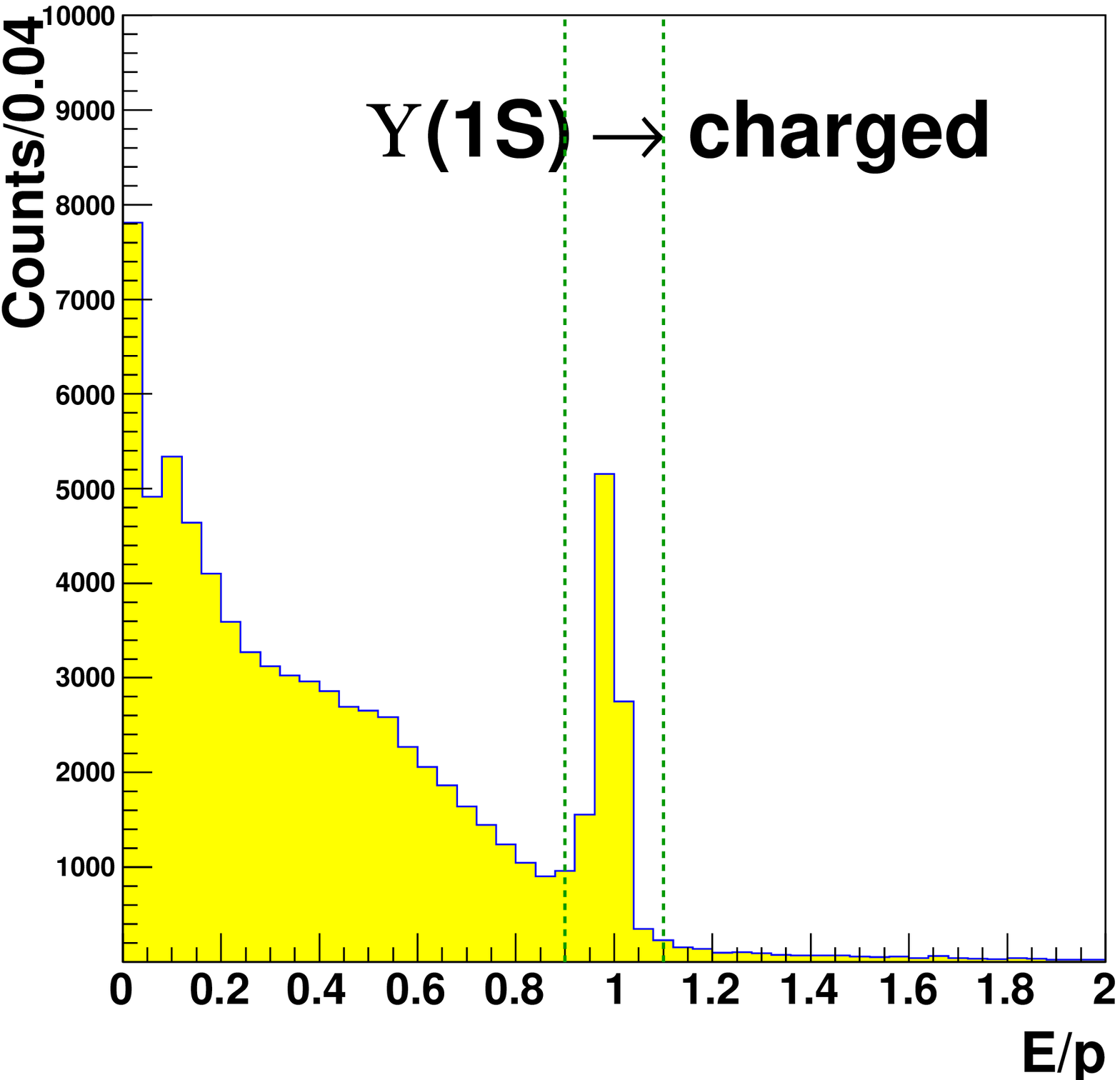}
\includegraphics[width=2.in]{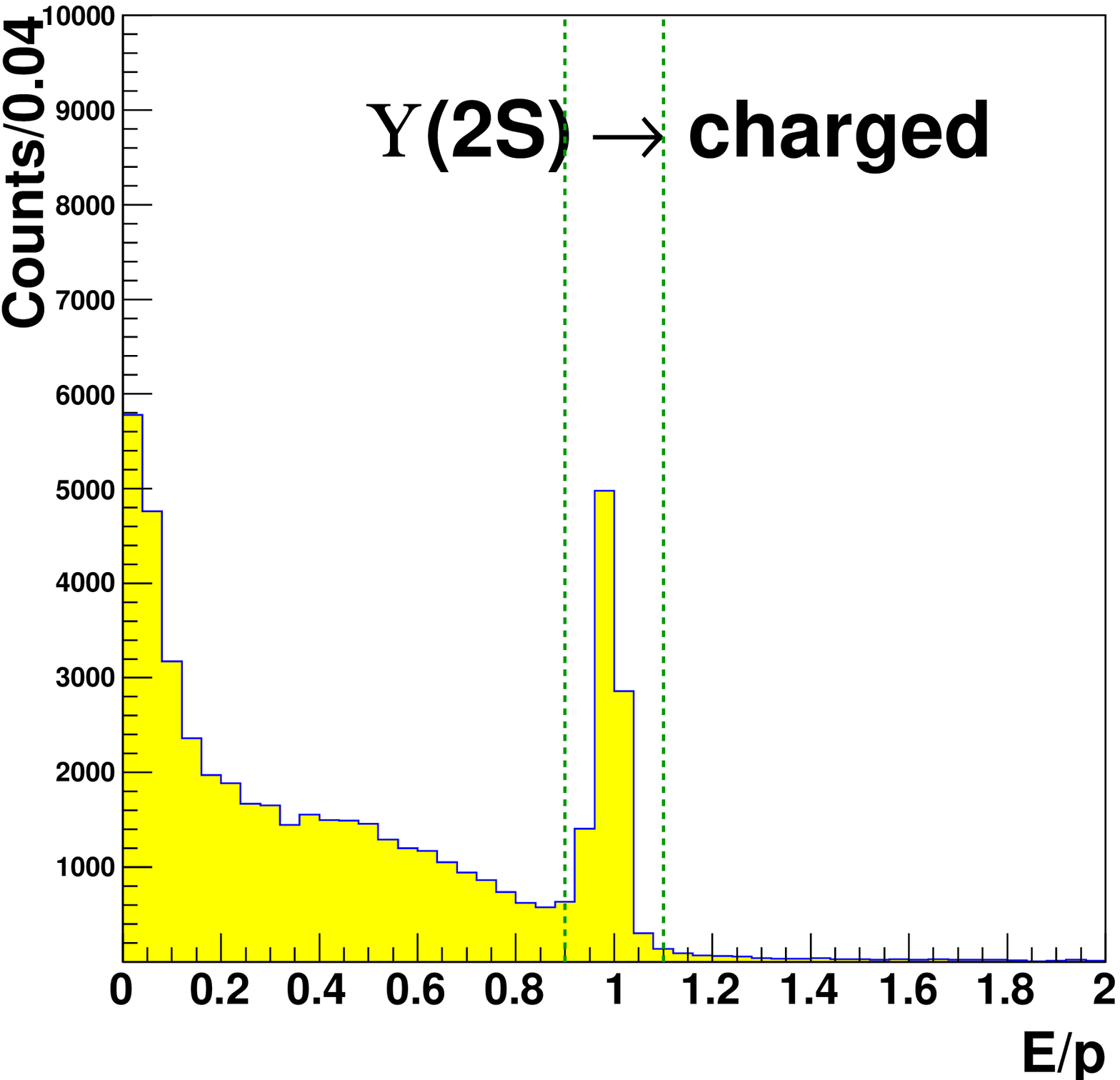}
\includegraphics[width=2.in]{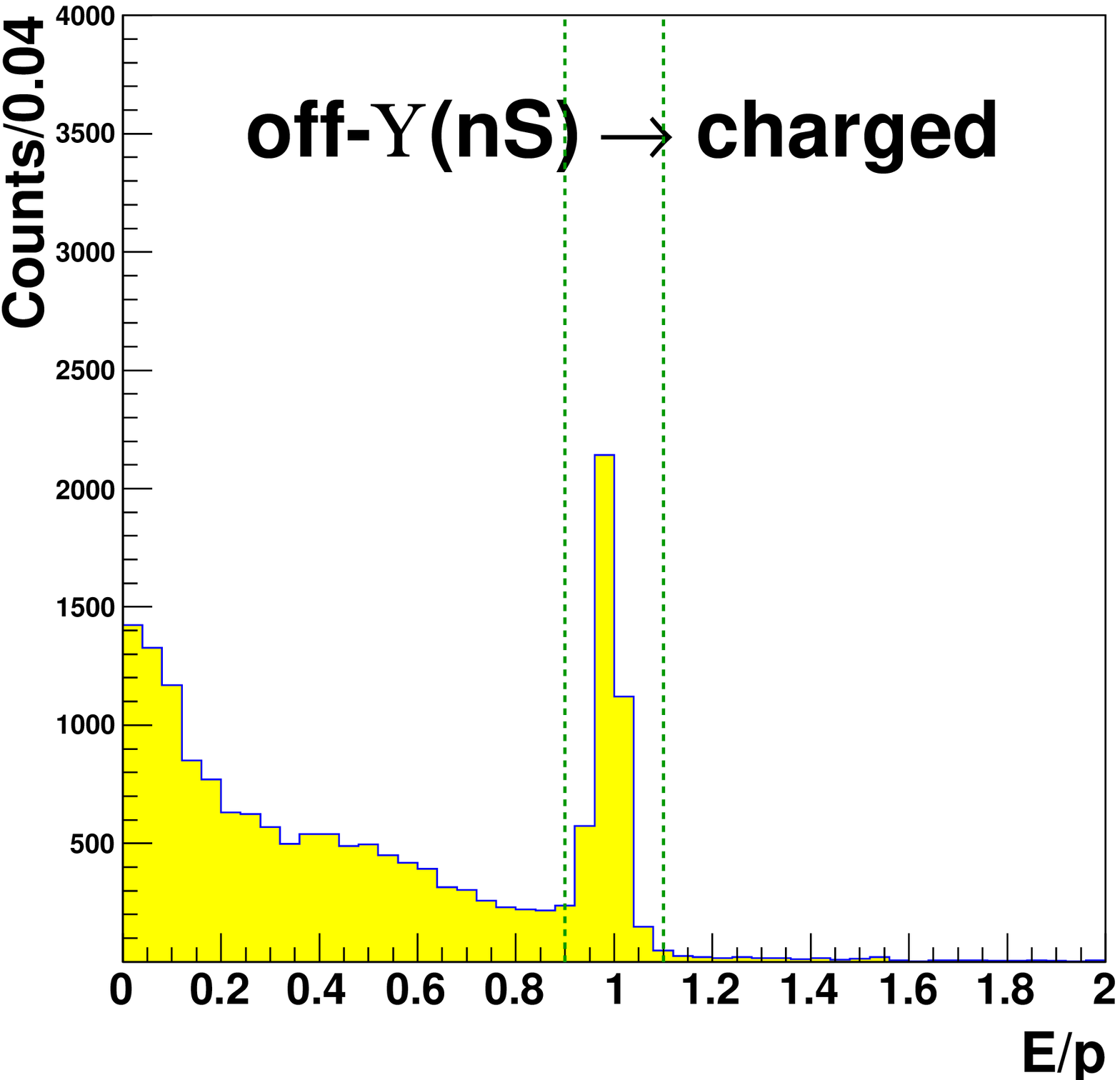}

\includegraphics[width=2.in]{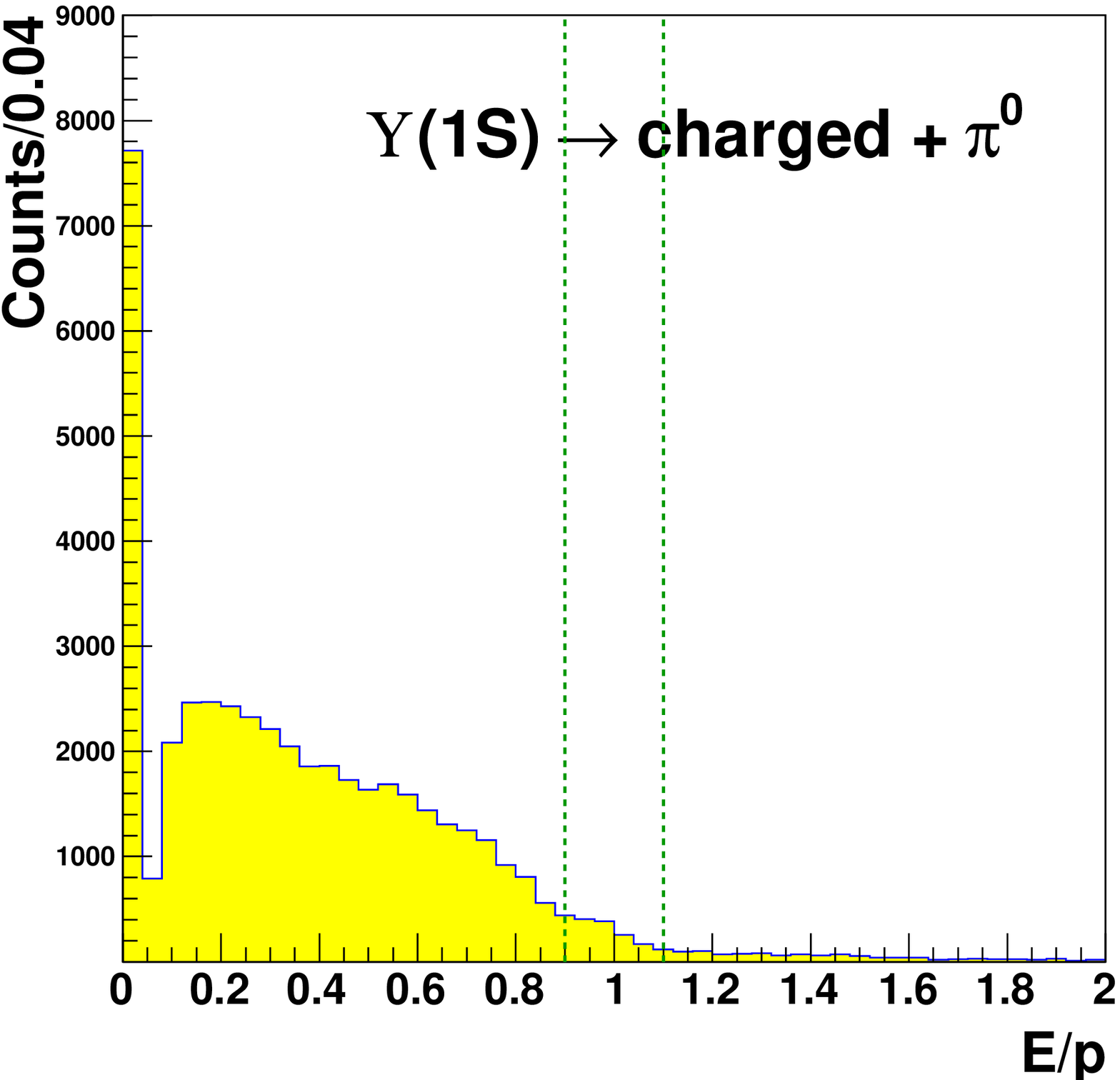}
\includegraphics[width=2.in]{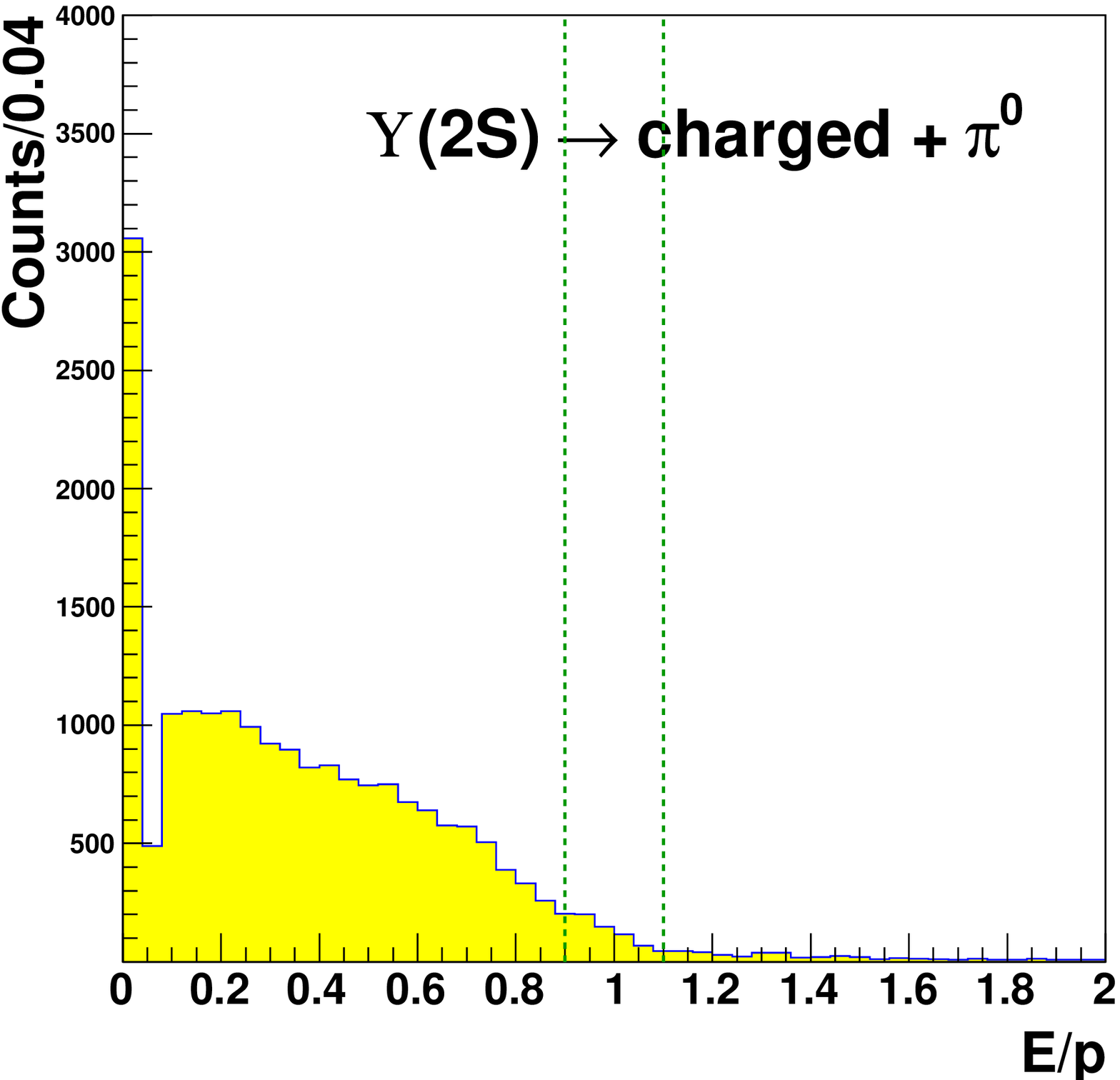}
\includegraphics[width=2.in]{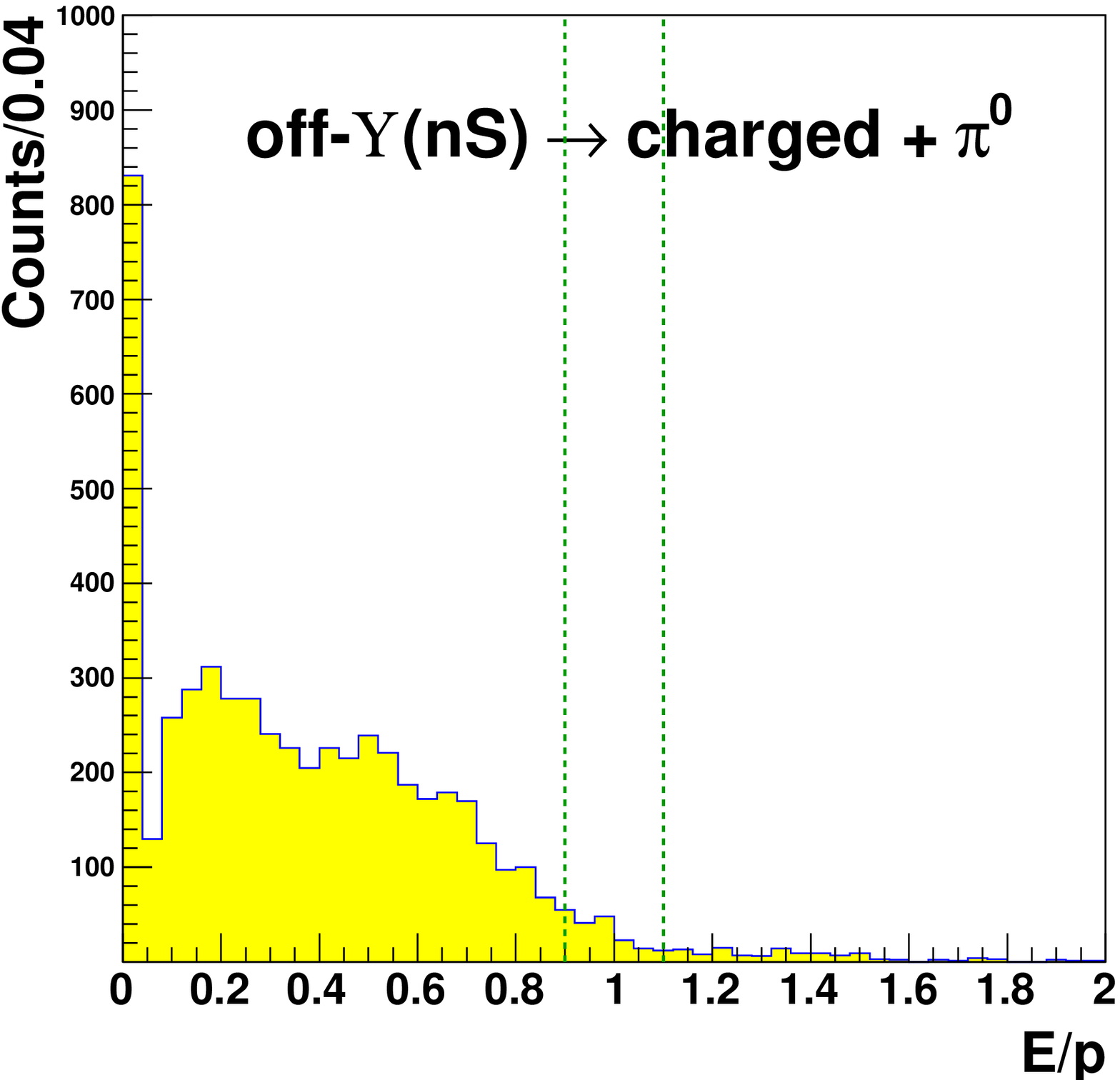}

\includegraphics[width=2.in]{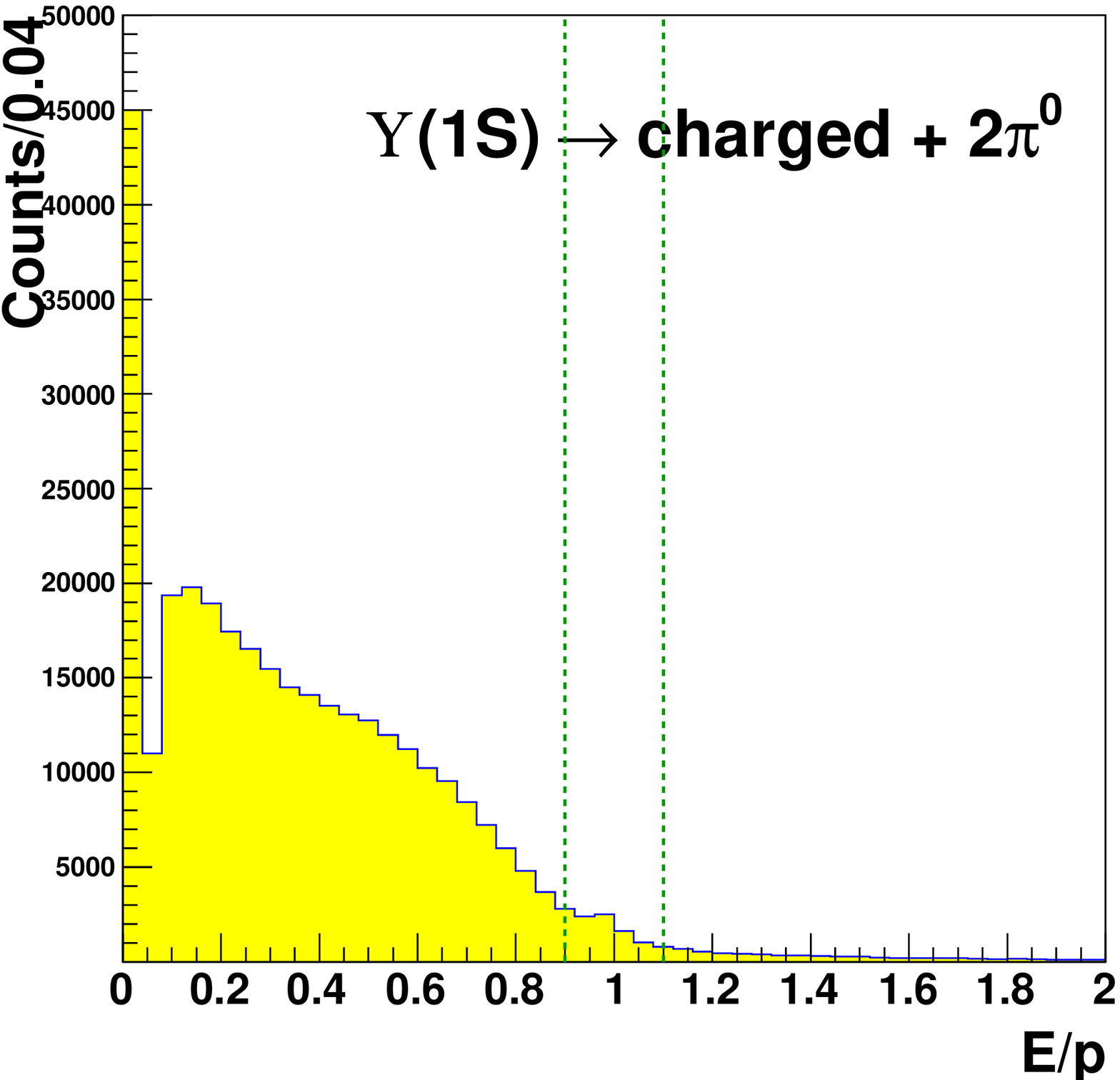}
\includegraphics[width=2.in]{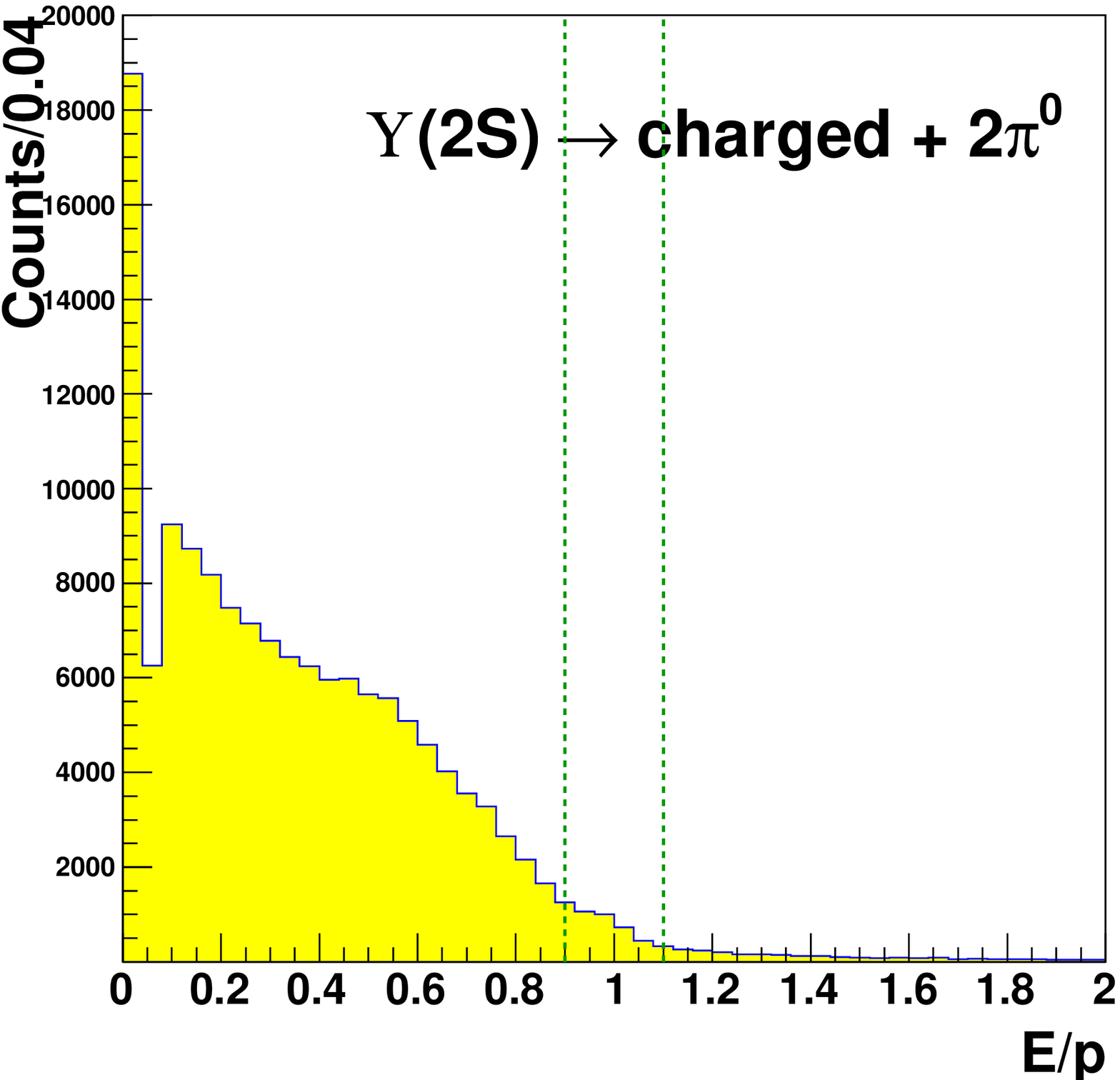}
\includegraphics[width=2.in]{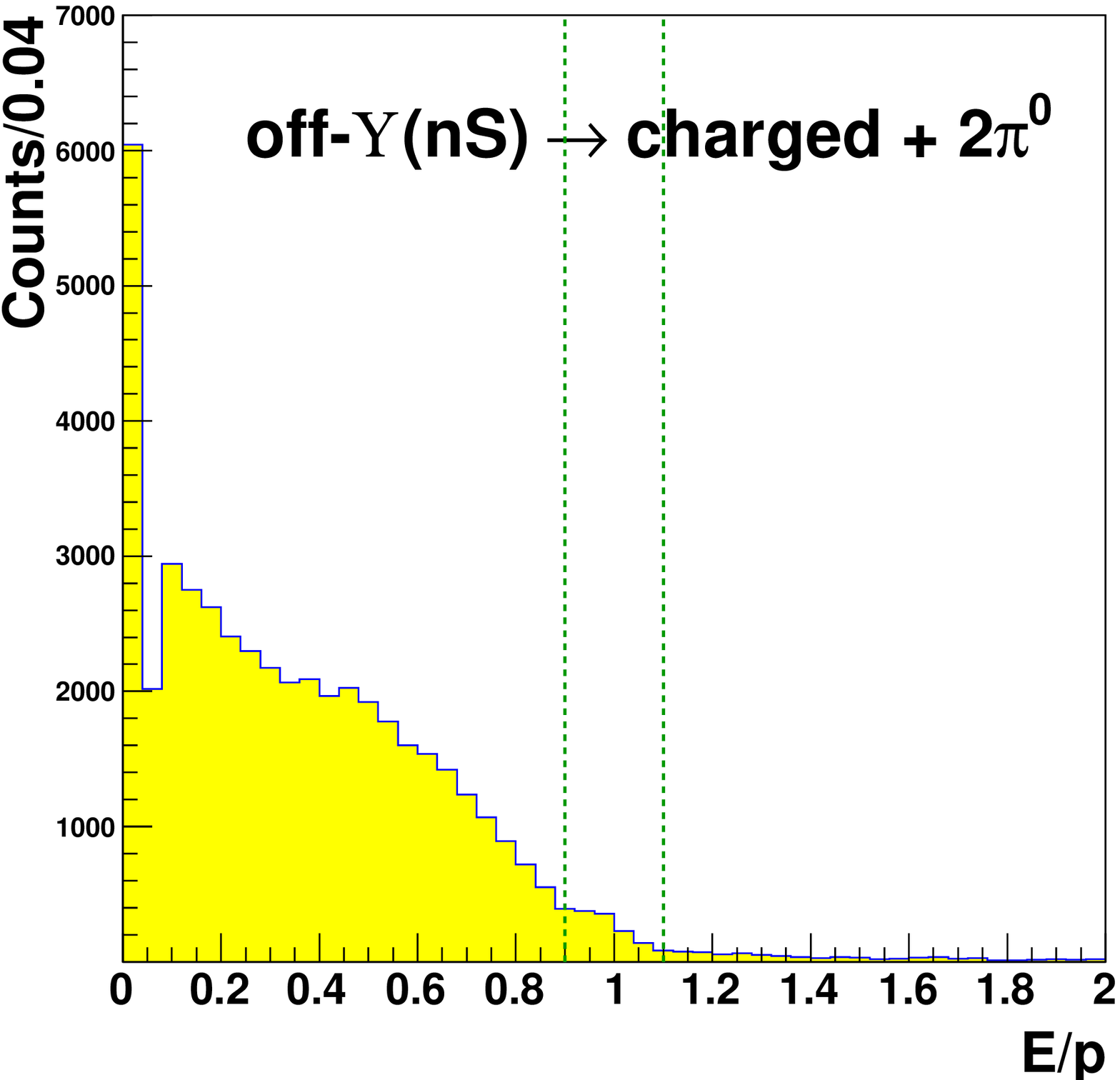}
\end{center}

\caption{Distributions of $E/p$ for reconstructed events for the on--$\Upsilon(1S)$ data (left column), for the on--$\Upsilon(2S)$ data (middle column),  and for the sum of the off--$\Upsilon(1S)$ and off--$\Upsilon(2S)$ data  (right column).  A significant contribution from events containing electrons ($E/p\approx1$) is seen in the all--charged hadrons decays. The vertical lines show the cut of $0.9<E/p<1.1$ used to reject events containing electrons in all cases.}
\label{fig:ep}
\end{figure*}

\begin{figure*}[!p]
\begin{center}
\includegraphics[width=2.in]{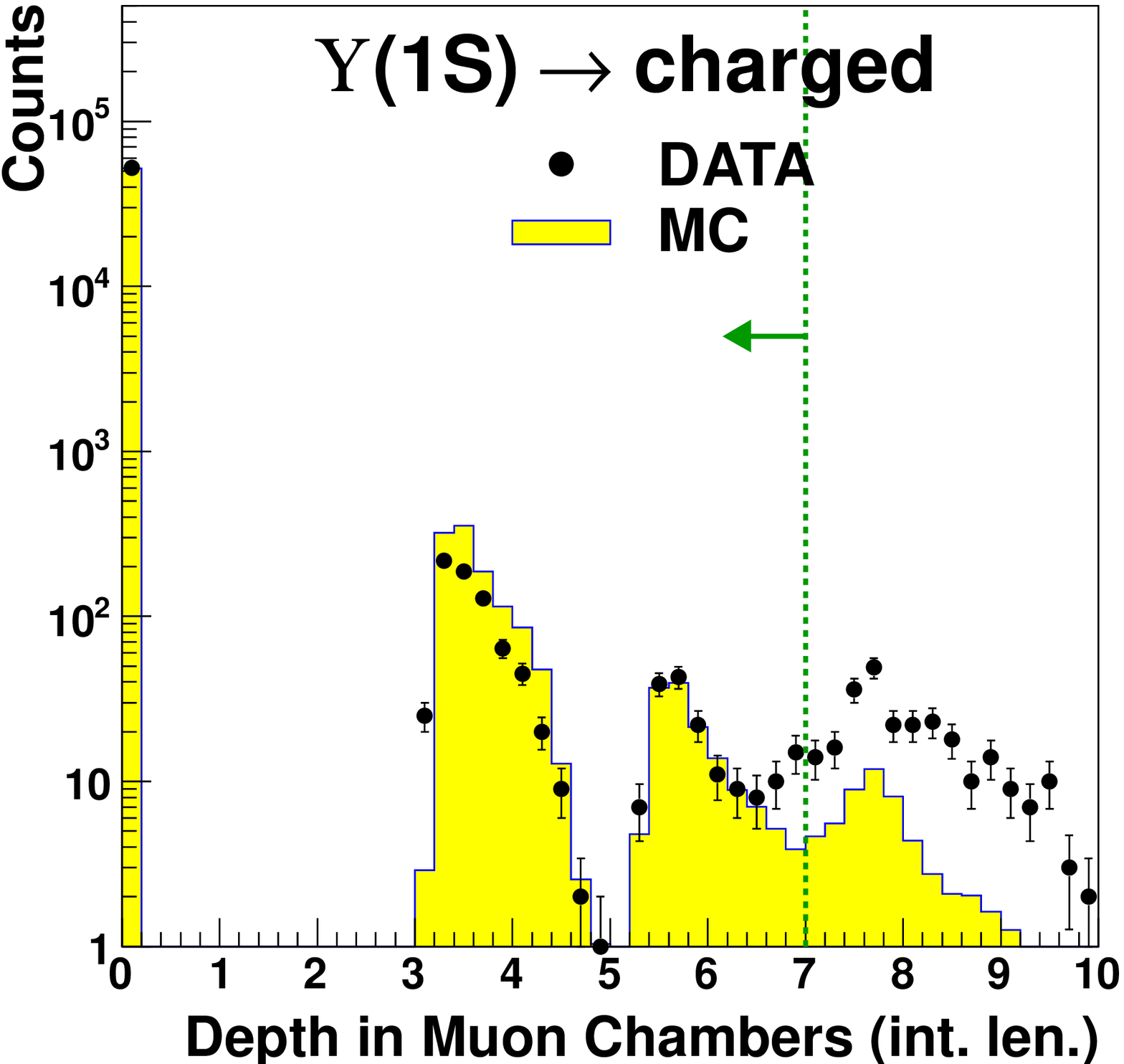}
\includegraphics[width=2.in]{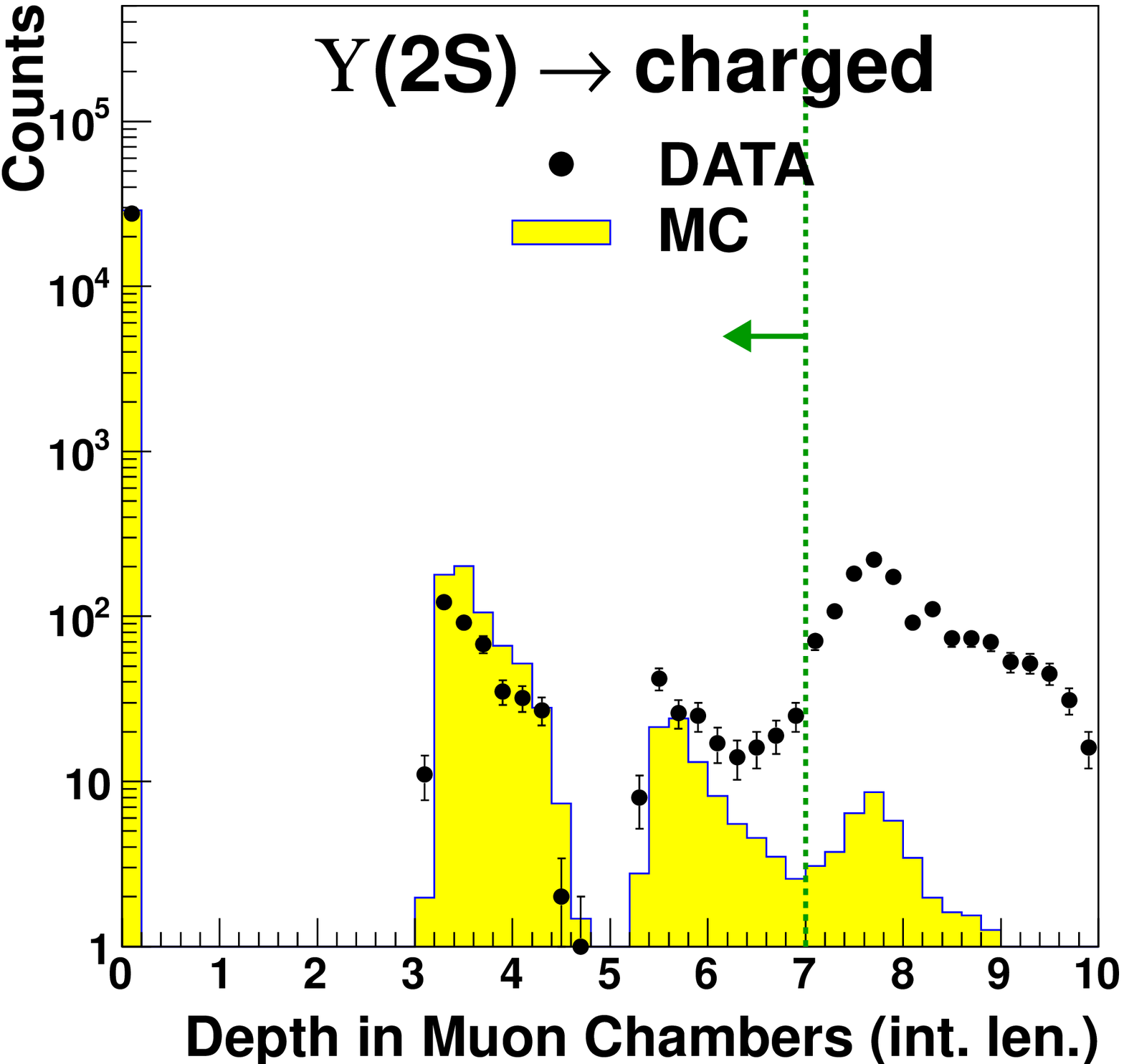}
\includegraphics[width=2.in]{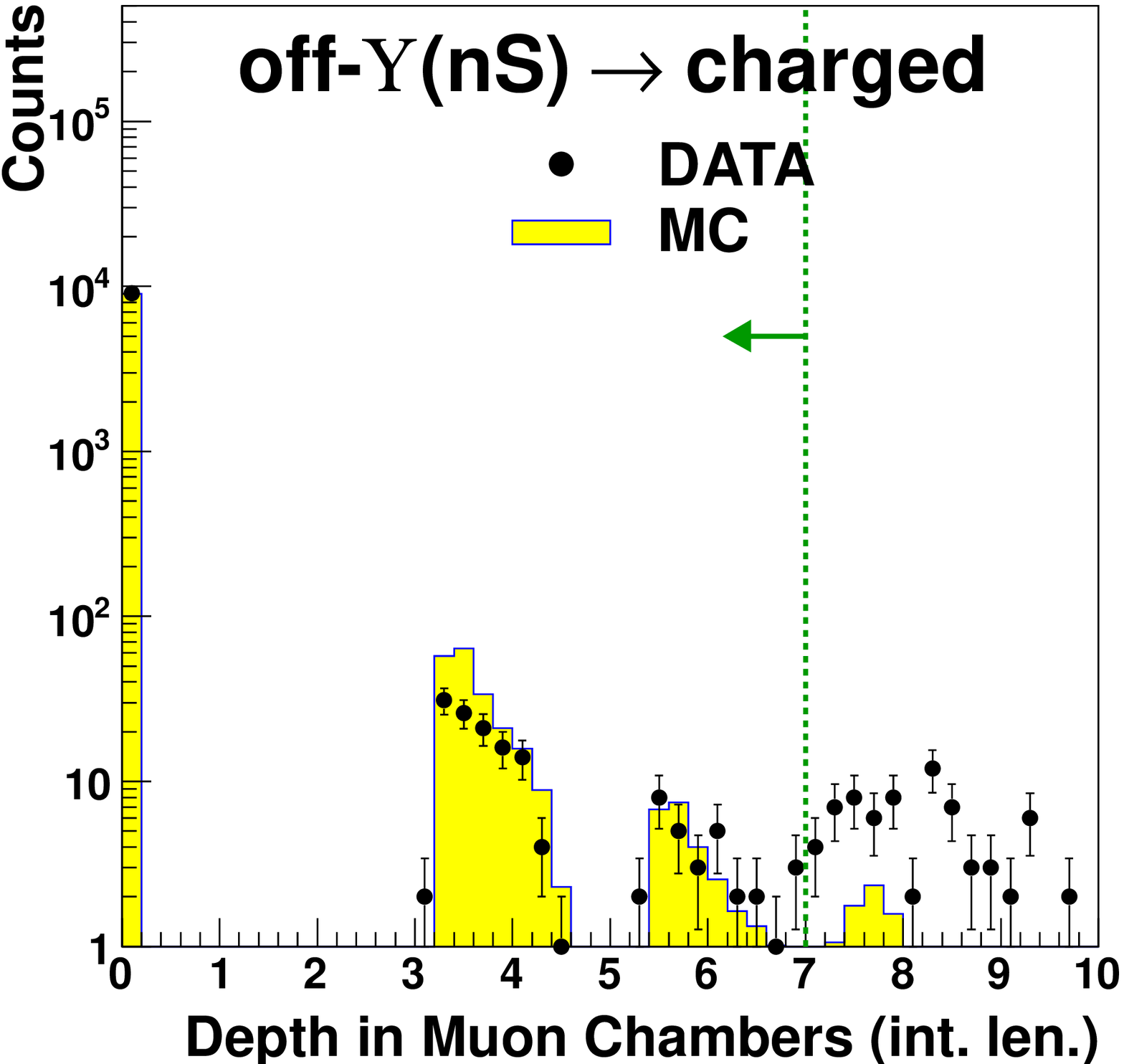}

\includegraphics[width=2.in]{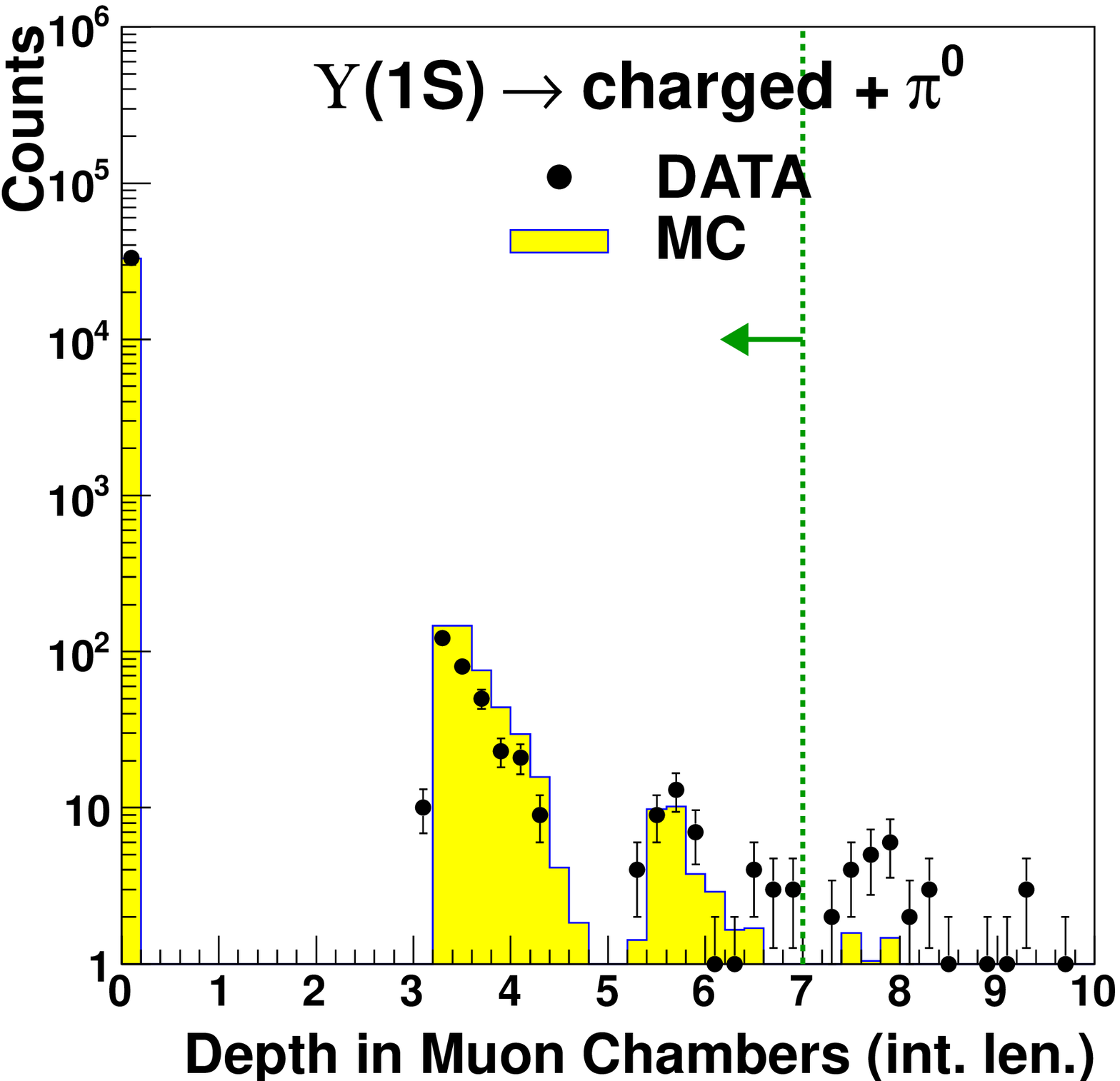}
\includegraphics[width=2.in]{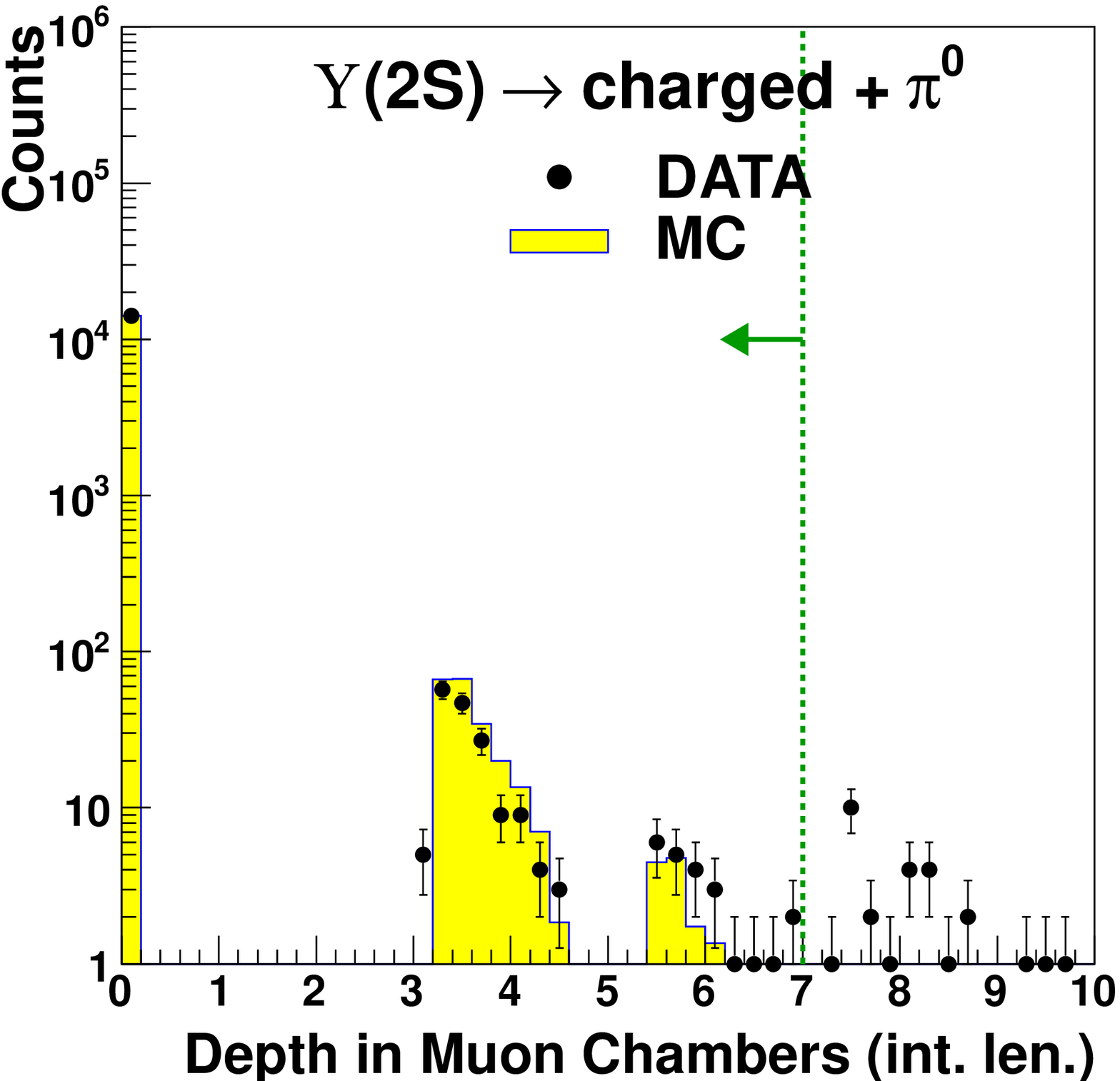}
\includegraphics[width=2.in]{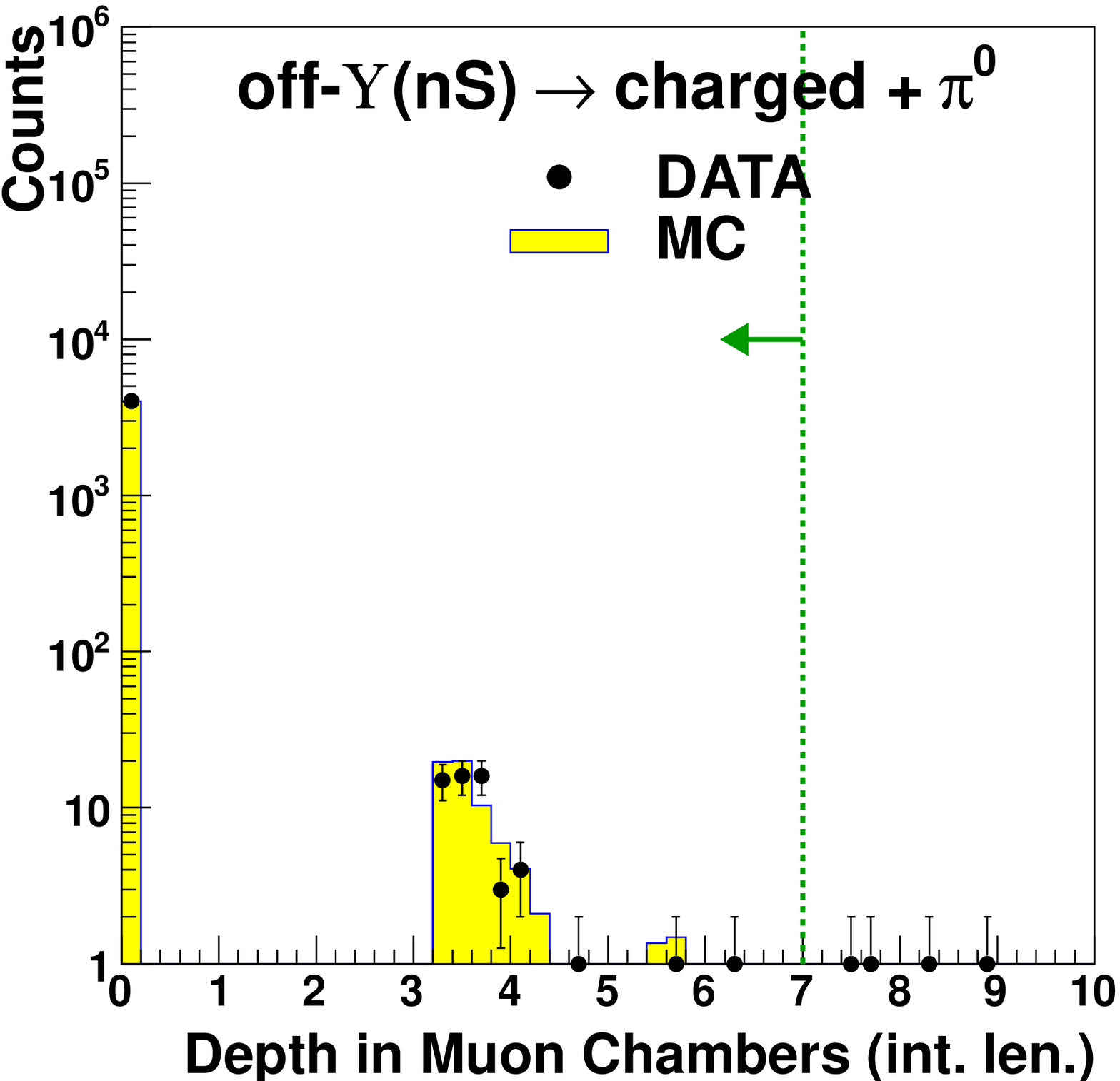}

\includegraphics[width=2.in]{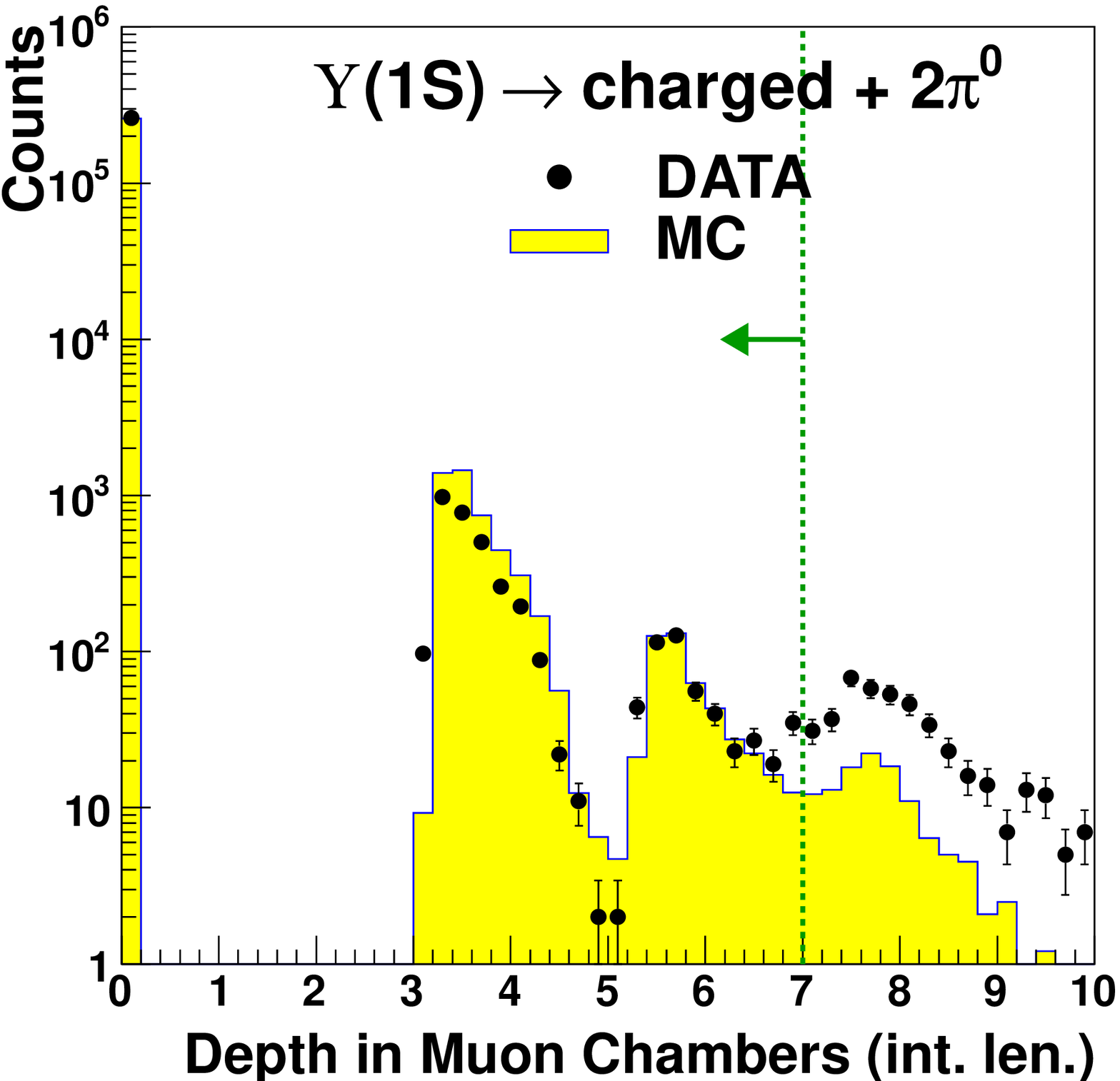}
\includegraphics[width=2.in]{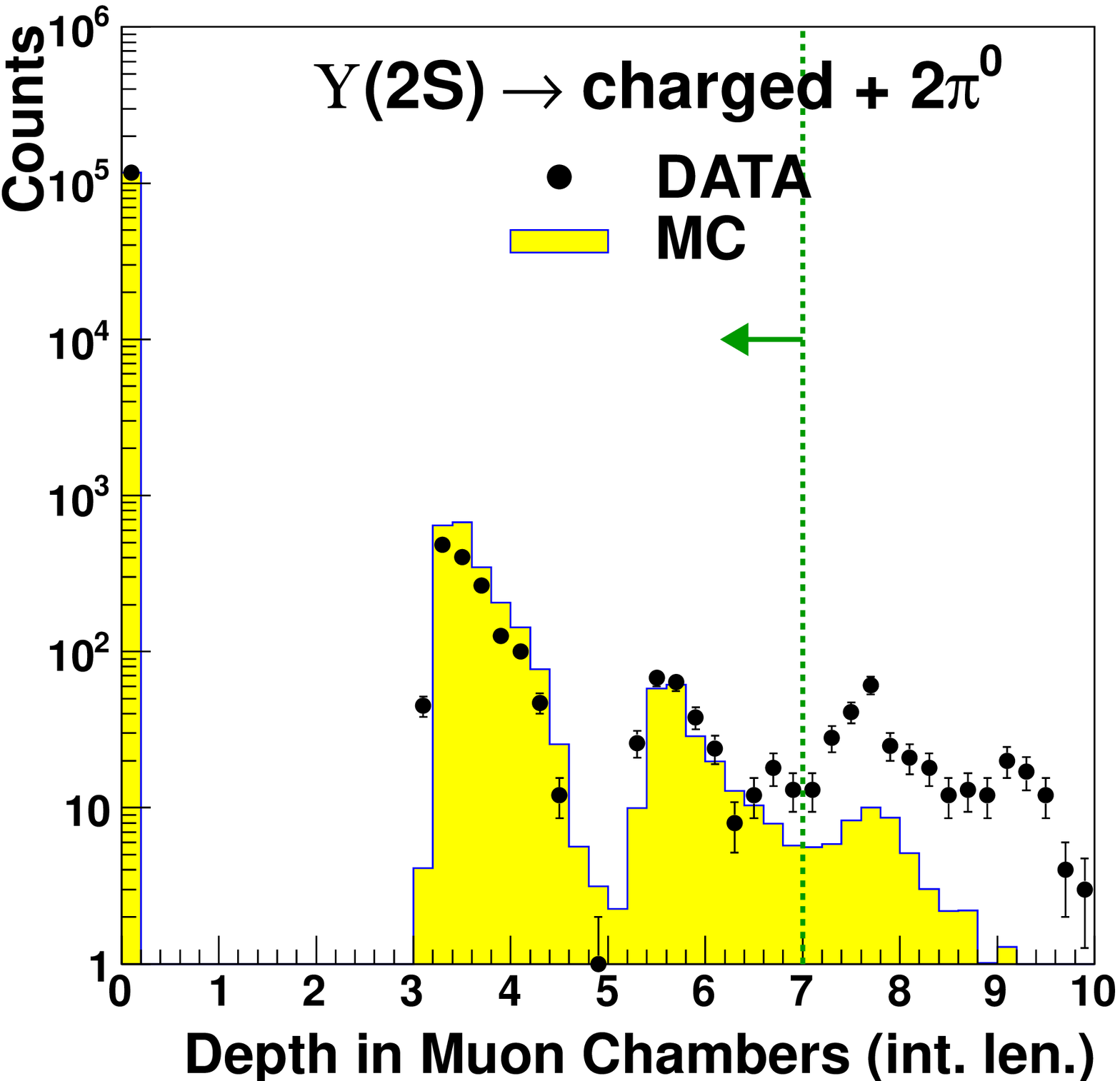}
\includegraphics[width=2.in]{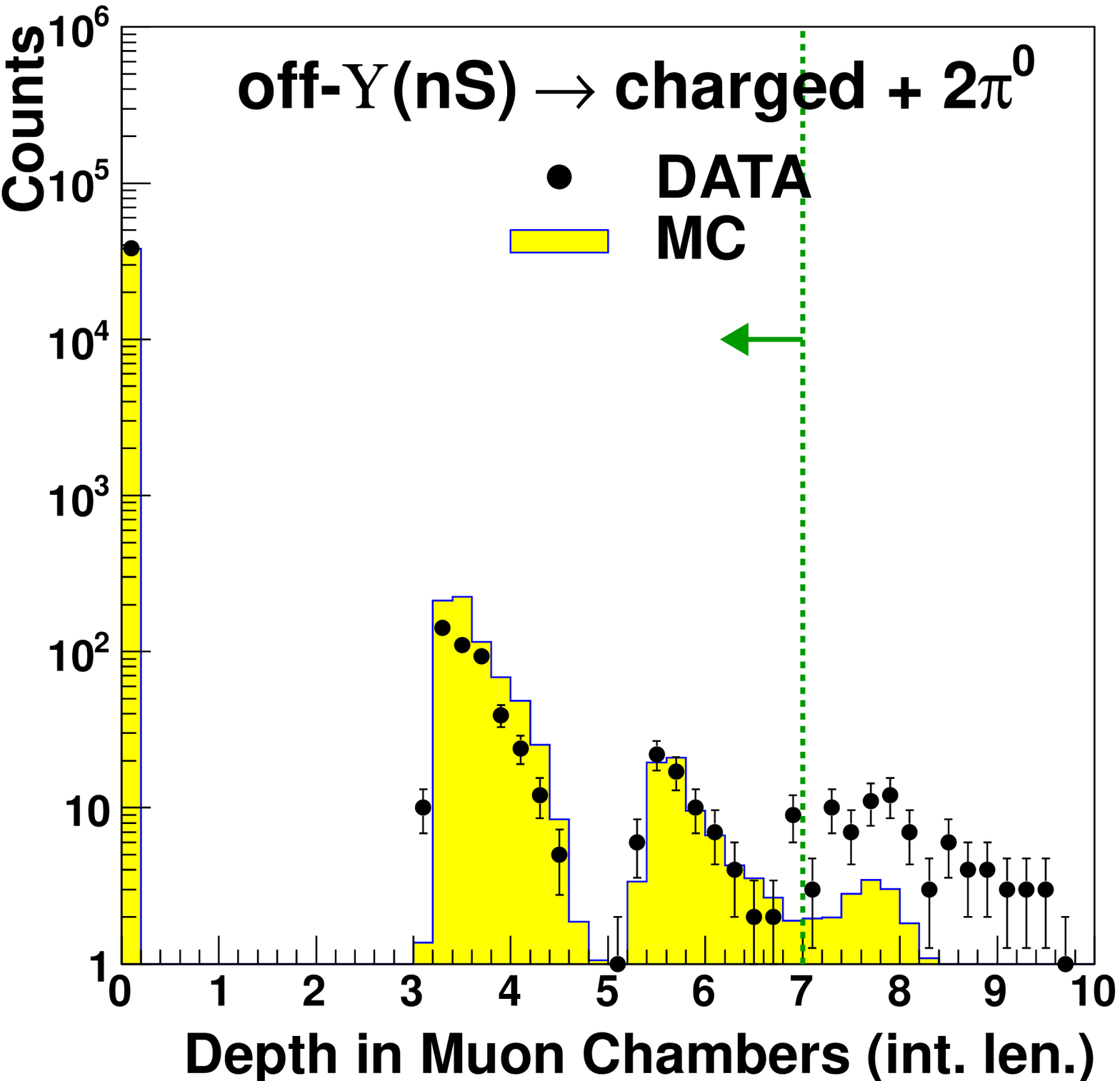}
\end{center}

\caption{Distributions of maximum penetration depth in the muon chambers for reconstructed events from the data  (points) and $\Upsilon(nS)\to\mathrm{hadrons}$~MC (shaded histogram): for the on--$\Upsilon(1S)$ data (left column), for the on--$\Upsilon(2S)$ data (middle column),  and for the sum of the off--$\Upsilon(1S)$ and off--$\Upsilon(2S)$ data  (right column).  In all cases, there is a clear excess in the data over MC predictions at large penetration depth.  The vertical line at a depth of 7 shows the cut used to reject events containing muons. }
\label{fig:mu}
\end{figure*}

Leptons can be generated in these data through a variety of processes, such as higher--order QED processes (e.g. $e^+e^- \to 2(e^+e^-,\mu^+\mu^-)$), weak decays of hadrons such as $D$--mesons, and Dalitz $\pi^0\to e^+e^-\gamma$ decays.

To examine possible electron contamination, we calculate the ratio $E/p$, where $E$ is the energy deposited in the calorimeter, and $p$ is the associated track momentum determined from the drift chambers.  For electrons we expect $E/p\approx1$.  The $E/p$ distributions for the on--resonance and off--resonance $\Upsilon(1S)$ and $\Upsilon(2S)$ data are shown in Fig.~\ref{fig:ep} for hadrons in the range $\Delta M\equiv M(\Upsilon(1S,2S))-M(\mathrm{hadrons})=\pm200$~MeV.  A significant signal at $E/p\approx1$ is only seen for charged--only events, with bare hints in the distributions for final states including $\pi^0$'s.  Since electrons arise from several different sources, we remove them from all decays by rejecting events which contain any track with $0.9<E/p<1.1$.

To study possible muon contamination, we examine the information in the muon detector.  Charged hadrons with $p\gtrsim2$~GeV/$c$  pass through the magnet and into the inner layers of the muon system, but rarely survive to reach the outer layers.  On the other hand, muons tend to reach the outer layers of the muon system.  The distribution of the depths reached in the muon system by charged particles is shown in Fig.~\ref{fig:mu}. For the $\Upsilon(1S)$ and $\Upsilon(2S)$ data, these distributions contain contributions from both hadrons and muons.  However, the MC distributions contains only the hadronic contributions.  In the figure, we see a clear enhancement of data over MC in the outer layers because of the presence of muons.  To remove this contamination, we reject events which have a track reaching $>7$ interaction lengths in the muon system.

The decay $K_S\to\pi^+\pi^-$ is reconstructed by fitting the two charged particle tracks to a common vertex which is required to be displaced from the $e^+e^-$ interaction point by $>3\sigma$. 

The decay $\pi^0\to\gamma\gamma$ is reconstructed using a 1C kinematic fit, constraining the mass of the photon pair to the known $M(\pi^0)$.  The $\pi^0$ candidates are required to have a mass within $3\sigma$ of $M(\pi^0)$.

\begin{figure*}[!p]
\begin{center}
\includegraphics[width=2.5in]{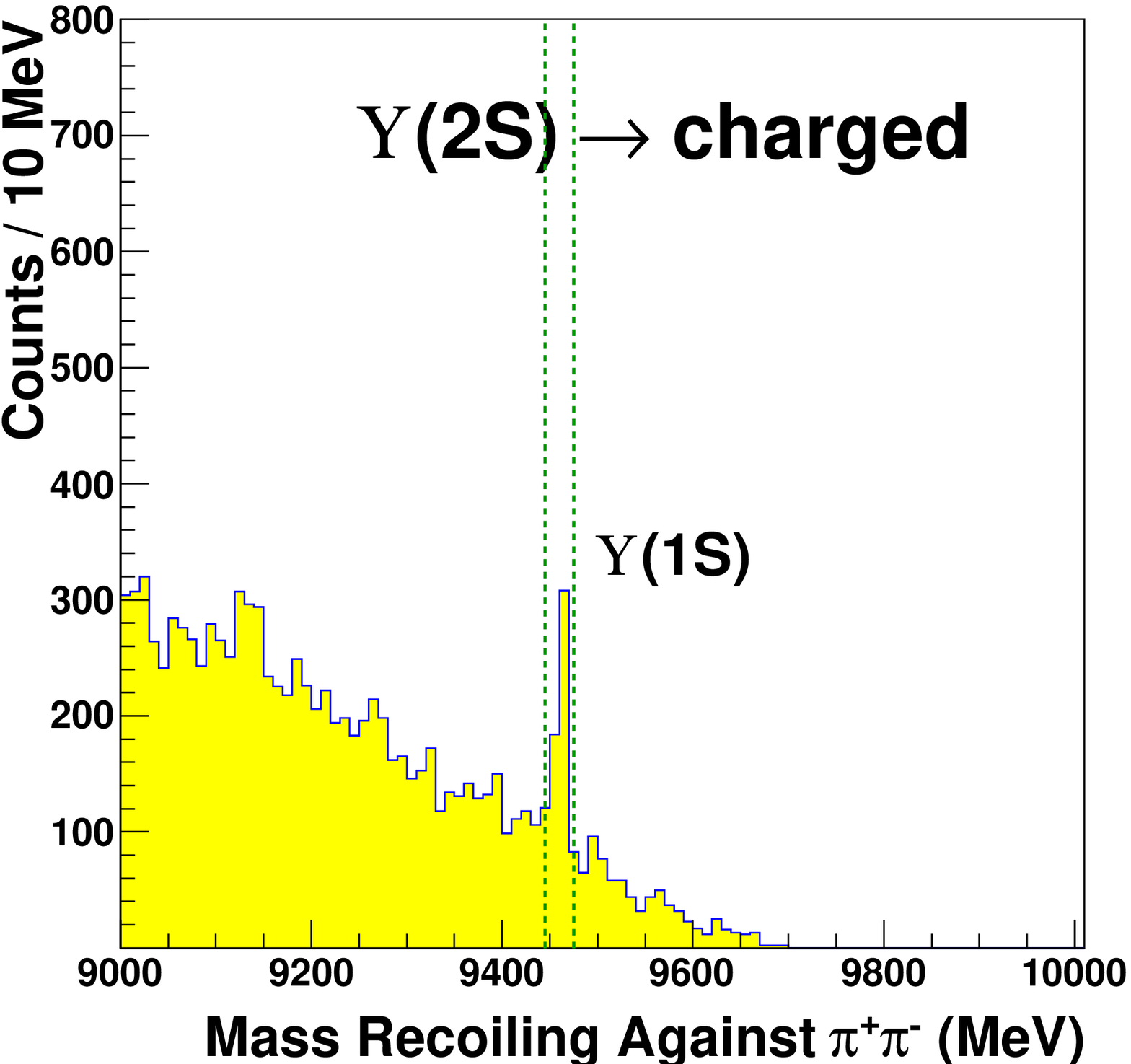}
\includegraphics[width=2.5in]{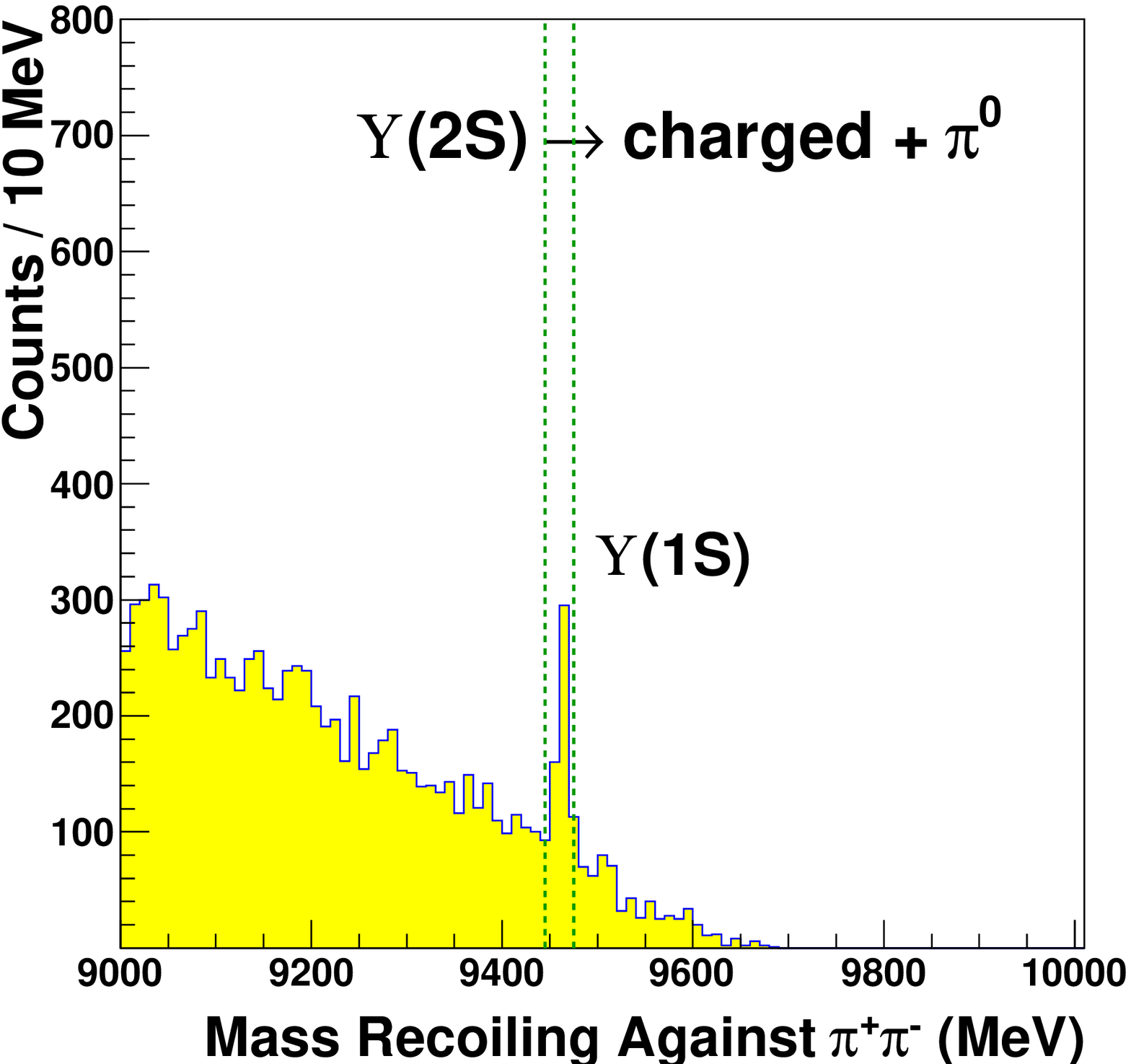}

\includegraphics[width=2.5in]{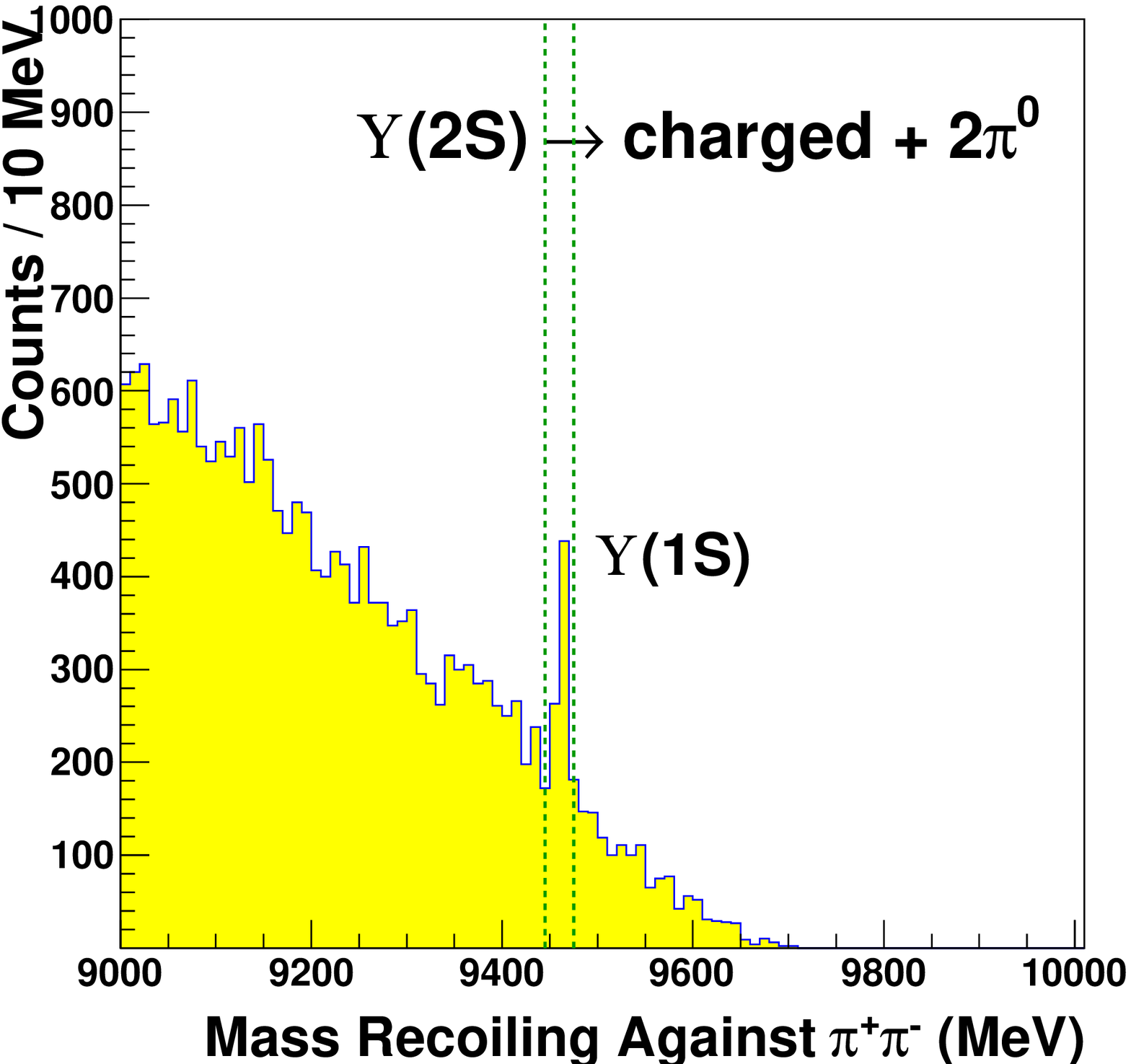}
\includegraphics[width=2.5in]{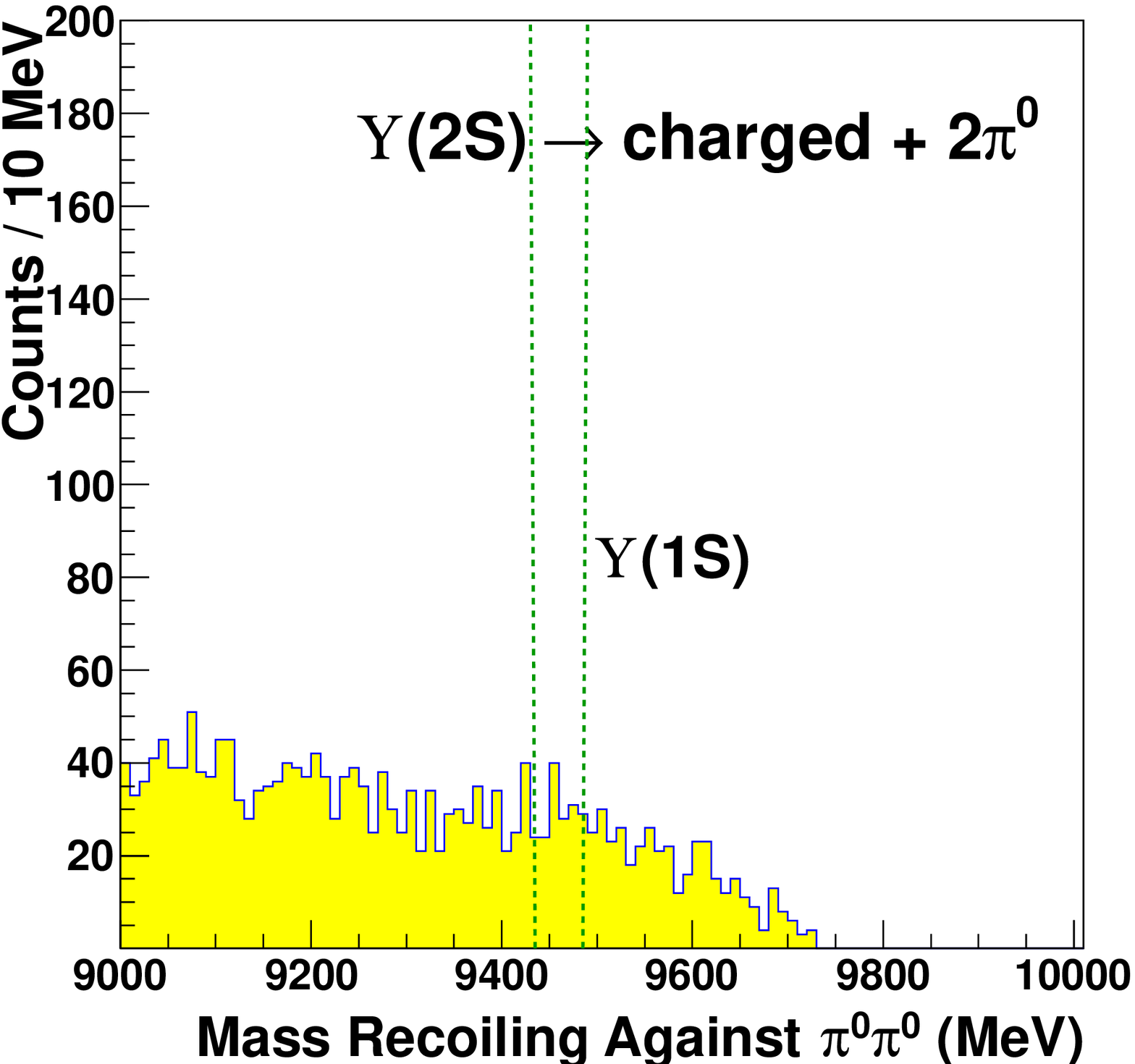}

\end{center}

\caption{Distributions of mass recoiling against pairs of pions in the $\Upsilon(2S)$ data.  In the three spectra with recoil against $\pi^+\pi^-$, signals for the decay $\Upsilon(2S)\to\pi^+\pi^-\Upsilon(1S)$ are seen, and events with a recoil mass of $M(\Upsilon(1S))\pm15$~MeV are rejected, as illustrated by the dashed vertical lines.  In the spectrum with recoils against $\pi^0\pi^0$ (bottom~right), the dashed vertical lines denote the region in which only 15--20 $\Upsilon(1S)$ counts are expected.}
\label{fig:pipirec}
\end{figure*}

\subsection{Kinematic Fits}

To select good, fully reconstructed $\Upsilon(1S,2S)\to\mathrm{hadrons}$ decays, we perform two kinematic fits.

We require all reconstructed hadrons to come from a common vertex and have  $\chi^2_{vtx}/d.o.f.<5$.
We also require the events to satisfy an energy and momentum conservation 4C kinematic fit with  $\chi^2_{FE}/d.o.f.<5$.
If there are multiple candidates for a particular decay in an event, the candidate with the smallest $\chi^2_{FE}$ is kept.

\subsection{Rejection of Dipion Transitions}

In the $\Upsilon(2S)$ data, final states containing pion pairs can have contributions from the decays $\Upsilon(2S)\to(\pi^+\pi^-,\pi^0\pi^0)\Upsilon(1S)$, $\Upsilon(1S)\to\mathrm{hadrons}$.  To identify these contributions, we construct the mass recoiling against $\pi^+\pi^-$ and $\pi^0\pi^0$ pairs.  We show the recoil mass distributions in Fig.~\ref{fig:pipirec}.  

Clear signals for the  $\Upsilon(2S)\to\pi^+\pi^-\Upsilon(1S)$ transitions are seen.
To reject these events in the $\Upsilon(2S)$ data, we reject any events with a $\pi^+\pi^-$ recoil mass of $M(\Upsilon(1S))\pm15$~MeV.

No signal for the $\Upsilon(2S)\to\pi^0\pi^0\Upsilon(1S)$ transitions is seen in recoils against $\pi^0\pi^0$, and no events are rejected.

\section{Results}

\begin{figure*}[!p]
\begin{center}
\includegraphics[width=2.5in]{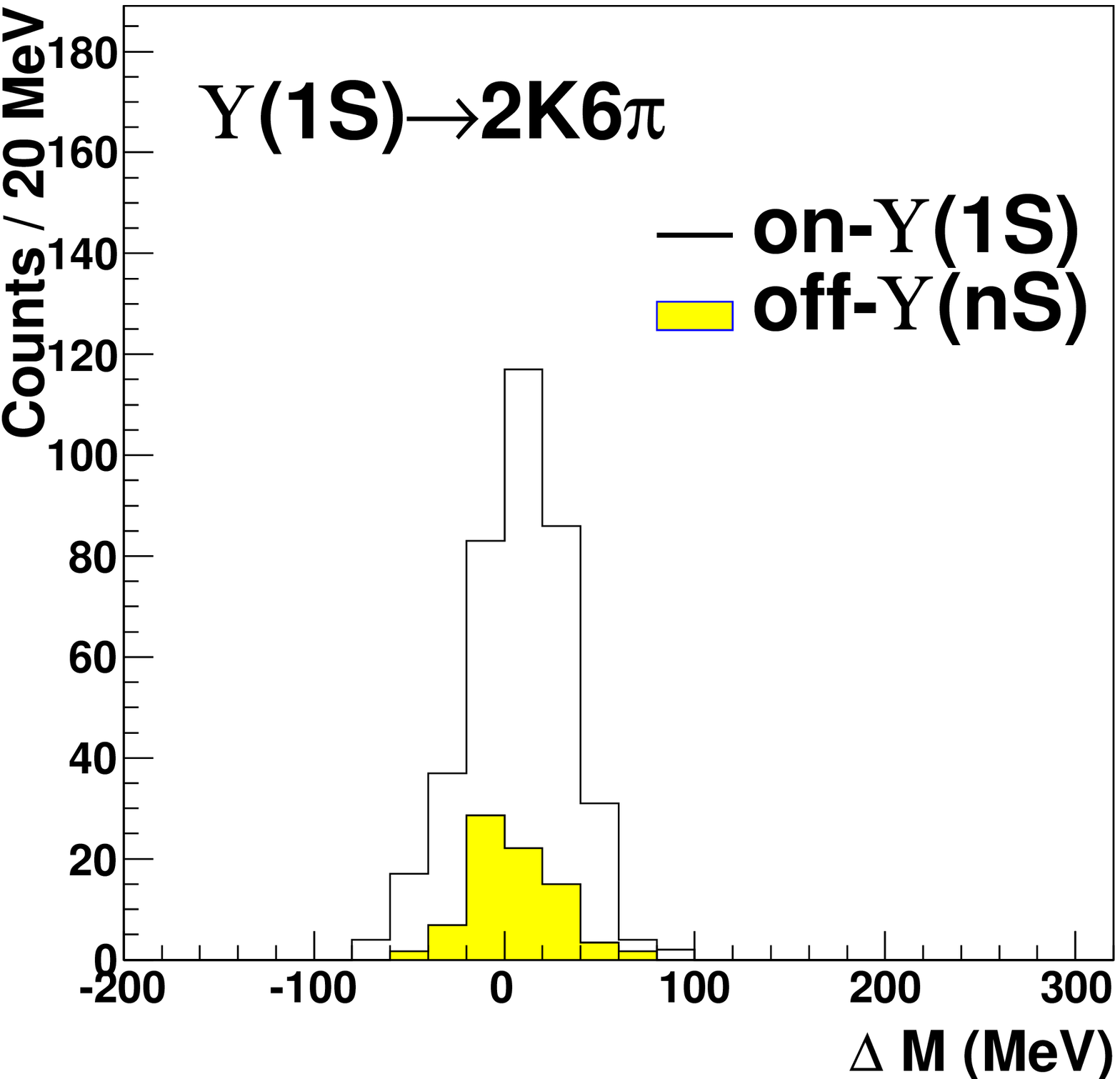}
\includegraphics[width=2.5in]{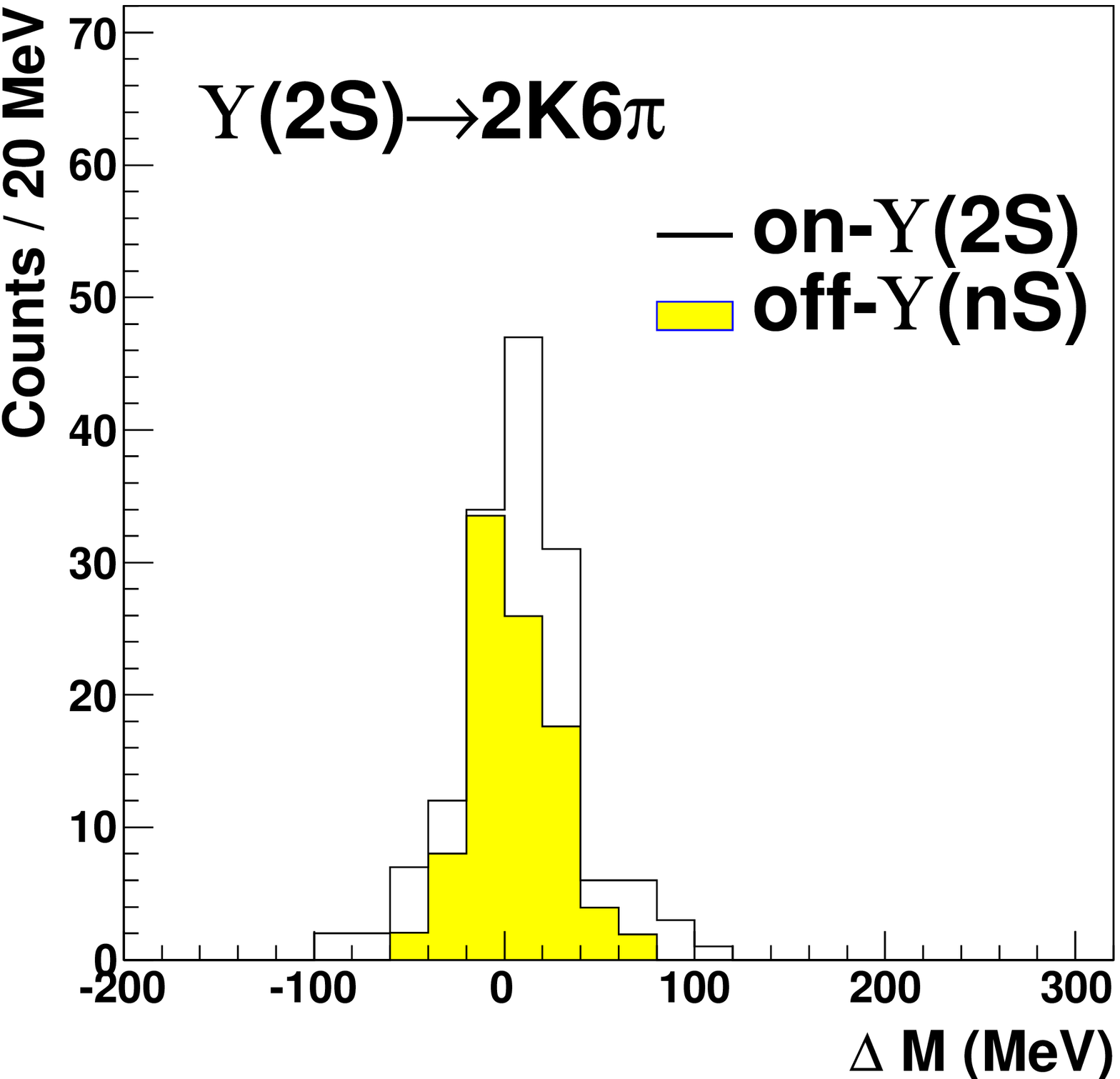}

\includegraphics[width=2.5in]{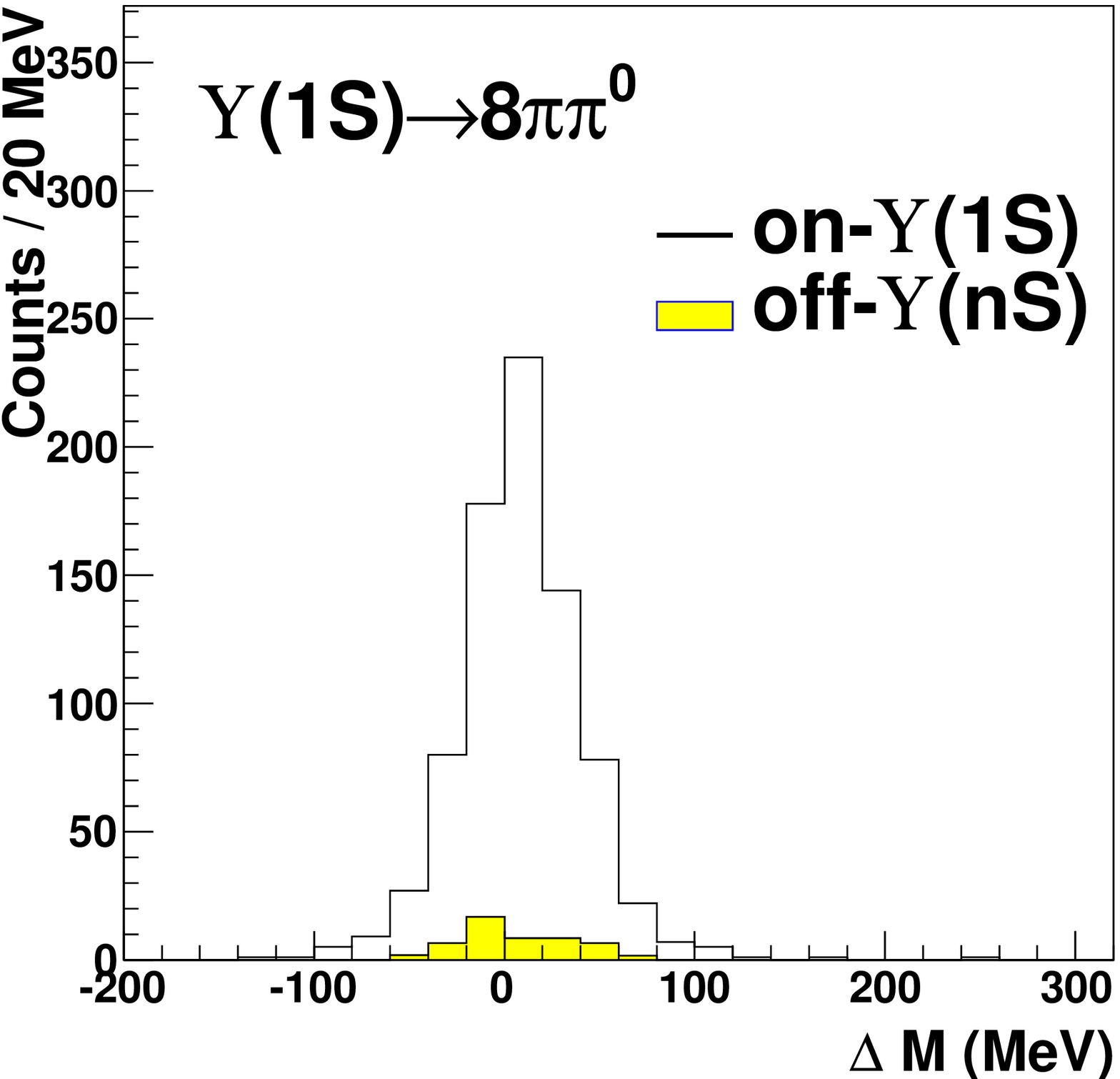}
\includegraphics[width=2.5in]{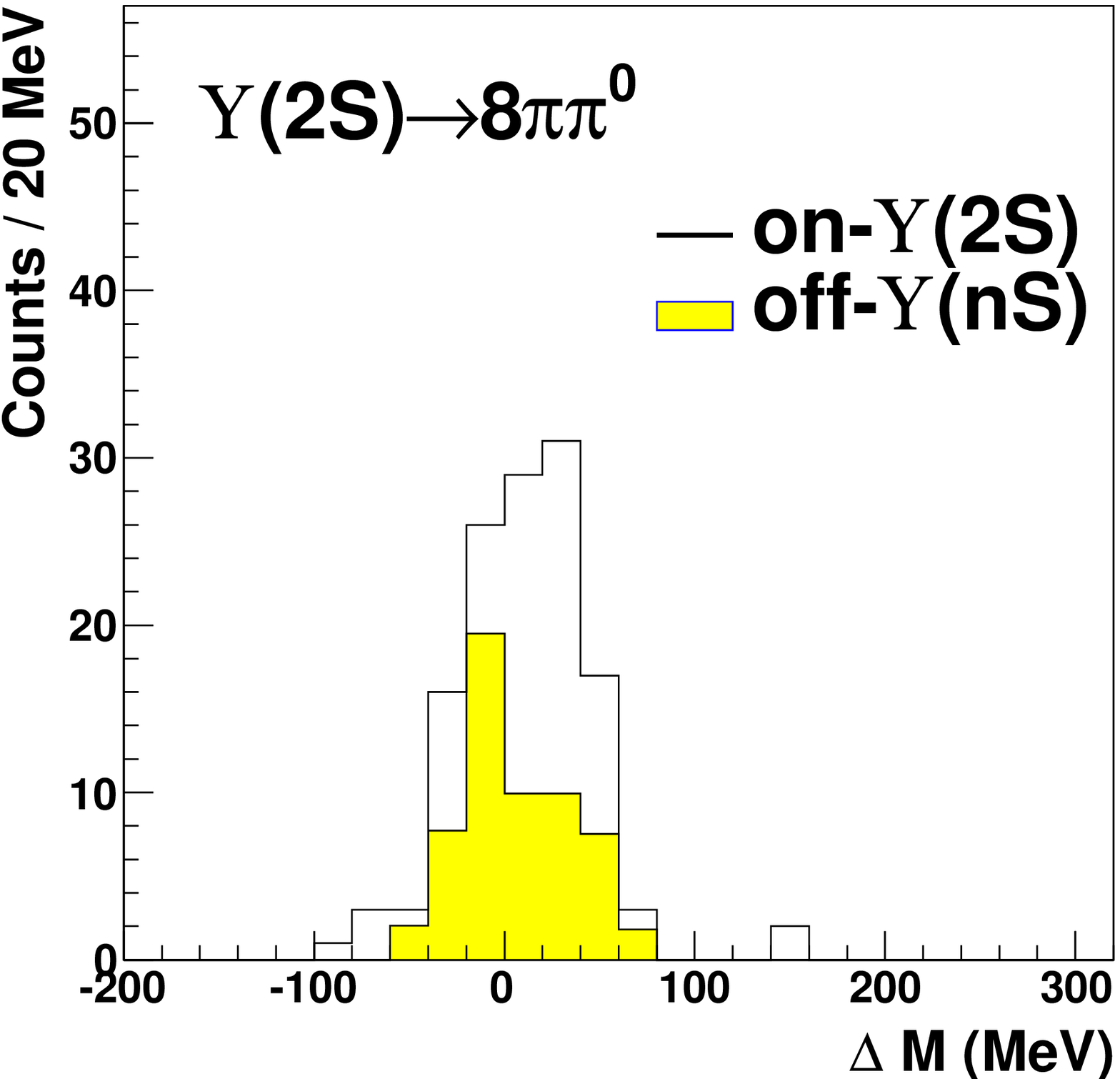}

\includegraphics[width=2.5in]{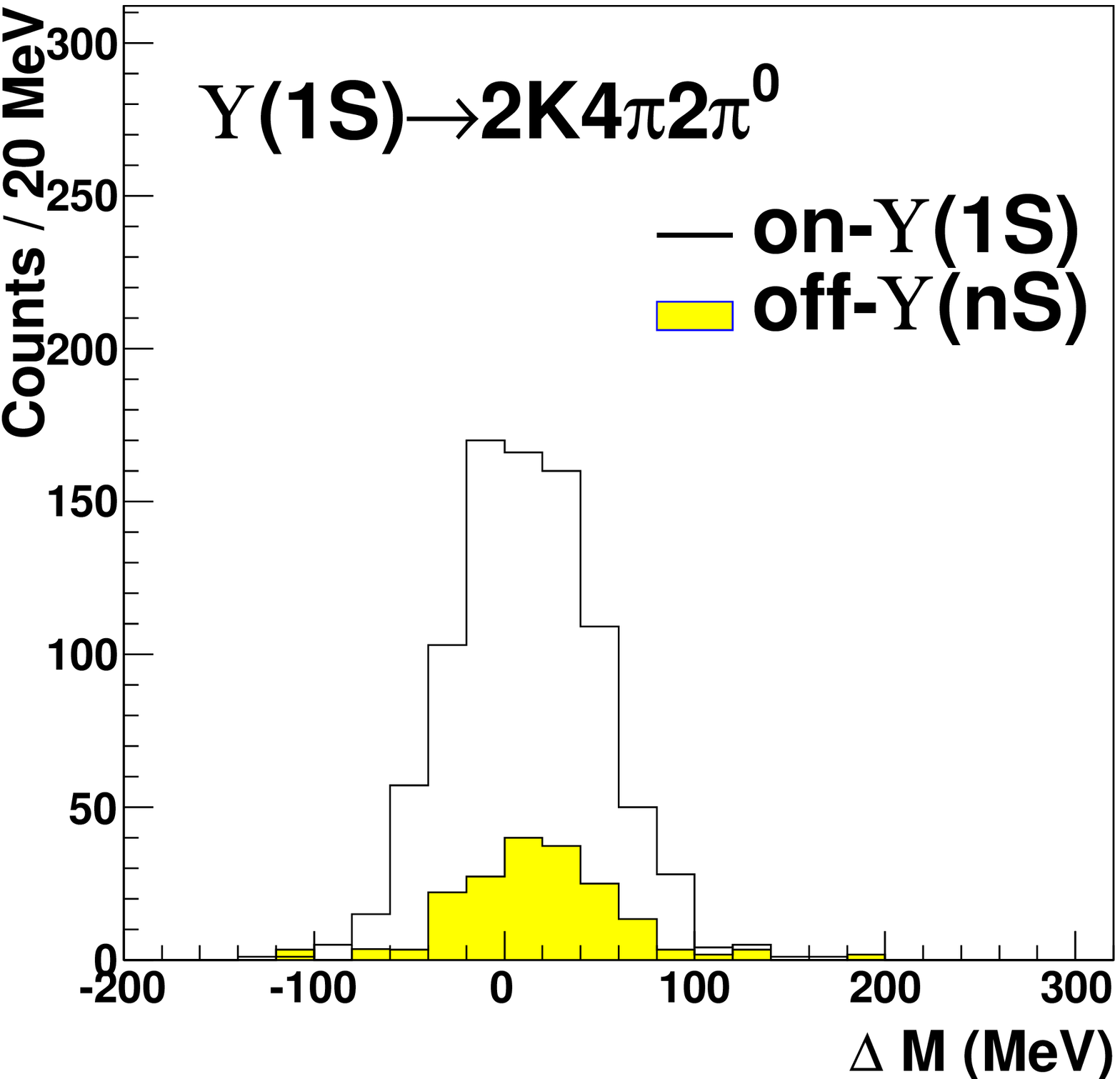}
\includegraphics[width=2.5in]{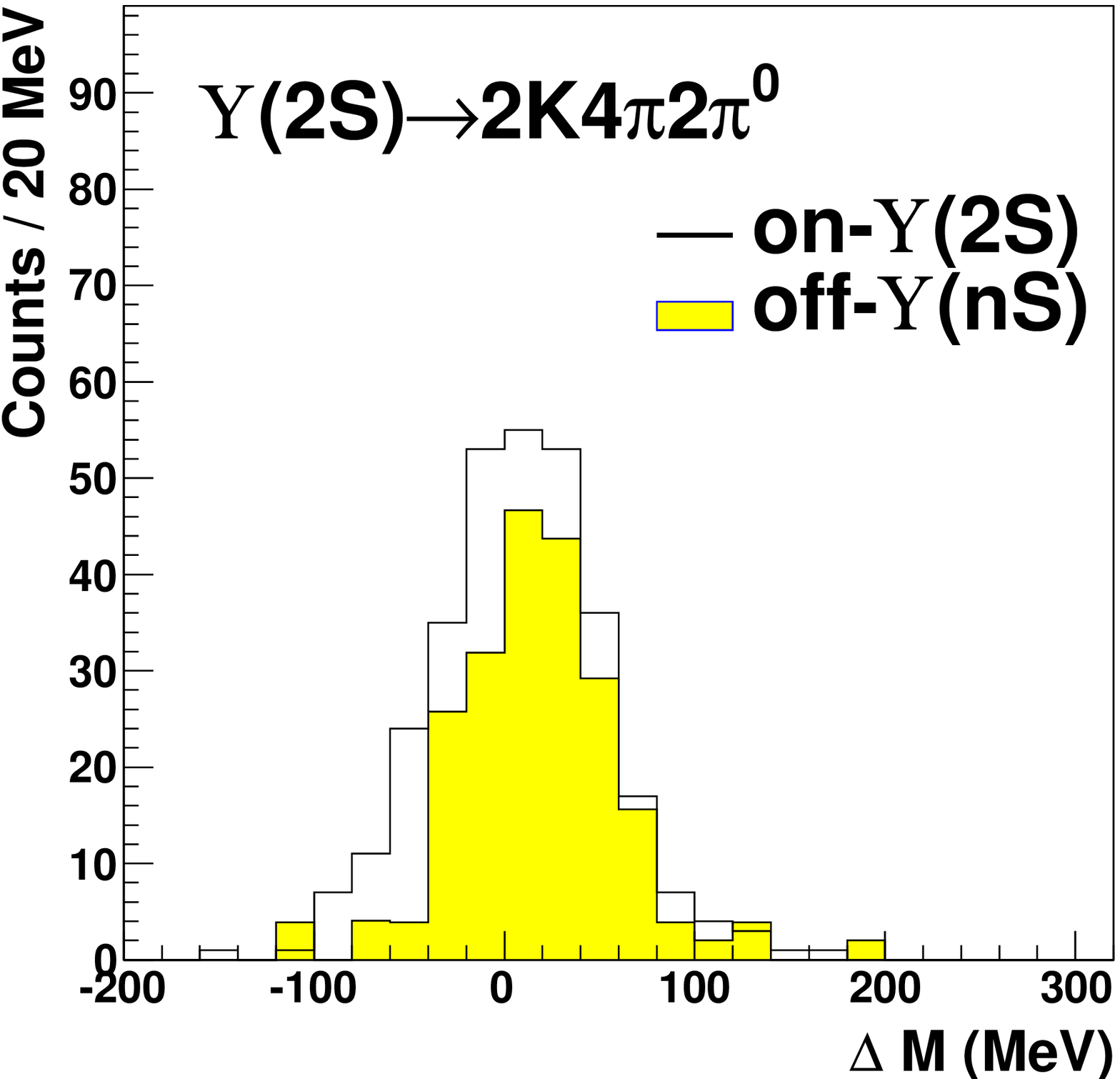}

\end{center}
\caption{ A selection of mass distributions $\Delta M\equiv M(\Upsilon(nS))-M(\mathrm{hadrons})$ for individual decay modes in the $\Upsilon(1S)$ and $\Upsilon(2S)$ data taken on--resonance (open histograms) and off--resonance (shaded histograms).
The off--resonance data have been scaled by the luminosity ratios $\mathcal{R}(1S)=\mathcal{L}_\mathrm{on}(1S)/\mathcal{L}_\mathrm{off} = 1.09/0.63 = 1.73$ for $\Upsilon(1S)$, and $\mathcal{R}(2S)=\mathcal{L}_\mathrm{on}(2S)/\mathcal{L}_\mathrm{off} = 1.28/0.63 = 2.03$ for $\Upsilon(2S)$.}
\label{fig:examplespectra}
\end{figure*}

The invariant mass distributions for all individual decay modes reconstructed in this analysis are shown in Appendix~A, Figs.~A1(a)--(e) for the $\Upsilon(1S)$ data, and Appendix~B, Figs.~B1(a)--(e) for the $\Upsilon(2S)$ data.  Some representative spectra are shown in Fig.~\ref{fig:examplespectra}.
The yields for the on--resonance data are shown as open histograms.  
The yields for the off--resonance data, scaled as described below, are shown as shaded histograms, and are superposed on the on--resonance distributions.

Significant peaks are seen in a large number of modes.  The hadron mass peaks are almost always contained within $\pm100$~MeV around $M(\Upsilon(1S,2S))$, with essentially no tails outside this region.

\clearpage

\subsection{Branching Fraction Calculation}

Hadronic events in the on--resonance $\Upsilon(1S)$ and $\Upsilon(2S)$ data  come from two sources:  resonance decays ($\Upsilon(1S,2S)\to ggg$), and non--resonant $e^+e^-\to q\bar{q}$ decays.  To determine the resonance yield for a given final state $X$, we measure the counts $N(X)_\mathrm{on}$ in the on--resonance data, and subtract from it the non--resonant contribution obtained as the yield observed in the off--resonance data, $N(X)_\mathrm{off}$. 
Because the amount of data taken off--resonance is much smaller than that taken on--resonance, we obtain the best measure of the non--resonant yield  by combining the off--$\Upsilon(1S)$ and off--$\Upsilon(2S)$ data sets, for a total off--resonance luminosity of $\mathcal{L}_\mathrm{off} = 0.63~\mathrm{fb}^{-1}$.  
The observed off--resonance counts $N^{1S}_\mathrm{off}$ and $N^{2S}_\mathrm{off}$ are combined after taking into account the different counting efficiencies $\epsilon(1S)$ and $\epsilon(2S)$, and the variation of the non--resonant cross section as $1/s\equiv 1/E_{cm}^2$.  Thus, the off--resonance counts for the summed luminosity, $\mathcal{L}_\mathrm{off}$, are
\begin{multline}
N(nS)_\mathrm{off} = \left[ \frac{N^{1S}_\mathrm{off}\,E^2(\mathrm{off}\!-\!1S)}{\epsilon(1S)}  \right. \\ 
\left. +  \frac{N^{2S}_\mathrm{off}\,E^2(\mathrm{off}\!-\!2S)}{\epsilon(2S)}  \right]  \frac{\epsilon(nS)}{E^2(\mathrm{off}\!-\!nS)}
\end{multline}
For a decay channel $X$, the branching fractions are then given by
\begin{equation}
\mathcal{B}(\Upsilon(nS) \to X) = \frac{ N(nS)_\mathrm{on} - [ N(nS)_\mathrm{off} \times \mathcal{R}(nS) ] } {\epsilon(nS) \times N(\Upsilon(nS)) } .
\end{equation}
The counts $N(nS)_\mathrm{on}$ and $N(nS)_\mathrm{off}$, and the efficiencies $\epsilon(nS)$ are for the decay $\Upsilon(nS)\to X$.
The total number of Upsilons produced are $N(\Upsilon(1S))=21.5\times10^6$ and $N(\Upsilon(2S))=9.32\times10^6$, and the luminosity ratios  $\mathcal{R}(nS)\equiv\mathcal{L}_\mathrm{on}(nS)/\mathcal{L}_\mathrm{off}$ are $\mathcal{R}(1S)=1.09/0.63 = 1.73$ for $\Upsilon(1S)$, and $\mathcal{R}(2S)=1.28/0.63=2.03$ for $\Upsilon(2S)$.

Because there is essentially no background, event yields in the individual decays, $N(nS)_\mathrm{on}$ and $N(nS)_\mathrm{off}$ are conservatively taken as the total number of counts in the region $\pm200$~MeV around the mass of $\Upsilon(1S,2S)$.

The MC--determined efficiencies for the various decay modes range from $\sim1\%$ to $\sim33\%$. 
The efficiencies for a given mode vary $\lesssim10\%$ between $\Upsilon(1S)$ and $\Upsilon(2S)$.
For each individual decay mode, the counts, $N(nS)_\mathrm{on}$, $N(nS)_\mathrm{off}\times\mathcal{R}(nS)$, $N(nS)_\mathrm{res}=N(nS)_\mathrm{on}-N(nS)_\mathrm{off}\times\mathcal{R}(nS)$, and the efficiencies, $\epsilon(nS)$,  are listed in Tables~I(a,b,c) and II(a,b,c).

The branching fraction results for individual modes are given in Table~I for $\Upsilon(1S)$ and Table~II for $\Upsilon(2S)$. 
The first uncertainties in the branching fractions are statistical, and the second uncertainties are systematic, as described in Sec.~6.
Upper limits at 90\% confidence level are also given for those which have a significance of $<2\sigma$.

As listed in Table~I(a,b,c), 73 different decay modes $\Upsilon(1S)$ are found to have branching fractions ranging from $0.3\times10^{-5}$ to $110\times10^{-5}$.  Upper limits at 90\% confidence level for the other 27 decay modes range from $0.2\times10^{-5}$ to $3.6\times10^{-5}$.  As listed in Table~II(a,b,c), resonance yields for $\Upsilon(2S)$ decays are generally smaller than those for $\Upsilon(1S)$, and only 17 decays have significantly large branching fractions.  These range from $1.3\times10^{-5}$ to $36\times10^{-5}$.  The upper limits for the other 83 decays range from $0.2\times10^{-5}$ to $34\times10^{-5}$.

\addtocounter{table}{1}

\begin{table*}[!p]

\flushleft

TABLE I(a): Branching fractions for $\Upsilon(1S)\to\mathrm{only~charged~hadrons}$. $N(1S)_\mathrm{res}\equiv N(1S)_\mathrm{on} - N(1S)_\mathrm{off} \times \mathcal{R}(1S)$, where $\mathcal{R}(1S)=\mathcal{L}_\mathrm{on}(1S)/\mathcal{L}_\mathrm{off} = 1.73$, and $\epsilon(1S)$ are the MC--calculated efficiencies.  Upper limits (UL) at 90\% confidence level are also given for modes with branching fractions which have a significance of $<2\sigma$.  
The modes marked with asterisks are used to construct the $\Upsilon(1S)/\Upsilon(2S)$ ratio.

\begin{ruledtabular}
\begin{tabular}{rlD{,}{\:\pm\:}{-1}D{,}{\:\pm\:}{-1}D{,}{\:\pm\:}{-1}dcc}
\# & modes & \multicolumn{1}{c}{$N(1S)_\mathrm{on}$} & \multicolumn{1}{c}{$N(1S)_\mathrm{off}\times\mathcal{R}(1S)$} & \multicolumn{1}{c}{$N(1S)_\mathrm{res}$} & \multicolumn{1}{c}{$\epsilon(1S) (\%)$} & $\mathcal{B}(1S)\times10^5$ & $\mathrm{UL}\times10^5$  \\
\hline

  1 & $           4\pi$ & 168 , 13 & 92.7 , 12.7  & 75.3 , 18.1 &  31.56 & $  1.11\pm 0.27\pm 0.13$ & --- \\
  2 & $           6\pi$ & 400 , 20 & 183.8 , 17.9  & 216.2 , 26.8 &  19.40 & $  5.18\pm 0.64\pm 0.69$ & --- \\
  3 & $           8\pi$ & 377 , 19 & 161.5 , 16.7  & 215.5 , 25.6 &  11.46 & $  8.74\pm 1.04\pm 1.33$ & --- \\
  4 & $          10\pi$ & 131 , 11 & 48.5 , 9.2  & 82.5 , 14.7 &   6.19 & $  6.20\pm 1.10\pm 1.08$ & --- \\
\\[-6pt]

  5 & $         2K2\pi$ & 116 , 11 & 36.2 , 7.9  & 79.8 , 13.4 &  26.92 & $  1.38\pm 0.23\pm 0.17$ & --- \\
  6 & $         2K4\pi$ & 414 , 20 & 127.8 , 14.9  & 286.2 , 25.2 &  17.20 & $  7.74\pm 0.68\pm 1.10$ & --- \\
  $*\:$7 & $         2K6\pi$ & 381 , 20 & 84.0 , 12.1  & 297.0 , 22.9 &   9.91 & $ 13.93\pm 1.08\pm 2.28$ & --- \\
  8 & $         2K8\pi$ & 179 , 13 & 31.9 , 7.4  & 147.1 , 15.3 &   5.14 & $ 13.30\pm 1.38\pm 2.48$ & --- \\
\\[-6pt]

  9 & $             4K$ &  36 ,  6 &  9.4 , 4.2  & 26.6 , 7.3 &  23.07 & $  0.54\pm 0.15\pm 0.07$ & --- \\
 10 & $         4K2\pi$ & 112 , 11 & 33.9 , 7.7  & 78.1 , 13.1 &  15.01 & $  2.42\pm 0.40\pm 0.37$ & --- \\
 11 & $         4K4\pi$ & 133 , 12 & 28.2 , 7.0  & 104.8 , 13.5 &   8.25 & $  5.91\pm 0.76\pm 1.04$ & --- \\
 12 & $         4K6\pi$ &  59 ,  8 &  8.7 , 4.8  & 50.3 , 9.1 &   3.93 & $  5.94\pm 1.07\pm 1.19$ & --- \\
\\[-6pt]

 13 & $             6K$ &   5 ,  3 &  4.5 , 2.8 &  0.5 , 4.0 &  13.05 & $  0.02\pm 0.14\pm 0.01$ & $<  0.23$ \\
 14 & $         6K2\pi$ &  13 ,  4 &  1.7 , 1.7  & 11.3 , 4.0 &   6.64 & $  0.79\pm 0.28\pm 0.15$ & --- \\
 15 & $         6K4\pi$ &   6 ,  3 &  1.7 , 1.7 &  4.3 , 3.5 &   3.03 & $  0.65\pm 0.54\pm 0.14$ & $<  1.90$ \\
\\[-6pt]

 16 & $             8K$ &   1 ,  2 & \multicolumn{1}{c}{0} &  1.0 , 2.8 &   5.17 & $  0.09\pm 0.25\pm 0.02$ & $<  0.49$ \\
 17 & $         8K2\pi$ & \multicolumn{1}{c}{0}  & \multicolumn{1}{c}{0} & \multicolumn{1}{c}{---} &   2.18 & --- & $<  0.68$ \\
\\[-6pt]

 18 & $            10K$ & \multicolumn{1}{c}{0}  & \multicolumn{1}{c}{0} & \multicolumn{1}{c}{---} &   1.42 & --- & $<  1.06$ \\
\\[-6pt]

 19 & $         K_{S}K\pi$ &   9 ,  4 &  3.5 , 3.9 &  5.5 , 5.4 &  18.49 & $  0.14\pm 0.14\pm 0.02$ & $<  0.34$ \\
 20 & $        K_{S}K3\pi$ & 145 , 12 & 41.2 , 8.5  & 103.8 , 14.7 &  12.36 & $  3.90\pm 0.55\pm 0.53$ & --- \\
 21 & $        K_{S}K5\pi$ & 231 , 15 & 50.5 , 9.4  & 180.5 , 17.8 &   6.67 & $ 12.58\pm 1.24\pm 1.89$ & --- \\
 22 & $        K_{S}K7\pi$ & 142 , 12 & 21.2 , 6.1  & 120.8 , 13.4 &   3.35 & $ 16.77\pm 1.86\pm 3.00$ & --- \\
\\[-6pt]

 23 & $        K_{S}3K\pi$ &  43 ,  7 &  9.0 , 4.5  & 34.0 , 8.0 &  10.97 & $  1.44\pm 0.34\pm 0.21$ & --- \\
 24 & $       K_{S}3K3\pi$ & 121 , 11 & 22.9 , 6.3  & 98.1 , 12.7 &   5.77 & $  7.90\pm 1.02\pm 1.27$ & --- \\
 25 & $       K_{S}3K5\pi$ &  71 ,  8 & 14.1 , 5.6  & 56.9 , 10.1 &   2.66 & $  9.94\pm 1.76\pm 1.89$ & --- \\
\\[-6pt]

 26 & $        K_{S}5K\pi$ &   7 ,  3 & \multicolumn{1}{c}{0}  &  7.0 , 3.9 &   4.56 & $  0.71\pm 0.40\pm 0.12$ & $<  1.54$ \\
 27 & $       K_{S}5K3\pi$ &  13 ,  4 & \multicolumn{1}{c}{0}  & 13.0 , 4.2 &   2.07 & $  2.92\pm 0.94\pm 0.59$ & --- \\
\\[-6pt]

 28 & $        K_{S}7K\pi$ &  \multicolumn{1}{c}{0} & \multicolumn{1}{c}{0}  & \multicolumn{1}{c}{---} &   1.44 & --- & $<  1.00$ \\
\\[-6pt]

 29 & $         2p2\pi$ &  44 ,  7 & 14.8 , 4.8  & 29.2 , 8.2 &  31.49 & $  0.43\pm 0.12\pm 0.05$ & --- \\
 30 & $         2p4\pi$ & 156 , 12 & 41.8 , 8.5  & 114.2 , 15.1 &  19.27 & $  2.76\pm 0.36\pm 0.39$ & --- \\
 31 & $         2p6\pi$ & 212 , 15 & 43.7 , 8.7  & 168.3 , 17.0 &  11.49 & $  6.81\pm 0.69\pm 1.11$ & --- \\
 $*\:$32 & $         2p8\pi$ & 109 , 10 &  7.2 , 4.5  & 101.8 , 11.4 &   6.15 & $  7.69\pm 0.86\pm 1.44$ & --- \\
\\[-6pt]

 33 & $       2p2K2\pi$ &  89 ,  9 & 16.3 , 5.9  & 72.7 , 11.1 &  17.13 & $  1.97\pm 0.30\pm 0.31$ & --- \\
 34 & $       2p2K4\pi$ & 129 , 11 & 12.5 , 5.3  & 116.5 , 12.5 &   9.50 & $  5.70\pm 0.61\pm 1.01$ & --- \\
 35 & $       2p2K6\pi$ &  66 ,  8 &  5.4 , 3.8  & 60.6 , 9.0 &   4.81 & $  5.86\pm 0.87\pm 1.17$ & --- \\
\\[-6pt]

 36 & $           2p4K$ &   2 ,  2 &  1.7 , 3.0 &  0.3 , 3.8 &  15.01 & $  0.01\pm 0.12\pm 0.01$ & $<  0.15$ \\
 37 & $       2p4K2\pi$ &  13 ,  4 &  1.8 , 3.0  & 11.2 , 4.7 &   7.82 & $  0.66\pm 0.28\pm 0.13$ & --- \\
 38 & $       2p4K4\pi$ &  10 ,  3 &  \multicolumn{1}{c}{0} & 10.0 , 3.8 &   3.59 & $  1.29\pm 0.49\pm 0.28$ & --- \\
\\[-6pt]

 39 & $           2p6K$ & \multicolumn{1}{c}{0}  &  \multicolumn{1}{c}{0} & \multicolumn{1}{c}{---} &   6.03 & --- & $<  0.24$ \\
 40 & $       2p6K2\pi$ & \multicolumn{1}{c}{0}  &  \multicolumn{1}{c}{0} & \multicolumn{1}{c}{---} &   2.96 & --- & $<  0.50$ \\
\\[-6pt]

 41 & $           2p8K$ & \multicolumn{1}{c}{0}  & \multicolumn{1}{c}{0}  & \multicolumn{1}{c}{---} &   1.54 & --- & $<  0.98$ \\
\\[-6pt]

 42 & $         4p2\pi$ &  13 ,  4 & \multicolumn{1}{c}{0} & 13.0 , 4.2 &  19.70 & $  0.31\pm 0.10\pm 0.04$ & --- \\
 43 & $         4p4\pi$ &  14 ,  4 & \multicolumn{1}{c}{0} & 14.0 , 4.3 &  11.64 & $  0.56\pm 0.17\pm 0.10$ & --- \\
 44 & $         4p6\pi$ &   5 ,  3 & \multicolumn{1}{c}{0} &  5.0 , 3.5 &   6.05 & $  0.38\pm 0.27\pm 0.08$ & $<  0.96$ \\

\hline

\multicolumn{6}{l}{Sum (without $\pi^0$)} & $164.68\pm 4.96$ & \\

\end{tabular}
\end{ruledtabular}

\end{table*}

\begin{table*}[!p]

\flushleft

TABLE I(b): Branching fractions for $\Upsilon(1S)\to\mathrm{charged~hadrons+one}~\pi^0$. $N(1S)_\mathrm{res}\equiv N(1S)_\mathrm{on} - N(1S)_\mathrm{off} \times \mathcal{R}(1S)$, where $\mathcal{R}(1S)=\mathcal{L}_\mathrm{on}(1S)/\mathcal{L}_\mathrm{off} = 1.73$, and $\epsilon(1S)$ are the MC--calculated efficiencies.  Upper limits (UL) at 90\% confidence level are also given for modes with branching fractions which have a significance of $<2\sigma$.  
The modes marked with asterisks are used to construct the $\Upsilon(1S)/\Upsilon(2S)$ ratio.

\begin{ruledtabular}
\begin{tabular}{rlD{,}{\:\pm\:}{-1}D{,}{\:\pm\:}{-1}D{,}{\:\pm\:}{-1}dcc}
\# & modes & \multicolumn{1}{c}{$N(1S)_\mathrm{on}$} & \multicolumn{1}{c}{$N(1S)_\mathrm{off}\times\mathcal{R}(1S)$} & \multicolumn{1}{c}{$N(1S)_\mathrm{res}$} & \multicolumn{1}{c}{$\epsilon(1S) (\%)$} & $\mathcal{B}(1S)\times10^5$ & $\mathrm{UL}\times10^5$  \\
\hline

 $*\:$45 & $      4\pi\pi^0$ & 264 , 16 & 20.5 , 6.0  & 243.5 , 17.3 &  18.57 & $  6.10\pm 0.43\pm 0.77$ & --- \\
 46 & $      6\pi\pi^0$ & 759 , 28 & 75.6 , 11.4  & 683.4 , 29.8 &  10.88 & $ 29.18\pm 1.27\pm 4.14$ & --- \\
 $*\:$47 & $      8\pi\pi^0$ & 795 , 28 & 55.5 , 9.8  & 739.5 , 29.9 &   6.19 & $ 55.49\pm 2.24\pm 8.88$ & --- \\
\\[-6pt]

 $*\:$48 & $    2K2\pi\pi^0$ & 265 , 16 & 70.9 , 11.1  & 194.1 , 19.7 &  16.62 & $  5.43\pm 0.55\pm 0.73$ & --- \\
 $*\:$49 & $    2K4\pi\pi^0$ & 850 , 29 & 217.5 , 19.4  & 632.5 , 35.0 &   9.54 & $ 30.81\pm 1.71\pm 4.67$ & --- \\
 50 & $    2K6\pi\pi^0$ & 806 , 28 & 214.1 , 19.3  & 591.9 , 34.3 &   5.03 & $ 54.65\pm 3.17\pm 9.34$ & --- \\
\\[-6pt]

 51 & $        4K\pi^0$ &  19 ,  4 &  7.6 , 4.2 & 11.4 , 6.0 &  14.37 & $  0.37\pm 0.19\pm 0.05$ & $<  0.75$ \\
 52 & $    4K2\pi\pi^0$ & 162 , 13 & 33.1 , 7.6  & 128.9 , 14.8 &   8.21 & $  7.30\pm 0.84\pm 1.19$ & --- \\
 53 & $    4K4\pi\pi^0$ & 193 , 14 & 30.9 , 7.3  & 162.1 , 15.7 &   3.91 & $ 19.26\pm 1.87\pm 3.53$ & --- \\
\\[-6pt]

 54 & $        6K\pi^0$ &   6 ,  3 &  3.5 , 3.9 &  2.5 , 5.0 &   6.93 & $  0.17\pm 0.33\pm 0.03$ & $<  0.65$ \\
 55 & $    6K2\pi\pi^0$ &  21 ,  5 &  0.0 , 2.1  & 21.0 , 5.1 &   2.98 & $  3.28\pm 0.79\pm 0.65$ & --- \\
\\[-6pt]

 56 & $        8K\pi^0$ &  \multicolumn{1}{c}{0}  & \multicolumn{1}{c}{0} & \multicolumn{1}{c}{---} &   2.20 & --- & $<  0.65$ \\
\\[-6pt]

 57 & $    K_{S}K\pi\pi^0$ &  55 ,  7 &  9.2 , 4.4  & 45.8 , 8.6 &  11.51 & $  1.85\pm 0.35\pm 0.25$ & --- \\
 58 & $   K_{S}K3\pi\pi^0$ & 356 , 19 & 107.6 , 13.7  & 248.4 , 23.3 &   7.06 & $ 16.36\pm 1.53\pm 2.35$ & --- \\
 59 & $   K_{S}K5\pi\pi^0$ & 466 , 22 & 113.0 , 14.0  & 353.0 , 25.7 &   3.43 & $ 47.89\pm 3.49\pm 7.56$ & --- \\
\\[-6pt]

 60 & $   K_{S}3K\pi\pi^0$ &  63 ,  8 & 16.1 , 6.0  & 46.9 , 9.9 &   5.97 & $  3.65\pm 0.77\pm 0.55$ & --- \\
 61 & $  K_{S}3K3\pi\pi^0$ & 151 , 12 & 35.2 , 7.8  & 115.8 , 14.6 &   2.84 & $ 18.99\pm 2.39\pm 3.18$ & --- \\
\\[-6pt]

 62 & $   K_{S}5K\pi\pi^0$ &   5 ,  3 & \multicolumn{1}{c}{0}  &  5.0 , 3.5 &   2.24 & $  1.04\pm 0.73\pm 0.19$ & $<  2.52$ \\
\\[-6pt]

 63 & $    2p2\pi\pi^0$ & 128 , 11 & 26.7 , 6.8  & 101.3 , 13.2 &  18.79 & $  2.51\pm 0.33\pm 0.34$ & --- \\
 $*\:$64 & $    2p4\pi\pi^0$ & 399 , 20 & 79.5 , 11.7  & 319.5 , 23.2 &  10.96 & $ 13.54\pm 0.98\pm 2.05$ & --- \\
 $*\:$65 & $    2p6\pi\pi^0$ & 466 , 22 & 49.3 , 9.2  & 416.7 , 23.5 &   5.90 & $ 32.82\pm 1.85\pm 5.61$ & --- \\
\\[-6pt]

 $*\:$66 & $  2p2K2\pi\pi^0$ & 120 , 11 & 16.4 , 5.8  & 103.6 , 12.4 &   9.30 & $  5.18\pm 0.62\pm 0.84$ & --- \\
 $*\:$67 & $  2p2K4\pi\pi^0$ & 168 , 13 & 14.2 , 5.4  & 153.8 , 14.0 &   4.76 & $ 15.03\pm 1.37\pm 2.75$ & --- \\
\\[-6pt]

 68 & $      2p4K\pi^0$ &   4 ,  3 &  1.9 , 2.9 &  2.1 , 4.0 &   7.65 & $  0.13\pm 0.24\pm 0.02$ & $<  0.50$ \\
 69 & $  2p4K2\pi\pi^0$ &  14 ,  4 &  1.8 , 3.0  & 12.2 , 4.8 &   3.67 & $  1.55\pm 0.61\pm 0.31$ & --- \\
\\[-6pt]

 70 & $      2p6K\pi^0$ & \multicolumn{1}{c}{0} & \multicolumn{1}{c}{0} & \multicolumn{1}{c}{---} &   2.58 & --- & $<  0.56$ \\
\\[-6pt]

 71 & $    4p2\pi\pi^0$ &  17 ,  4 &  9.1 , 4.4 &  7.9 , 6.1 &  10.75 & $  0.34\pm 0.26\pm 0.05$ & $<  0.81$ \\
 72 & $    4p4\pi\pi^0$ &  14 ,  4 &  \multicolumn{1}{c}{0} & 14.0 , 4.3 &   5.67 & $  1.15\pm 0.35\pm 0.21$ & --- \\

\hline

\multicolumn{6}{l}{Sum (with one $\pi^0$)} & $374.07\pm 7.37$ & \\

\end{tabular}
\end{ruledtabular}

\end{table*}

\begin{table*}[!p]

\flushleft

TABLE I(c): Branching fractions for $\Upsilon(1S)\to\mathrm{charged~hadrons}+2\pi^0$. $N(1S)_\mathrm{res}\equiv N(1S)_\mathrm{on} - N(1S)_\mathrm{off} \times \mathcal{R}(1S)$, where $\mathcal{R}(1S)=\mathcal{L}_\mathrm{on}(1S)/\mathcal{L}_\mathrm{off} = 1.73$, and $\epsilon(1S)$ are the MC--calculated efficiencies.  Upper limits (UL) at 90\% confidence level are also given for modes with branching fractions which have a significance of $<2\sigma$.  
The modes marked with asterisks are used to construct the $\Upsilon(1S)/\Upsilon(2S)$ ratio.

\begin{ruledtabular}
\begin{tabular}{rlD{,}{\:\pm\:}{-1}D{,}{\:\pm\:}{-1}D{,}{\:\pm\:}{-1}dcc}
\# & modes & \multicolumn{1}{c}{$N(1S)_\mathrm{on}$} & \multicolumn{1}{c}{$N(1S)_\mathrm{off}\times\mathcal{R}(1S)$} & \multicolumn{1}{c}{$N(1S)_\mathrm{res}$} & \multicolumn{1}{c}{$\epsilon(1S) (\%)$} & $\mathcal{B}(1S)\times10^5$ & $\mathrm{UL}\times10^5$  \\
\hline

 73 & $     4\pi2\pi^0$ & 779 , 28 & 355.6 , 24.8  & 423.4 , 37.4 &  10.93 & $ 18.01\pm 1.59\pm 2.77$ & --- \\
 74 & $     6\pi2\pi^0$ & 1546 , 39 & 706.8 , 35.0  & 839.2 , 52.6 &   6.05 & $ 64.50\pm 4.05\pm10.70$ & --- \\
 75 & $     8\pi2\pi^0$ & 935 , 31 & 421.3 , 27.0  & 513.7 , 40.8 &   3.08 & $ 77.46\pm 6.15\pm14.14$ & --- \\
\\[-6pt]

 76 & $   2K2\pi2\pi^0$ & 293 , 17 & 124.5 , 14.7  & 168.5 , 22.6 &   9.52 & $  8.23\pm 1.10\pm 1.32$ & --- \\
 $*\:$77 & $   2K4\pi2\pi^0$ & 876 , 30 & 200.4 , 18.6  & 675.6 , 35.0 &   5.09 & $ 61.67\pm 3.19\pm10.74$ & --- \\
 $*\:$78 & $   2K6\pi2\pi^0$ & 757 , 28 & 158.0 , 16.5  & 599.0 , 32.1 &   2.54 & $109.53\pm 5.87\pm21.04$ & --- \\
\\[-6pt]

 79 & $       4K2\pi^0$ &  10 ,  3 &  8.9 , 4.6 &  1.1 , 5.6 &   8.25 & $  0.06\pm 0.32\pm 0.01$ & $<  0.52$ \\
 80 & $   4K2\pi2\pi^0$ & 114 , 11 & 18.0 , 5.6  & 96.0 , 12.1 &   3.49 & $ 12.77\pm 1.60\pm 2.35$ & --- \\
 81 & $   4K4\pi2\pi^0$ & 133 , 12 & 16.1 , 6.0  & 116.9 , 13.0 &   2.01 & $ 27.09\pm 3.01\pm 5.50$ & --- \\
\\[-6pt]

 82 & $       6K2\pi^0$ &   3 ,  2 &  \multicolumn{1}{c}{0} &  3.0 , 3.1 &   3.43 & $  0.41\pm 0.42\pm 0.08$ & $<  1.25$ \\
 83 & $   6K2\pi2\pi^0$ &   9 ,  4 & \multicolumn{1}{c}{0}  &  9.0 , 4.3 &   1.32 & $  3.17\pm 1.52\pm 0.68$ & --- \\
\\[-6pt]

 84 & $       8K2\pi^0$ &   \multicolumn{1}{c}{0} & \multicolumn{1}{c}{0}  & \multicolumn{1}{c}{---} &   0.93 & --- & $<  1.58$ \\
\\[-6pt]

 85 & $   K_{S}K\pi2\pi^0$ &  41 ,  6 & 12.9 , 5.0  & 28.1 , 8.1 &   6.50 & $  2.01\pm 0.58\pm 0.32$ & --- \\
 86 & $  K_{S}K3\pi2\pi^0$ & 308 , 18 & 85.8 , 12.2  & 222.2 , 21.4 &   3.65 & $ 28.29\pm 2.72\pm 4.74$ & --- \\
 $*\:$87 & $  K_{S}K5\pi2\pi^0$ & 471 , 22 & 105.3 , 13.5  & 365.7 , 25.6 &   1.68 & $101.43\pm 7.09\pm18.31$ & --- \\
\\[-6pt]

 88 & $  K_{S}3K\pi2\pi^0$ &  44 ,  7 &  9.8 , 3.7  & 34.2 , 7.6 &   2.87 & $  5.53\pm 1.23\pm 0.97$ & --- \\
 $*\:$89 & $ K_{S}3K3\pi2\pi^0$ &  93 , 10 & 13.8 , 4.1  & 79.2 , 10.5 &   1.25 & $ 29.36\pm 3.88\pm 5.56$ & --- \\
\\[-6pt]

 90 & $  K_{S}5K\pi2\pi^0$ &   2 ,  2 & \multicolumn{1}{c}{0}  &  2.0 , 3.1 &   0.96 & $  0.97\pm 1.50\pm 0.19$ & $<  3.59$ \\
\\[-6pt]

 91 & $   2p2\pi2\pi^0$ & 136 , 12 & 40.7 , 8.4  & 95.3 , 14.4 &  10.71 & $  4.13\pm 0.62\pm 0.66$ & --- \\
 $*\:$92 & $   2p4\pi2\pi^0$ & 368 , 19 & 78.6 , 11.7  & 289.4 , 22.5 &   5.93 & $ 22.68\pm 1.76\pm 3.95$ & --- \\
 $*\:$93 & $   2p6\pi2\pi^0$ & 335 , 18 & 35.6 , 7.8  & 299.4 , 19.9 &   2.81 & $ 49.55\pm 3.30\pm 9.52$ & --- \\
\\[-6pt]

 94 & $ 2p2K2\pi2\pi^0$ &  89 ,  9 & 20.1 , 5.9  & 68.9 , 11.1 &   4.72 & $  6.78\pm 1.09\pm 1.25$ & --- \\
 $*\:$95 & $ 2p2K4\pi2\pi^0$ & 113 , 11 &  7.0 , 4.7  & 106.0 , 11.6 &   2.18 & $ 22.58\pm 2.48\pm 4.59$ & --- \\
\\[-6pt]

 96 & $     2p4K2\pi^0$ &   7 ,  3 & \multicolumn{1}{c}{0}  &  7.0 , 3.9 &   3.65 & $  0.89\pm 0.50\pm 0.17$ & $<  1.98$ \\
 97 & $ 2p4K2\pi2\pi^0$ &   7 ,  3 &  1.7 , 3.0 &  5.3 , 4.5 &   1.62 & $  1.52\pm 1.29\pm 0.33$ & $<  3.96$ \\
\\[-6pt]

 98 & $     2p6K2\pi^0$ &  \multicolumn{1}{c}{0} & \multicolumn{1}{c}{0}  & \multicolumn{1}{c}{---} &   1.04 & --- & $<  1.41$ \\
\\[-6pt]

 99 & $   4p2\pi2\pi^0$ &  12 ,  3 &  \multicolumn{1}{c}{0}  & 12.0 , 4.1 &   5.60 & $  1.00\pm 0.34\pm 0.17$ & --- \\
100 & $   4p4\pi2\pi^0$ &  10 ,  3 &  \multicolumn{1}{c}{0}  & 10.0 , 3.8 &   2.65 & $  1.75\pm 0.67\pm 0.36$ & --- \\

\hline

\multicolumn{6}{l}{Sum (with two $\pi^0$)} & $661.38\pm14.85$ & \\

\end{tabular}
\end{ruledtabular}

\end{table*}

\addtocounter{table}{1}

\begin{table*}[!p]

\flushleft

TABLE~II(a): Branching fractions for $\Upsilon(2S)\to\mathrm{only~charged~hadrons}$. $N(2S)_\mathrm{res}\equiv N(2S)_\mathrm{on} - N(2S)_\mathrm{off} \times \mathcal{R}(2S)$, where $\mathcal{R}(2S)=\mathcal{L}_\mathrm{on}(2S)/\mathcal{L}_\mathrm{off} = 2.03$, and $\epsilon(2S)$ are the MC--calculated efficiencies.  Upper limits (UL) at 90\% confidence level are also given for modes with branching fractions which have a significance of $<2\sigma$.  
The modes marked with asterisks are used to construct the $\Upsilon(1S)/\Upsilon(2S)$ ratio.

\begin{ruledtabular}
\begin{tabular}{rlD{,}{\:\pm\:}{-1}D{,}{\:\pm\:}{-1}D{,}{\:\pm\:}{-1}dcc}
\# & modes & \multicolumn{1}{c}{$N(2S)_\mathrm{on}$} & \multicolumn{1}{c}{$N(2S)_\mathrm{off}\times\mathcal{R}(2S)$} & \multicolumn{1}{c}{$N(2S)_\mathrm{res}$} & \multicolumn{1}{c}{$\epsilon(2S) (\%)$} & $\mathcal{B}(2S)\times10^5$ & $\mathrm{UL}\times10^5$  \\
\hline

  1 & $           4\pi$ &  91 , 10 & 100.0 , 14.3 & -9.0 , 17.2 &  32.63 & $ -0.30\pm 0.56\pm 0.03$ & $<  0.43$ \\
  2 & $           6\pi$ & 213 , 15 & 203.3 , 20.3 &  9.7 , 25.0 &  20.55 & $  0.51\pm 1.31\pm 0.07$ & $<  2.18$ \\
  3 & $           8\pi$ & 192 , 14 & 176.7 , 18.9 & 15.3 , 23.5 &  12.01 & $  1.37\pm 2.10\pm 0.21$ & $<  4.06$ \\
  4 & $          10\pi$ &  48 ,  7 & 51.9 , 10.3 & -3.9 , 12.4 &   6.36 & $ -0.67\pm 2.09\pm 0.12$ & $<  2.02$ \\
\\[-6pt]

  5 & $         2K2\pi$ &  46 ,  7 & 38.9 , 8.9 &  7.1 , 11.2 &  27.70 & $  0.27\pm 0.43\pm 0.03$ & $<  0.83$ \\
  6 & $         2K4\pi$ & 149 , 12 & 140.4 , 16.9 &  8.6 , 20.8 &  18.10 & $  0.51\pm 1.24\pm 0.07$ & $<  2.09$ \\
  $*\:$7 & $         2K6\pi$ & 151 , 12 & 93.0 , 13.7  & 58.0 , 18.4 &  10.51 & $  5.92\pm 1.88\pm 0.97$ & --- \\
  8 & $         2K8\pi$ &  43 ,  7 & 36.1 , 8.6 &  6.9 , 10.8 &   5.58 & $  1.33\pm 2.07\pm 0.25$ & $<  4.01$ \\
\\[-6pt]

  9 & $             4K$ &   8 ,  3 & 10.0 , 3.8 & -2.0 , 5.0 &  23.53 & $ -0.09\pm 0.23\pm 0.01$ & $<  0.22$ \\
 10 & $         4K2\pi$ &  37 ,  6 & 37.5 , 8.7 & -0.5 , 10.6 &  15.91 & $ -0.03\pm 0.72\pm 0.01$ & $<  0.89$ \\
 11 & $         4K4\pi$ &  45 ,  7 & 28.0 , 7.5 & 17.0 , 10.1 &   7.87 & $  2.32\pm 1.38\pm 0.41$ & $<  4.15$ \\
 12 & $         4K6\pi$ &  13 ,  4 & 10.1 , 3.7 &  2.9 , 5.1 &   4.37 & $  0.71\pm 1.26\pm 0.14$ & $<  3.06$ \\
\\[-6pt]

 13 & $             6K$ &   5 ,  3 &  4.1 , 4.6 &  0.9 , 5.4 &  11.18 & $  0.09\pm 0.52\pm 0.01$ & $<  0.68$ \\
 14 & $         6K2\pi$ &   3 ,  2 &  2.0 , 3.6 &  1.0 , 4.2 &   7.58 & $  0.14\pm 0.60\pm 0.03$ & $<  0.94$ \\
 15 & $         6K4\pi$ &   2 ,  2 &  2.0 , 3.6 & 0.0 , 4.2 &   3.41 & $ -0.01\pm 1.33\pm 0.01$ & $<  1.56$ \\
\\[-6pt]

 16 & $             8K$ &  \multicolumn{1}{c}{0}  &  \multicolumn{1}{c}{0}  & \multicolumn{1}{c}{---} &   5.79 & --- & $<  0.57$ \\
 17 & $         8K2\pi$ &  \multicolumn{1}{c}{0}  &  \multicolumn{1}{c}{0}  & \multicolumn{1}{c}{---} &   2.53 & --- & $<  1.34$ \\
\\[-6pt]

 18 & $            10K$ &  \multicolumn{1}{c}{0}  &  \multicolumn{1}{c}{0}  & \multicolumn{1}{c}{---} &   1.76 & --- & $<  1.98$ \\
\\[-6pt]

 19 & $         K_{S}K\pi$ &   4 ,  3 &  3.8 , 2.0 &  0.2 , 3.4 &  19.38 & $  0.01\pm 0.19\pm 0.01$ & $<  0.30$ \\
 20 & $        K_{S}K3\pi$ &  61 ,  8 & 45.6 , 9.6 & 15.4 , 12.4 &  13.10 & $  1.26\pm 1.02\pm 0.17$ & $<  2.58$ \\
 21 & $        K_{S}K5\pi$ &  66 ,  8 & 58.6 , 10.9 &  7.4 , 13.6 &   7.42 & $  1.06\pm 1.97\pm 0.16$ & $<  3.59$ \\
 22 & $        K_{S}K7\pi$ &  36 ,  6 & 24.2 , 7.0 & 11.8 , 9.2 &   3.68 & $  3.43\pm 2.69\pm 0.61$ & $<  6.97$ \\
\\[-6pt]

 23 & $        K_{S}3K\pi$ &  13 ,  4 &  9.6 , 4.2 &  3.4 , 5.5 &  11.13 & $  0.33\pm 0.53\pm 0.05$ & $<  1.18$ \\
 24 & $       K_{S}3K3\pi$ &  37 ,  6 & 26.2 , 7.3 & 10.8 , 9.5 &   6.34 & $  1.82\pm 1.61\pm 0.29$ & $<  3.91$ \\
 25 & $       K_{S}3K5\pi$ &  15 ,  4 & 16.0 , 4.9 & -1.0 , 6.2 &   2.91 & $ -0.38\pm 2.30\pm 0.07$ & $<  2.97$ \\
\\[-6pt]

 26 & $        K_{S}5K\pi$ &   1 ,  2 &  \multicolumn{1}{c}{0}  &  1.0 , 3.0 &   5.28 & $  0.20\pm 0.62\pm 0.04$ & $<  1.07$ \\
 27 & $       K_{S}5K3\pi$ &   4 ,  3 &  \multicolumn{1}{c}{0}  &  4.0 , 3.7 &   2.25 & $  1.91\pm 1.78\pm 0.39$ & $<  5.14$ \\
\\[-6pt]

 28 & $        K_{S}7K\pi$ &   \multicolumn{1}{c}{0}  &  \multicolumn{1}{c}{0}  & \multicolumn{1}{c}{---} &   1.66 & --- & $<  2.02$ \\
\\[-6pt]

 29 & $         2p2\pi$ &  19 ,  4 & 15.9 , 5.0 &  3.1 , 6.7 &  32.29 & $  0.10\pm 0.22\pm 0.01$ & $<  0.44$ \\
 30 & $         2p4\pi$ &  63 ,  8 & 45.6 , 9.6 & 17.4 , 12.5 &  20.15 & $  0.93\pm 0.66\pm 0.13$ & $<  1.79$ \\
 31 & $         2p6\pi$ &  67 ,  8 & 47.8 , 9.9 & 19.2 , 12.8 &  12.03 & $  1.71\pm 1.14\pm 0.28$ & $<  3.22$ \\
 $*\:$32 & $         2p8\pi$ &  27 ,  5 &  7.8 , 2.9  & 19.2 , 6.0 &   6.40 & $  3.21\pm 1.00\pm 0.60$ & --- \\
\\[-6pt]

 33 & $       2p2K2\pi$ &  28 ,  5 & 17.7 , 5.3 & 10.3 , 7.5 &  17.88 & $  0.62\pm 0.45\pm 0.10$ & $<  1.20$ \\
 34 & $       2p2K4\pi$ &  19 ,  4 & 14.0 , 4.4 &  5.0 , 6.2 &  10.19 & $  0.52\pm 0.65\pm 0.09$ & $<  1.72$ \\
 35 & $       2p2K6\pi$ &  15 ,  4 &  6.0 , 2.6 &  9.0 , 4.7 &   5.17 & $  1.87\pm 0.97\pm 0.37$ & $<  4.29$ \\
\\[-6pt]

 36 & $           2p4K$ &   1 ,  2 &  1.9 , 2.0 & -0.9 , 2.7 &  15.54 & $ -0.06\pm 0.19\pm 0.01$ & $<  0.22$ \\
 37 & $       2p4K2\pi$ &   1 ,  2 &  2.0 , 3.6 & -1.0 , 4.0 &   8.37 & $ -0.13\pm 0.51\pm 0.03$ & $<  0.40$ \\
 38 & $       2p4K4\pi$ &   4 ,  3 & \multicolumn{1}{c}{0}  &  4.0 , 3.7 &   3.94 & $  1.09\pm 1.02\pm 0.23$ & $<  2.98$ \\
\\[-6pt]

 39 & $           2p6K$ &    \multicolumn{1}{c}{0} & \multicolumn{1}{c}{0}  & \multicolumn{1}{c}{---} &   6.62 & --- & $<  0.50$ \\
 40 & $       2p6K2\pi$ &   1 ,  2 & \multicolumn{1}{c}{0}  &  1.0 , 3.0 &   2.92 & $  0.37\pm 1.12\pm 0.08$ & $<  2.08$ \\
\\[-6pt]

 41 & $           2p8K$ &    \multicolumn{1}{c}{0} & \multicolumn{1}{c}{0}  & \multicolumn{1}{c}{---} &   1.91 & --- & $<  1.81$ \\
\\[-6pt]

 42 & $         4p2\pi$ &   3 ,  2 & \multicolumn{1}{c}{0}  &  3.0 , 3.4 &  20.41 & $  0.16\pm 0.18\pm 0.02$ & $<  0.45$ \\
 43 & $         4p4\pi$ &   \multicolumn{1}{c}{0} & \multicolumn{1}{c}{0}  & \multicolumn{1}{c}{---} &  12.21 & --- & $<  0.26$ \\
 44 & $         4p6\pi$ &   2 ,  2 & \multicolumn{1}{c}{0}  &  2.0 , 3.4 &   6.48 & $  0.33\pm 0.56\pm 0.07$ & $<  1.22$ \\

\hline

\multicolumn{6}{l}{Sum (without $\pi^0$)} & $32.45\pm 8.27$ & \\

\end{tabular}
\end{ruledtabular}

\end{table*}

\begin{table*}[!p]

\flushleft

TABLE~II(b): Branching fractions for $\Upsilon(2S)\to\mathrm{charged~hadrons+one}~\pi^0$. $N(2S)_\mathrm{res}\equiv N(2S)_\mathrm{on} - N(2S)_\mathrm{off} \times \mathcal{R}(2S)$, where $\mathcal{R}(2S)=\mathcal{L}_\mathrm{on}(2S)/\mathcal{L}_\mathrm{off} = 2.03$, and $\epsilon(2S)$ are the MC--calculated efficiencies.  Upper limits (UL) at 90\% confidence level are also given for modes with branching fractions which have a significance of $<2\sigma$.  
The modes marked with asterisks are used to construct the $\Upsilon(1S)/\Upsilon(2S)$ ratio.

\begin{ruledtabular}
\begin{tabular}{rlD{,}{\:\pm\:}{-1}D{,}{\:\pm\:}{-1}D{,}{\:\pm\:}{-1}dcc}
\# & modes & \multicolumn{1}{c}{$N(2S)_\mathrm{on}$} & \multicolumn{1}{c}{$N(2S)_\mathrm{off}\times\mathcal{R}(2S)$} & \multicolumn{1}{c}{$N(2S)_\mathrm{res}$} & \multicolumn{1}{c}{$\epsilon(2S) (\%)$} & $\mathcal{B}(2S)\times10^5$ & $\mathrm{UL}\times10^5$  \\
\hline

 $*\:$45 & $      4\pi\pi^0$ &  44 ,  7 & 21.5 , 6.6  & 22.5 , 9.4 &  18.70 & $  1.29\pm 0.54\pm 0.16$ & --- \\
 46 & $      6\pi\pi^0$ & 115 , 11 & 83.0 , 13.0 & 32.0 , 16.8 &  11.45 & $  3.00\pm 1.58\pm 0.43$ & $<  5.09$ \\
 $*\:$47 & $      8\pi\pi^0$ & 131 , 11 & 58.5 , 10.9  & 72.5 , 15.8 &   6.25 & $ 12.45\pm 2.71\pm 1.99$ & --- \\
\\[-6pt]

 $*\:$48 & $    2K2\pi\pi^0$ & 122 , 11 & 73.2 , 12.2  & 48.8 , 16.5 &  16.43 & $  3.19\pm 1.07\pm 0.43$ & --- \\
 $*\:$49 & $    2K4\pi\pi^0$ & 312 , 18 & 241.8 , 22.2  & 70.2 , 28.3 &  10.16 & $  7.42\pm 2.99\pm 1.12$ & --- \\
 50 & $    2K6\pi\pi^0$ & 260 , 16 & 235.0 , 21.8 & 25.0 , 27.2 &   5.29 & $  5.07\pm 5.50\pm 0.87$ & $< 12.20$ \\
\\[-6pt]

 51 & $        4K\pi^0$ &   5 ,  3 &  7.9 , 2.9 & -2.9 , 4.0 &  14.31 & $ -0.22\pm 0.30\pm 0.03$ & $<  0.23$ \\
 52 & $    4K2\pi\pi^0$ &  40 ,  6 & 35.5 , 8.5 &  4.5 , 10.6 &   8.44 & $  0.57\pm 1.35\pm 0.09$ & $<  2.29$ \\
 53 & $    4K4\pi\pi^0$ &  53 ,  7 & 36.9 , 8.7 & 16.1 , 11.3 &   4.46 & $  3.88\pm 2.72\pm 0.71$ & $<  7.47$ \\
\\[-6pt]

 54 & $        6K\pi^0$ &   \multicolumn{1}{c}{0} &  3.9 , 2.0 & -3.9 , 2.4 &   7.46 & $ -0.56\pm 0.34\pm 0.10$ & $<  0.10$ \\
 55 & $    6K2\pi\pi^0$ &   7 ,  3 &  \multicolumn{1}{c}{0}  &  7.0 , 4.1 &   3.49 & $  2.15\pm 1.27\pm 0.42$ & $<  4.80$ \\
\\[-6pt]

 56 & $        8K\pi^0$ &   \multicolumn{1}{c}{0}  &   \multicolumn{1}{c}{0} & \multicolumn{1}{c}{---} &   2.55 & --- & $<  1.30$ \\
\\[-6pt]

 57 & $    K_{S}K\pi\pi^0$ &  16 ,  4 &  9.4 , 4.4 &  6.6 , 5.9 &  11.24 & $  0.63\pm 0.57\pm 0.08$ & $<  1.61$ \\
 58 & $   K_{S}K3\pi\pi^0$ & 115 , 11 & 112.9 , 15.1 &  2.1 , 18.6 &   7.10 & $  0.31\pm 2.81\pm 0.05$ & $<  3.91$ \\
 59 & $   K_{S}K5\pi\pi^0$ & 143 , 12 & 125.7 , 16.0 & 17.3 , 20.0 &   3.65 & $  5.10\pm 5.87\pm 0.80$ & $< 12.68$ \\
\\[-6pt]

 60 & $   K_{S}3K\pi\pi^0$ &  26 ,  5 & 17.1 , 5.8 &  8.9 , 7.8 &   6.07 & $  1.56\pm 1.37\pm 0.24$ & $<  3.34$ \\
 61 & $  K_{S}3K3\pi\pi^0$ &  62 ,  8 & 40.4 , 9.1 & 21.6 , 12.0 &   3.12 & $  7.44\pm 4.13\pm 1.25$ & $< 12.95$ \\
\\[-6pt]

 62 & $   K_{S}5K\pi\pi^0$ &   1 ,  2 &  \multicolumn{1}{c}{0}  &  1.0 , 3.0 &   2.45 & $  0.44\pm 1.33\pm 0.08$ & $<  2.33$ \\
\\[-6pt]

 63 & $    2p2\pi\pi^0$ &  43 ,  7 & 27.7 , 7.5 & 15.3 , 10.0 &  18.71 & $  0.87\pm 0.57\pm 0.12$ & $<  1.62$ \\
 $*\:$64 & $    2p4\pi\pi^0$ & 125 , 11 & 84.5 , 13.1  & 40.5 , 17.2 &  11.15 & $  3.90\pm 1.66\pm 0.59$ & --- \\
 $*\:$65 & $    2p6\pi\pi^0$ & 113 , 11 & 53.0 , 10.4  & 60.0 , 14.9 &   6.09 & $ 10.58\pm 2.62\pm 1.81$ & --- \\
\\[-6pt]

 $*\:$66 & $  2p2K2\pi\pi^0$ &  50 ,  7 & 17.9 , 5.1  & 32.1 , 8.7 &   9.74 & $  3.54\pm 0.96\pm 0.58$ & --- \\
 $*\:$67 & $  2p2K4\pi\pi^0$ &  46 ,  7 & 16.0 , 4.9  & 30.0 , 8.4 &   5.12 & $  6.29\pm 1.76\pm 1.15$ & --- \\
\\[-6pt]

 68 & $      2p4K\pi^0$ &   2 ,  2 &  2.0 , 3.6 & -0.0 , 4.2 &   8.00 & $ -0.00\pm 0.57\pm 0.01$ & $<  0.63$ \\
 69 & $  2p4K2\pi\pi^0$ &   4 ,  3 &  2.0 , 3.6 &  2.0 , 4.5 &   4.01 & $  0.53\pm 1.21\pm 0.10$ & $<  2.19$ \\
\\[-6pt]

 70 & $      2p6K\pi^0$ &   \multicolumn{1}{c}{0}  &  \multicolumn{1}{c}{0}  & \multicolumn{1}{c}{---} &   2.98 & --- & $<  1.11$ \\
\\[-6pt]

 71 & $    4p2\pi\pi^0$ &   4 ,  3 &  9.9 , 3.9 & -5.9 , 4.8 &  11.16 & $ -0.56\pm 0.46\pm 0.09$ & $<  0.16$ \\
 72 & $    4p4\pi\pi^0$ &   4 ,  3 &  \multicolumn{1}{c}{0}  &  4.0 , 3.7 &   5.99 & $  0.72\pm 0.67\pm 0.13$ & $<  1.89$ \\

\hline

\multicolumn{6}{l}{Sum (with one $\pi^0$)} & $79.57\pm11.98$ & \\

\end{tabular}
\end{ruledtabular}

\end{table*}

\begin{table*}[!p]

\flushleft

TABLE~II(c): Branching fractions for $\Upsilon(2S)\to\mathrm{charged~hadrons}+2\pi^0$. $N(2S)_\mathrm{res}\equiv N(2S)_\mathrm{on} - N(2S)_\mathrm{off} \times \mathcal{R}(2S)$, where $\mathcal{R}(2S)=\mathcal{L}_\mathrm{on}(2S)/\mathcal{L}_\mathrm{off} = 2.03$, and $\epsilon(2S)$ are the MC--calculated efficiencies.  Upper limits (UL) at 90\% confidence level are also given for modes with branching fractions which have a significance of $<2\sigma$.  
The modes marked with asterisks are used to construct the $\Upsilon(1S)/\Upsilon(2S)$ ratio.

\begin{ruledtabular}
\begin{tabular}{rlD{,}{\:\pm\:}{-1}D{,}{\:\pm\:}{-1}D{,}{\:\pm\:}{-1}dcc}
\# & modes & \multicolumn{1}{c}{$N(2S)_\mathrm{on}$} & \multicolumn{1}{c}{$N(2S)_\mathrm{off}\times\mathcal{R}(2S)$} & \multicolumn{1}{c}{$N(2S)_\mathrm{res}$} & \multicolumn{1}{c}{$\epsilon(2S) (\%)$} & $\mathcal{B}(2S)\times10^5$ & $\mathrm{UL}\times10^5$  \\
\hline

 73 & $     4\pi2\pi^0$ & 360 , 19 & 370.3 , 27.4 & -10.3 , 33.3 &  10.90 & $ -1.01\pm 3.28\pm 0.16$ & $<  3.19$ \\
 74 & $     6\pi2\pi^0$ & 825 , 29 & 755.3 , 39.2 & 69.7 , 48.6 &   6.19 & $ 12.08\pm 8.42\pm 2.00$ & $< 23.15$ \\
 75 & $     8\pi2\pi^0$ & 521 , 23 & 467.1 , 30.8 & 53.9 , 38.3 &   3.28 & $ 17.65\pm12.56\pm 3.22$ & $< 34.25$ \\
\\[-6pt]

 76 & $   2K2\pi2\pi^0$ & 131 , 11 & 130.5 , 16.3 &  0.5 , 19.9 &   9.56 & $  0.06\pm 2.23\pm 0.01$ & $<  2.92$ \\
 $*\:$77 & $   2K4\pi2\pi^0$ & 309 , 18 & 216.6 , 21.0  & 92.4 , 27.4 &   5.27 & $ 18.80\pm 5.57\pm 3.27$ & --- \\
 $*\:$78 & $   2K6\pi2\pi^0$ & 250 , 16 & 164.5 , 18.3  & 85.5 , 24.2 &   2.54 & $ 36.18\pm10.22\pm 6.95$ & --- \\
\\[-6pt]

 79 & $       4K2\pi^0$ &   6 ,  3 &  9.2 , 4.5 & -3.2 , 5.5 &   8.22 & $ -0.42\pm 0.71\pm 0.07$ & $<  0.45$ \\
 80 & $   4K2\pi2\pi^0$ &  36 ,  6 & 22.6 , 6.8 & 13.4 , 9.1 &   4.21 & $  3.41\pm 2.31\pm 0.63$ & $<  6.47$ \\
 81 & $   4K4\pi2\pi^0$ &  31 ,  6 & 17.8 , 5.2 & 13.2 , 7.6 &   2.13 & $  6.64\pm 3.83\pm 1.35$ & $< 11.83$ \\
\\[-6pt]

 82 & $       6K2\pi^0$ &  \multicolumn{1}{c}{0} &  \multicolumn{1}{c}{0} & \multicolumn{1}{c}{---} &   3.67 & --- & $<  0.89$ \\
 83 & $   6K2\pi2\pi^0$ &   1 ,  2 & \multicolumn{1}{c}{0}  &  1.0 , 3.0 &   1.44 & $  0.75\pm 2.27\pm 0.16$ & $<  4.15$ \\
\\[-6pt]

 84 & $       8K2\pi^0$ &  \multicolumn{1}{c}{0}  & \multicolumn{1}{c}{0}  & \multicolumn{1}{c}{---} &   1.12 & --- & $<  3.04$ \\
\\[-6pt]

 85 & $   K_{S}K\pi2\pi^0$ &  20 ,  4 & 13.6 , 4.8 &  6.4 , 6.6 &   6.58 & $  1.04\pm 1.07\pm 0.17$ & $<  2.42$ \\
 86 & $  K_{S}K3\pi2\pi^0$ & 122 , 11 & 89.7 , 13.5 & 32.3 , 17.4 &   3.66 & $  9.47\pm 5.12\pm 1.59$ & $< 16.33$ \\
 $*\:$87 & $  K_{S}K5\pi2\pi^0$ & 161 , 13 & 114.9 , 15.3  & 46.1 , 19.9 &   1.75 & $ 28.26\pm12.16\pm 5.10$ & --- \\
\\[-6pt]

 88 & $  K_{S}3K\pi2\pi^0$ &  13 ,  4 &  9.4 , 4.4 &  3.6 , 5.7 &   2.64 & $  1.46\pm 2.30\pm 0.26$ & $<  5.25$ \\
 $*\:$89 & $ K_{S}3K3\pi2\pi^0$ &  37 ,  6 & 16.3 , 6.7  & 20.7 , 9.1 &   1.42 & $ 15.68\pm 6.85\pm 2.97$ & --- \\
\\[-6pt]

 90 & $  K_{S}5K\pi2\pi^0$ &  \multicolumn{1}{c}{0}  &  \multicolumn{1}{c}{0} & \multicolumn{1}{c}{---} &   1.19 & --- & $<  2.74$ \\
\\[-6pt]

 91 & $   2p2\pi2\pi^0$ &  46 ,  7 & 42.0 , 9.2 &  4.0 , 11.5 &  10.59 & $  0.40\pm 1.16\pm 0.06$ & $<  1.89$ \\
 $*\:$92 & $   2p4\pi2\pi^0$ & 119 , 11 & 80.4 , 12.8  & 38.6 , 16.8 &   5.82 & $  7.12\pm 3.10\pm 1.24$ & --- \\
 $*\:$93 & $   2p6\pi2\pi^0$ &  78 ,  9 & 40.2 , 9.0  & 37.8 , 12.6 &   3.04 & $ 13.34\pm 4.46\pm 2.56$ & --- \\
\\[-6pt]

 94 & $ 2p2K2\pi2\pi^0$ &  39 ,  6 & 21.9 , 6.7 & 17.1 , 9.1 &   4.94 & $  3.71\pm 1.99\pm 0.68$ & $<  6.40$ \\
 $*\:$95 & $ 2p2K4\pi2\pi^0$ &  19 ,  4 &  7.8 , 3.0  & 11.2 , 5.3 &   2.31 & $  5.23\pm 2.46\pm 1.06$ & --- \\
\\[-6pt]

 96 & $     2p4K2\pi^0$ &  \multicolumn{1}{c}{0}  & \multicolumn{1}{c}{0}  & \multicolumn{1}{c}{---} &   3.87 & --- & $<  0.84$ \\
 97 & $ 2p4K2\pi2\pi^0$ &   1 ,  2 &  1.9 , 2.0 & -0.9 , 2.7 &   1.69 & $ -0.57\pm 1.70\pm 0.12$ & $<  2.11$ \\
\\[-6pt]

 98 & $     2p6K2\pi^0$ & \multicolumn{1}{c}{0}   & \multicolumn{1}{c}{0}  & \multicolumn{1}{c}{---} &   1.27 & --- & $<  2.67$ \\
\\[-6pt]

 99 & $   4p2\pi2\pi^0$ &   5 ,  3 & \multicolumn{1}{c}{0}  &  5.0 , 3.8 &   5.98 & $  0.90\pm 0.67\pm 0.16$ & $<  2.17$ \\
100 & $   4p4\pi2\pi^0$ &   2 ,  2 & \multicolumn{1}{c}{0}  &  2.0 , 3.4 &   2.82 & $  0.76\pm 1.28\pm 0.15$ & $<  2.82$ \\

\hline

\multicolumn{6}{l}{Sum (with two $\pi^0$)} & $180.93\pm26.44$ & \\

\end{tabular}
\end{ruledtabular}

\end{table*}

With 100 different decay modes measured it is difficult to comment on individual decays.  However, we note certain characteristics.

\begin{figure*}[!tb]
\begin{center}

\includegraphics[width=4.8in]{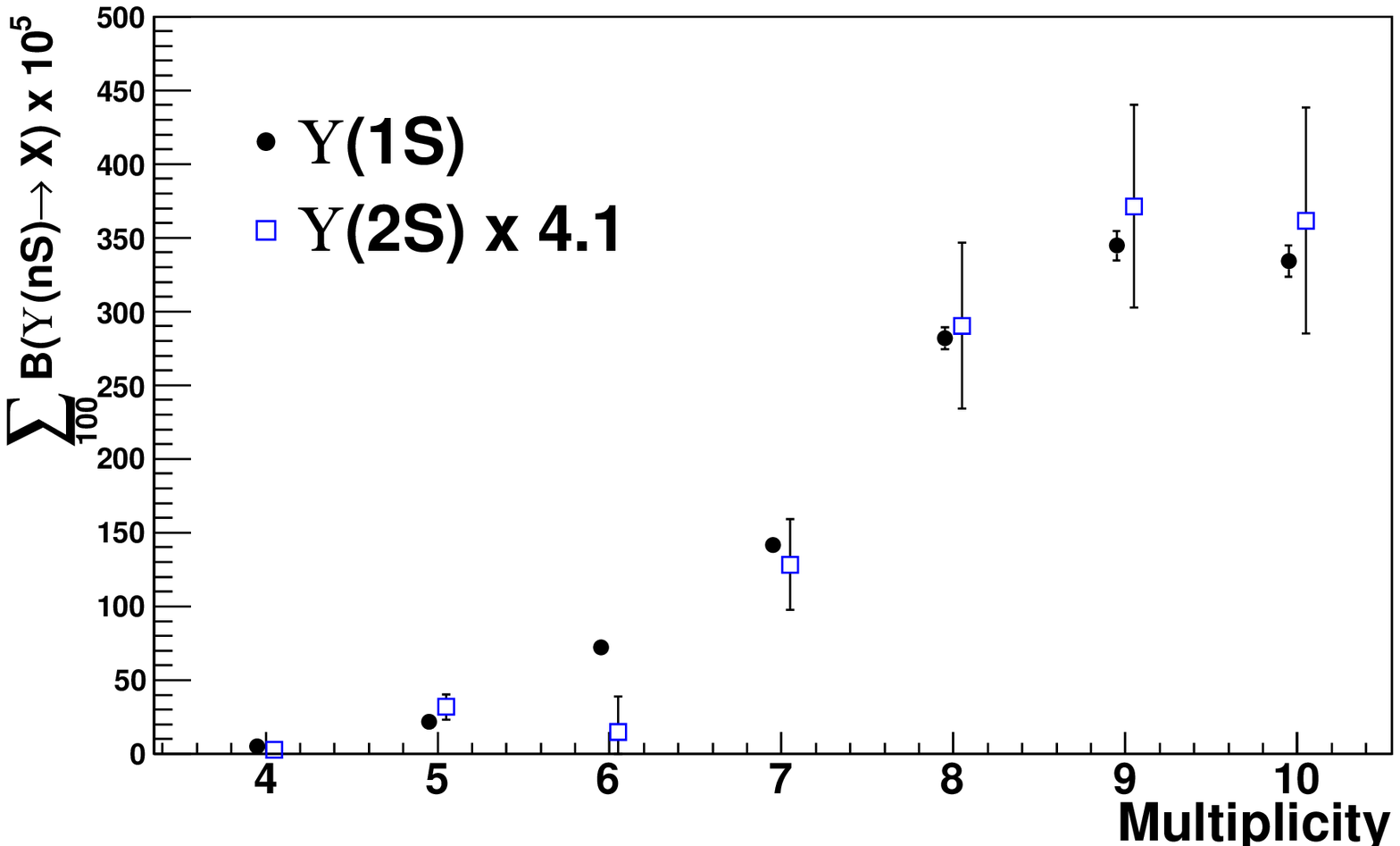}

\end{center}

\caption{Sums of branching fractions for $\sum_{100}\mathcal{B}(\Upsilon(nS)\to X)$ as a function of hadron multiplicity.  Only statistical errors are shown.  The $\Upsilon(2S)\times4.1$ values are plotted slightly displaced for clarity.}
\label{fig:brmulti}
\end{figure*}

The first notable feature is the multiplicity dependence of the branching fractions.  Table~\ref{tbl:brmulti} and Fig.~\ref{fig:brmulti} illustrate that the multiplicity dependence of the sum of branching fractions are very similar for $\Upsilon(1S)$ and $\Upsilon(2S)$.   Both increase with increasing multiplicity, and reach a maximum at a multiplicity of 9 or 10. The ratio $\sum_{100}\mathcal{B}(\Upsilon(1S)\to X)/\sum_{100}\mathcal{B}(\Upsilon(2S)\to X)=4.1\pm0.4$.
For the 17 modes with finite branching fractions common to both $\Upsilon(1S)$ and $\Upsilon(2S)$ (marked with asterisks in Tables I and II), the ratio $\sum_{17} \mathcal{B}(\Upsilon(1S)\to X) / \sum_{17} \mathcal{B}(\Upsilon(2S)\to X) = 3.2\pm0.4$, which is consistent with the above ratio for all decays.
We note that these ratios differ from $\mathcal{B}(\Upsilon(1S)\to e^+e^-)/\mathcal{B}(\Upsilon(2S)\to e^+e^-)=1.3\pm0.1$.

\begin{table}[!tb]
\caption{Summed branching fractions for different multiplicity decays.  The ratio $\mathcal{B}(\Upsilon(1S)\to X)/\mathcal{B}(\Upsilon(2S)\to X)=4.1\pm0.4$.}

\begin{ruledtabular}
\begin{tabular}{lD{,}{\:\pm\:}{-1}D{,}{\:\pm\:}{-1}}
 Decay Modes    & \multicolumn{1}{c}{$\mathcal{B}(\Upsilon(1S)\to X)\times10^5$} & \multicolumn{1}{c}{$\mathcal{B}(\Upsilon(2S)\to X)\times10^5$}  \\
\hline
      4 hadrons &   5.30 ,  0.53  &   0.62 ,  0.96 \\
      5 hadrons &  21.75 ,  1.18  &   7.77 ,  2.08 \\
      6 hadrons &  70.84 ,  2.91  &   3.63 ,  5.67 \\
      7 hadrons & 141.67 ,  4.30  &  31.31 ,  7.48 \\
      8 hadrons & 281.80 ,  7.51  &  70.83 , 13.73 \\
      9 hadrons & 344.63 , 10.05  &  90.56 , 16.76 \\
     10 hadrons & 333.98 , 10.65  &  88.21 , 18.66 \\
\hline

$\text{Sum}(100~\text{modes})$    & 1200.1 , 17.3   & 292.95 , 30.18 \\

\end{tabular}
\end{ruledtabular}
\label{tbl:brmulti}
\end{table}

The other notable feature concerns the relative strengths of branching fractions for decays containing 0, 1, and 2 $\pi^0$'s.  As noted in Tables~I and II, the relative strengths of decays containing only~$\pi^\pm:\pi^\pm+\mathrm{one}~\pi^0:\pi^\pm+\mathrm{two}~\pi^0$ are $1:2.3(1):4.0(2)$ for $\Upsilon(1S)$ and $1:2.5(7):5.5(16)$ for $\Upsilon(2S)$.  This feature results from the isospin dependence of the summed branching fractions.  Long ago, Pais~\cite{pais} had calculated the isospin dependence of the decays containing neutral and charged pions, in particular the ratios of decays $(n-2)\pi^\pm 2\pi^0/n\pi^\pm$ for $n=2-8$.  In Table~\ref{tbl:paiscomp}, we present the results from our measurements of $\Upsilon(1S)$ decays.  It is interesting to note that not only do the results for pure pionic decays agree with Pais' predictions, but that the predictions hold even when the pions are accompanied by even numbers of kaons and protons.

\begin{table}[!tb]

\caption{Measured branching fraction ratios $\mathcal{B}(\Upsilon(1S)\to(n-2)\pi^\pm2\pi^0+X)/\mathcal{B}(\Upsilon(1S)\to n\pi^\pm+X)$, where $X$ denotes kaons and/or protons in the decays. Isospin--based predictions of Pais~\cite{pais} for pure pionic decays are listed in the last column.}

\begin{ruledtabular}
\begin{tabular}{lD{,}{\:\pm\:}{-1}d}
\multirow{2}{*}{\large $\frac{\mathcal{B}((n\!-\!2)\pi^\pm2\pi^0+X)}{\mathcal{B}(n\pi^\pm+X)}$}
   & \multicolumn{1}{c}{$\Upsilon(1S)$} & \multicolumn{1}{c}{Theoretical} \\
 & & \multicolumn{1}{c}{Pais~\cite{pais}} \\

\hline

$4\pi^\pm2\pi^0~/~6\pi^\pm $  &   3.48 , 0.53      & 3.3 \\
$6\pi^\pm2\pi^0~/~8\pi^\pm $  &    7.38 , 0.99     & 6.3 \\ 
$8\pi^\pm2\pi^0~/~10\pi^\pm $  &   12.49 , 2.43    & \multicolumn{1}{c}{---} \\ 
 & &   \\
$2K^\pm+2\pi^\pm2\pi^0~/~4\pi^\pm $  &  1.06 , 0.17     & 1.3 \\
$\hspace*{31pt}4\pi^\pm2\pi^0~/~6\pi^\pm $  &   4.43 , 0.41    & 3.3 \\
$\hspace*{31pt}6\pi^\pm2\pi^0~/~8\pi^\pm $  &   8.23 , 0.96    & 6.3  \\
 & &   \\
$4K^\pm+2\pi^\pm2\pi^0~/~4\pi^\pm $  &   2.16 , 0.39   &  1.3 \\
$\hspace*{31pt}4\pi^\pm2\pi^0~/~6\pi^\pm $  &   4.56 , 0.97     & 3.3 \\
 & &   \\
$6K^\pm+2\pi^\pm2\pi^0~/~4\pi^\pm $  &   4.85 , 4.63    & 1.3 \\
 & &   \\
  $p\bar{p}+2\pi^\pm2\pi^0~/~4\pi^\pm $  &   1.50 , 0.30    & 1.3 \\
  $\hspace*{20.5pt}4\pi^\pm2\pi^0~/~6\pi^\pm $  &   3.33 , 0.42   & 3.3 \\
  $\hspace*{20.5pt}6\pi^\pm2\pi^0~/~8\pi^\pm $  &   6.44 , 0.84   & 6.3  \\
 & &   \\
  $p\bar{p}2K^\pm+2\pi^\pm2\pi^0~/~4\pi^\pm $  &   1.19 , 0.23   & 1.3 \\ 
  $\hspace*{40pt}4\pi^\pm2\pi^0~/~6\pi^\pm $  &   3.86 , 0.71   & 3.3 \\
 & &  \\
 $K_{S}K^\mp+~\,\pi^\pm2\pi^0~/~3\pi^\pm $  &   0.52 , 0.17   & 0.7 \\
  $\hspace*{40pt}3\pi^\pm2\pi^0~/~5\pi^\pm $  &   2.25 , 0.31    & 2.2 \\
  $\hspace*{40pt}5\pi^\pm2\pi^0~/~7\pi^\pm $  &   6.05 , 0.79   & 4.7 \\
 & &   \\
  $K_{S}3K^\mp+~\,\pi^\pm2\pi^0~/~3\pi^\pm$ &  0.70 , 0.18    & 0.7 \\
 $\hspace*{44pt}3\pi^\pm2\pi^0~/~5\pi^\pm$ &  2.95 , 0.65   & 2.2 \\
\end{tabular}
\end{ruledtabular}

\label{tbl:paiscomp}
\end{table}

\section{Systematic Uncertainties} 

We evaluate systematic uncertainties in the branching fractions due to the following sources.

The uncertainty in the number of $\Upsilon(1S)$ and $\Upsilon(2S)$ produced is $\pm2\%$~\cite{etabcleoiii}.
The uncertainty in the luminosity ratios $\mathcal{L}_\mathrm{on}/\mathcal{L}_\mathrm{off}$ is $\pm2\%$.

The accuracy of the Monte Carlo calculated efficiencies for particle reconstruction and identification in the CLEO~III detector have been extensively studied.  Since we use standard CLEO selection criteria in this analysis, we use standard systematic uncertainties for the branching fraction determination due to the following sources:
\begin{itemize}
\item Track reconstruction: $\pm1\%$ per track,
\item Charged particle ID: $\pm1\%$ per charged pion and $\pm2\%$ per charged kaon or proton,
\item $K_S$ reconstruction \& ID: $\pm6\%$ per $K_S$,
\item $\pi^0$ reconstruction \& ID: $\pm5\%$ per $\pi^0$.
\end{itemize}
The reconstruction and identification uncertainties for multiple tracks are added linearly.  For instance, the $2K2\pi$ mode has a total $\pm(4\times1)\%=\pm4\%$ uncertainty due to track reconstruction and $\pm((2\times1)+(2\times2))\% = 6\%$ uncertainty due to charged particle ID. 

The event selection efficiencies in Tables~I and II have been calculated using MC simulations in which $\Upsilon(1S)$ and $\Upsilon(2S)$ are assumed to decay into hadrons according to phase space.  To investigate the effect of intermediate resonances which decay to the same final states on our efficiency calculations we have investigated possible contributions from the following intermediate states: $\rho^0\to\pi^+\pi^-$, $\rho^\pm\to\pi^\pm\pi^0$, $\omega\to\pi^+\pi^-\pi^0$, $K^{*\pm}\to K^\pm\pi^0$, $K^{*0}\to K^\pm\pi^\mp$.  Signal MC samples were generated including these intermediate resonances, and analyzed in the same way as the phase space signal MC.  
We find that the effective change in efficiency due to the inclusion of intermediate states is less than $\pm10\%$ in all cases.  For all decays we therefore assign a systematic uncertainty of $\pm10\%$ due to this source.

We add the systematic uncertainties  due to all above sources in quadrature.
These total systematic uncertainties are listed in Tables~I and II, and range from 12\%~to~23\%.

\section{Summary of Results}

We have made the first measurements of exclusive hadronic decays of $\Upsilon(1S)$ and $\Upsilon(2S)$, each, into different final states containing 4 to 10  light hadrons ($\pi$, $K$, and $p$), charged hadrons only, charged hadrons plus one $\pi^0$, and charged hadrons plus two $\pi^0$'s.  Branching fractions for 73~decays of $\Upsilon(1S)$ and 17 decays of $\Upsilon(2S)$, ranging from $0.3\times10^{-5}$ to $110\times10^{-5}$ have been determined, and upper limits at 90\% confidence levels are presented for the others.  These measurements represent the first measurements of any exclusive decays of the Upsilon resonances.

Multiplicity distributions for the decays and pion isospin--dependent effects are noted.

\begin{acknowledgments}
We thank the CLEO Collaboration for the use of the data.  This research was supported by the U.S. Department of Energy.
\end{acknowledgments}

\clearpage

\onecolumngrid

\appendix

\section{Individual Mass Spectra for the $\bm{\Upsilon(1S)}$ Decays}

\begin{figure*}[!htb]
\begin{center}

\includegraphics[width=5.in]{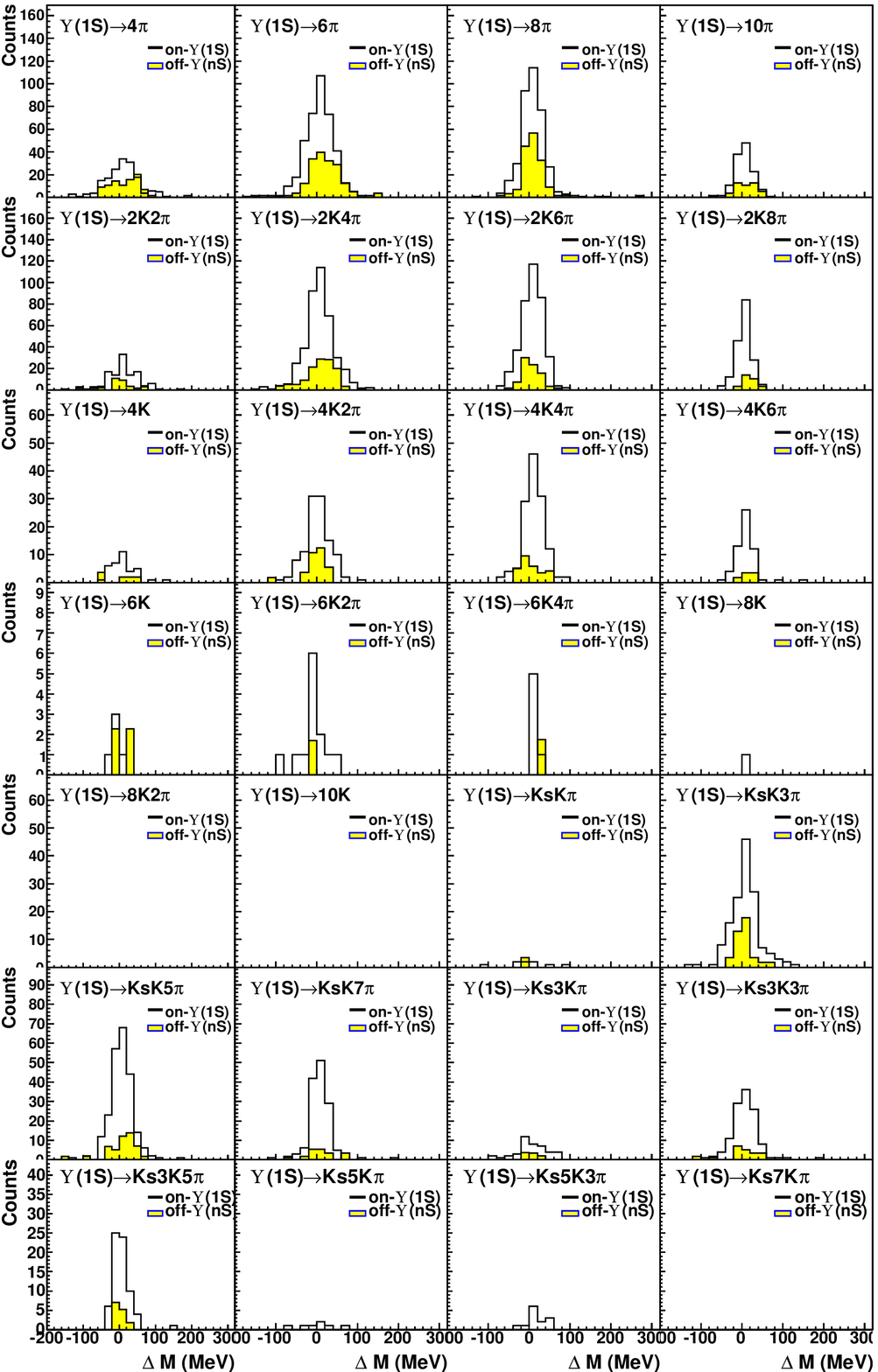}

\end{center}

\flushleft

FIG. A1(a): Mass distributions $\Delta M\equiv M(\Upsilon(1S))-M(\mathrm{hadrons})$ for individual decay modes in on--resonance $\Upsilon(1S)$ data and the sum of off--resonance $\Upsilon(1S)$ and $\Upsilon(2S)$ data.
The off--resonance data has been scaled by the luminosity ratio $\mathcal{L}_\mathrm{on}/\mathcal{L}_\mathrm{off} = 1.73$.
The ordinate scale is counts/20~MeV.

\end{figure*}

\begin{figure*}[!p]
\begin{center}

\includegraphics[width=5.in]{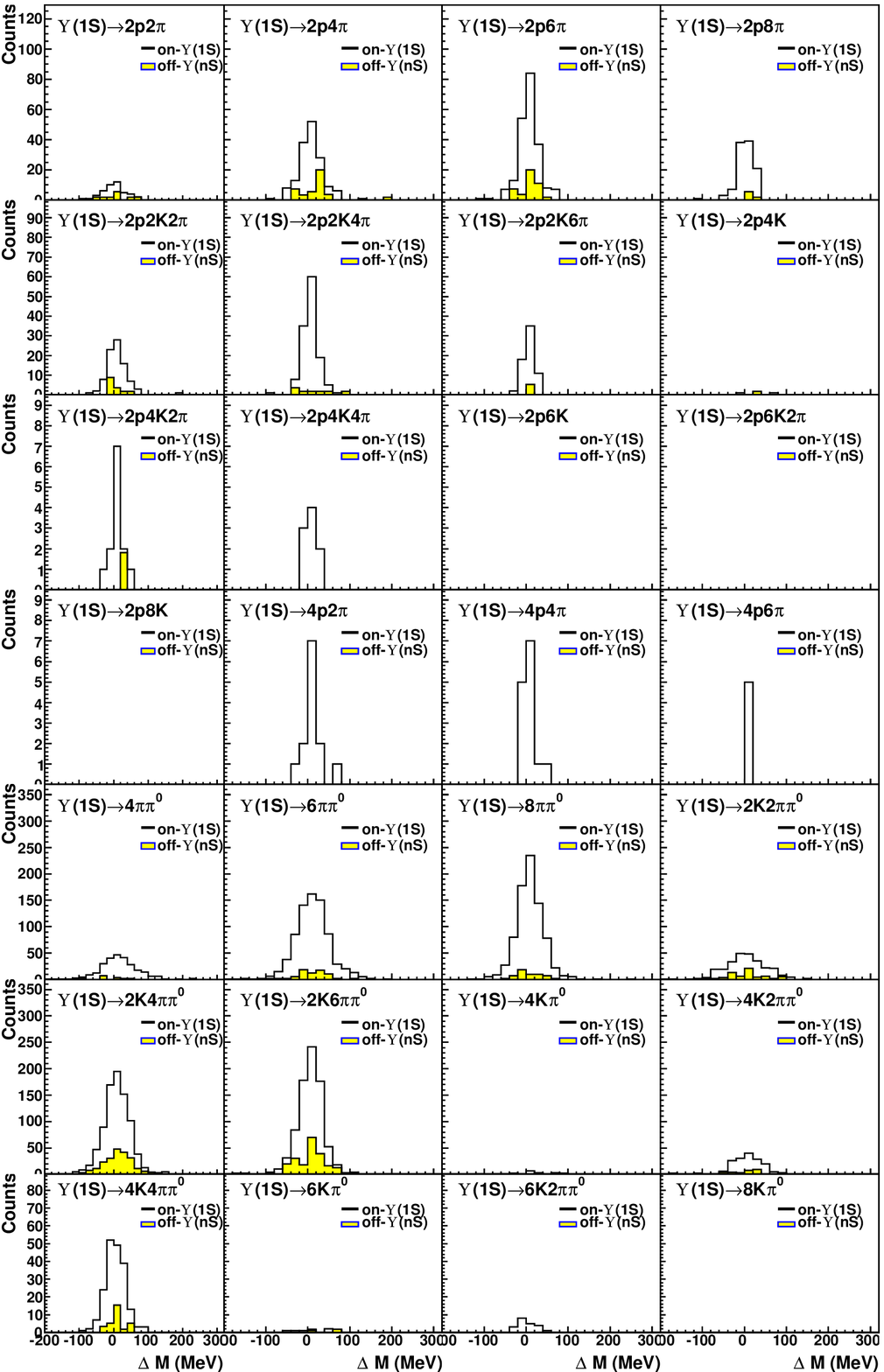}

\end{center}

\flushleft

FIG. A1(b), cont'd: Mass distributions $\Delta M\equiv M(\Upsilon(1S))-M(\mathrm{hadrons})$ for individual decay modes in on--resonance $\Upsilon(1S)$ data and the sum of off--resonance $\Upsilon(1S)$ and $\Upsilon(2S)$ data.
The off--resonance data has been scaled by the luminosity ratio $\mathcal{L}_\mathrm{on}/\mathcal{L}_\mathrm{off} = 1.73$.
The ordinate scale is counts/20~MeV.

\end{figure*}

\begin{figure*}[!p]
\begin{center}

\includegraphics[width=5.in]{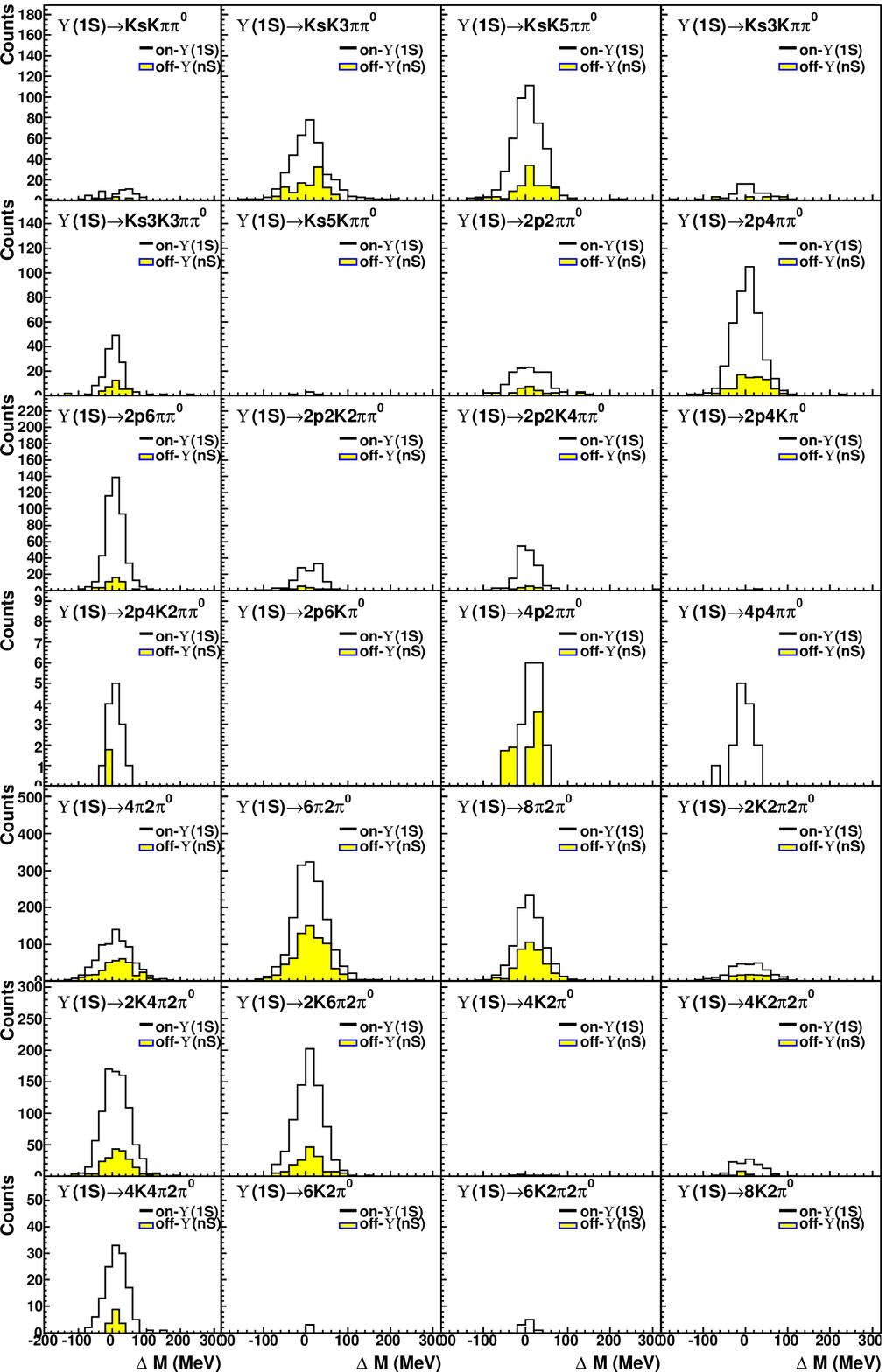}

\end{center}

\flushleft

FIG. A1(c), cont'd: Mass distributions $\Delta M\equiv M(\Upsilon(1S))-M(\mathrm{hadrons})$ for individual decay modes in on--resonance $\Upsilon(1S)$ data and the sum of off--resonance $\Upsilon(1S)$ and $\Upsilon(2S)$ data.
The off--resonance data has been scaled by the luminosity ratio $\mathcal{L}_\mathrm{on}/\mathcal{L}_\mathrm{off} = 1.73$.
The ordinate scale is counts/20~MeV.

\end{figure*}

\begin{figure*}[!p]
\begin{center}

\includegraphics[width=5.in]{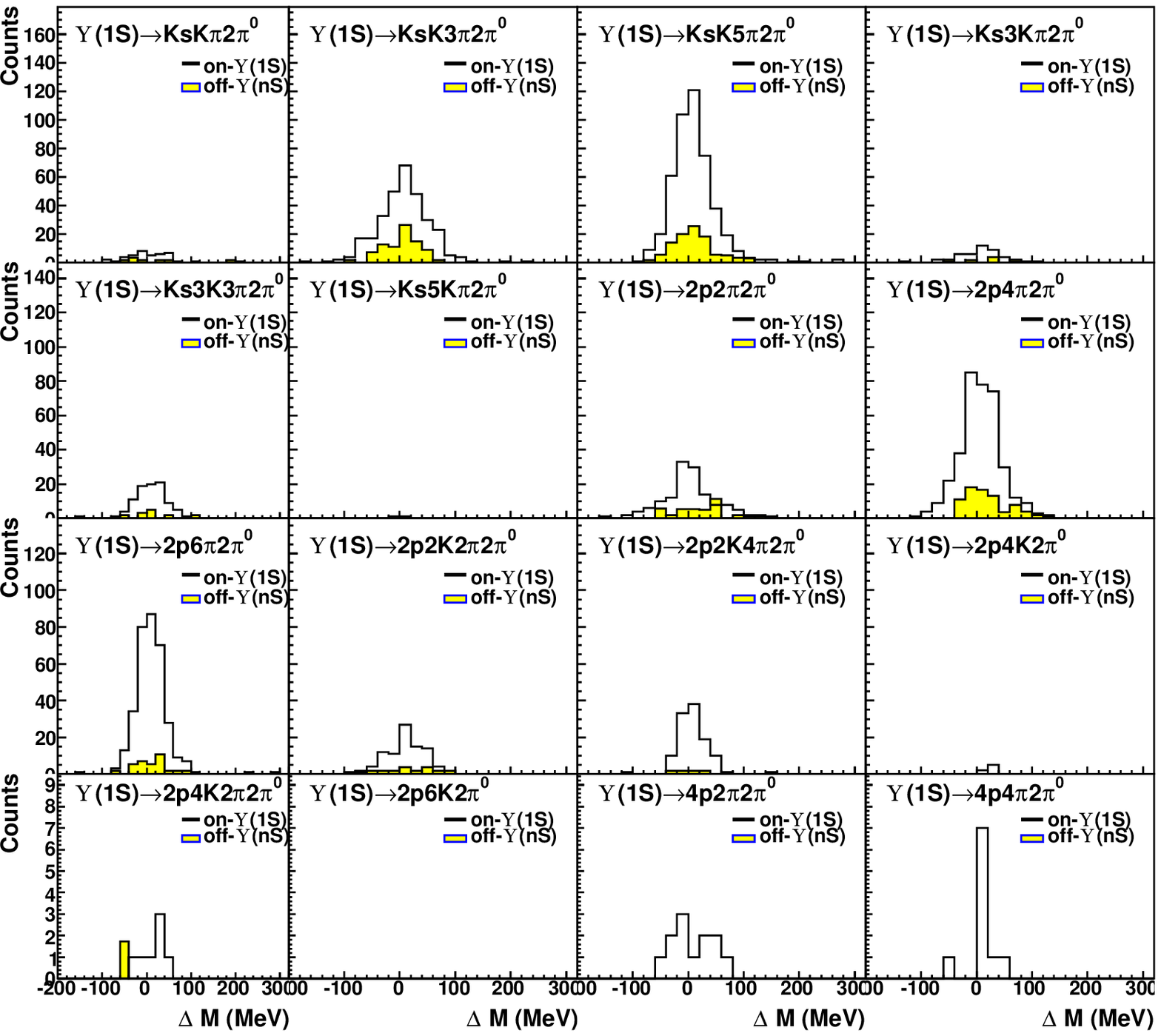}

\end{center}

\vspace*{-3.2in}

\flushleft

FIG. A1(d), cont'd: Mass distributions $\Delta M\equiv M(\Upsilon(1S))-M(\mathrm{hadrons})$ for individual decay modes in on--resonance $\Upsilon(1S)$ data and the sum of off--resonance $\Upsilon(1S)$ and $\Upsilon(2S)$ data.
The off--resonance data has been scaled by luminosity ratio $\mathcal{L}_\mathrm{on}/\mathcal{L}_\mathrm{off} = 1.73$.
The ordinate scale is counts/20~MeV.

\end{figure*}

\addtocounter{figure}{1}

\clearpage

\section{Individual Mass Spectra for the $\bm{\Upsilon(2S)}$ Decays}

\begin{figure*}[!h]
\begin{center}

\includegraphics[width=5.in]{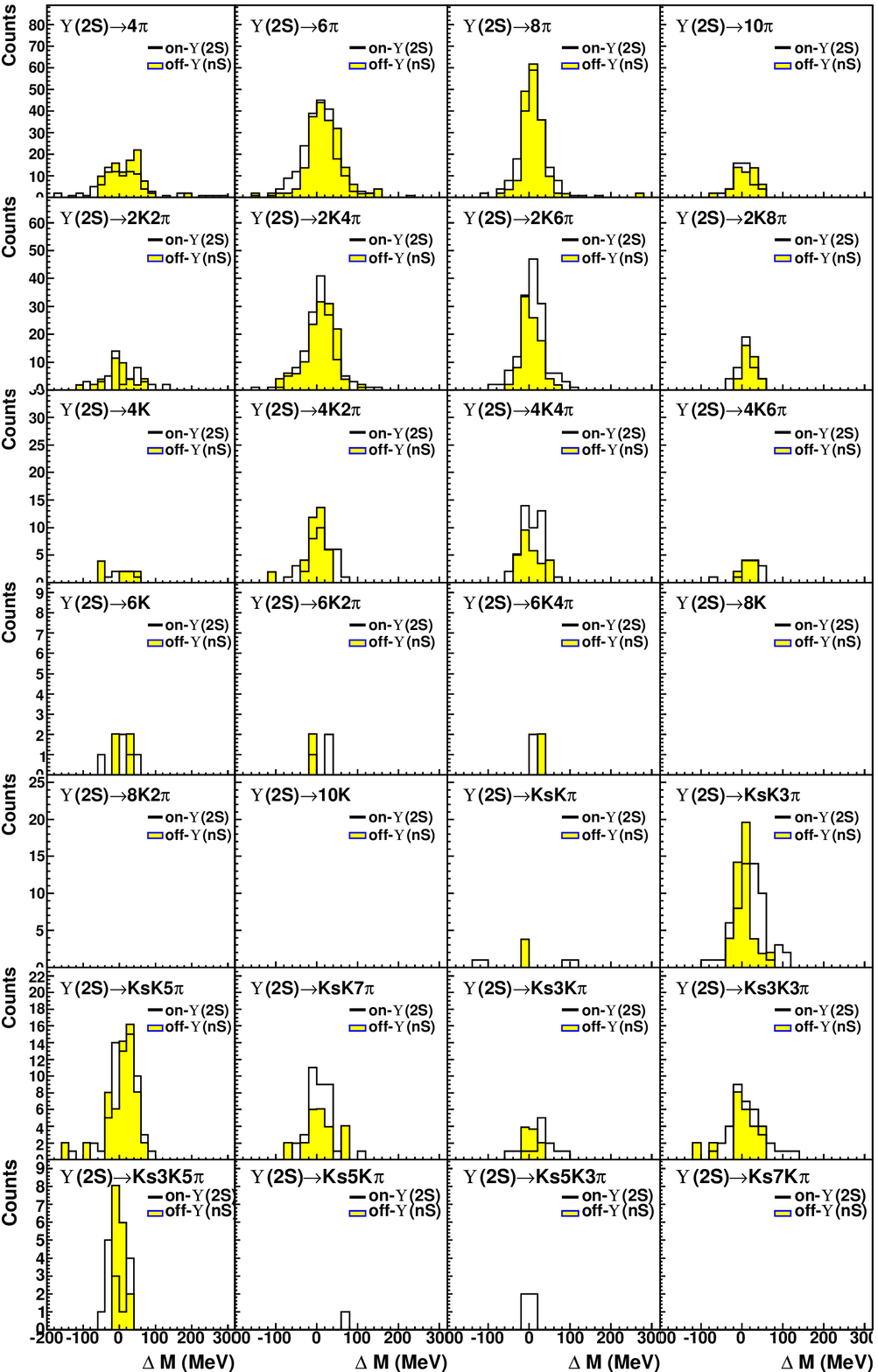}

\end{center}

\flushleft

FIG. B1(a): Mass distributions $\Delta M\equiv M(\Upsilon(2S))-M(\mathrm{hadrons})$ for individual decay modes in data taken on--resonance $\Upsilon(2S)$ data and the sum of off--resonance $\Upsilon(1S)$ and $\Upsilon(2S)$ data.
The off--resonance data has been scaled by the luminosity ratio $\mathcal{L}_\mathrm{on}/\mathcal{L}_\mathrm{off} = 2.03$.
The ordinate scale is counts/20~MeV.

\end{figure*}

\begin{figure*}[!p]
\begin{center}

\includegraphics[width=5.in]{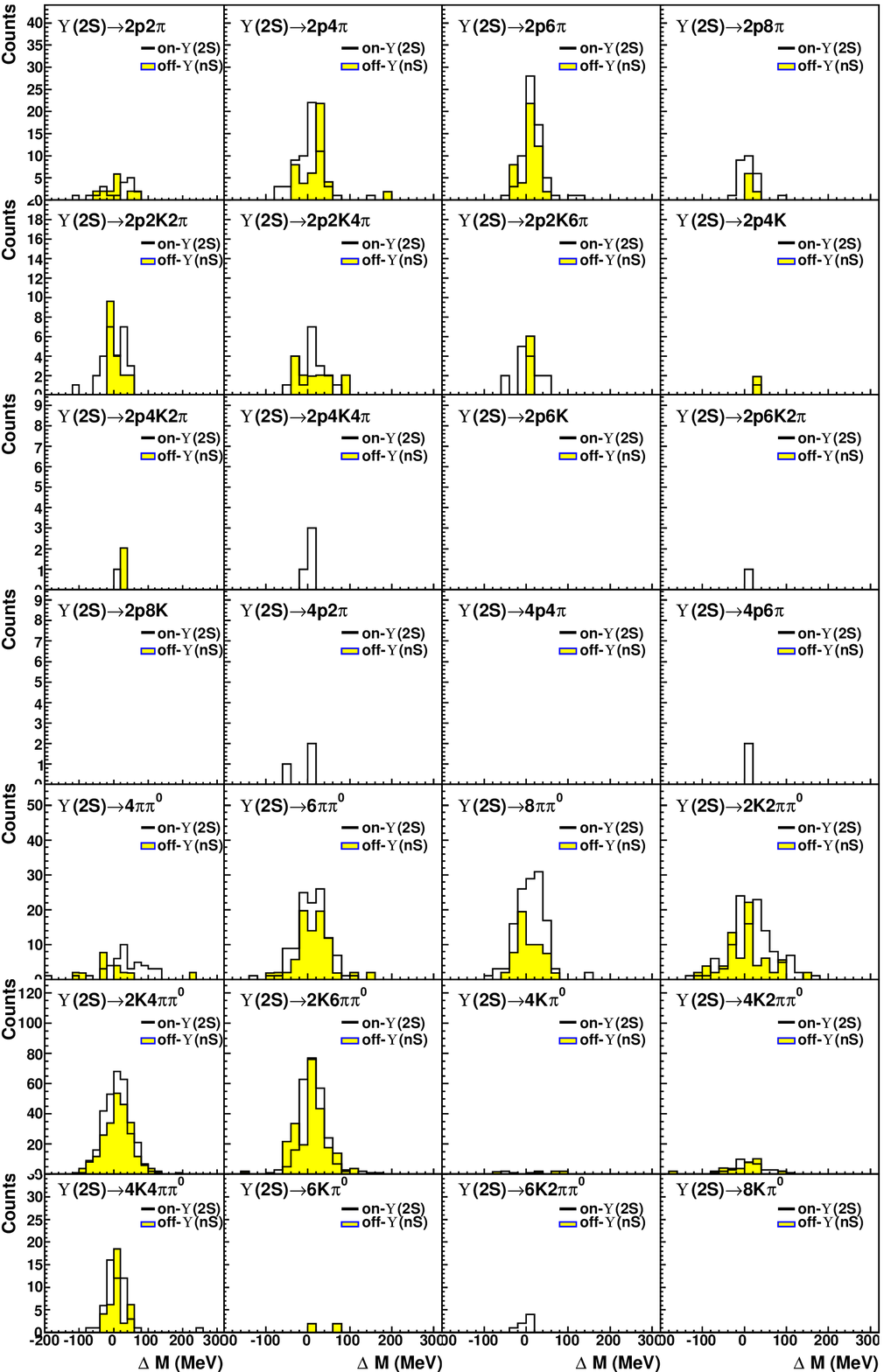}

\end{center}

\flushleft

FIG. B1(b), cont'd: Mass distributions $\Delta M\equiv M(\Upsilon(2S))-M(\mathrm{hadrons})$ for individual decay modes in on--resonance $\Upsilon(2S)$ data and the sum of off--resonance $\Upsilon(1S)$ and $\Upsilon(2S)$ data.
The off--resonance data has been scaled by the luminosity ratio $\mathcal{L}_\mathrm{on}/\mathcal{L}_\mathrm{off} = 2.03$.
The ordinate scale is counts/20~MeV.

\end{figure*}

\begin{figure*}[!p]
\begin{center}

\includegraphics[width=5.in]{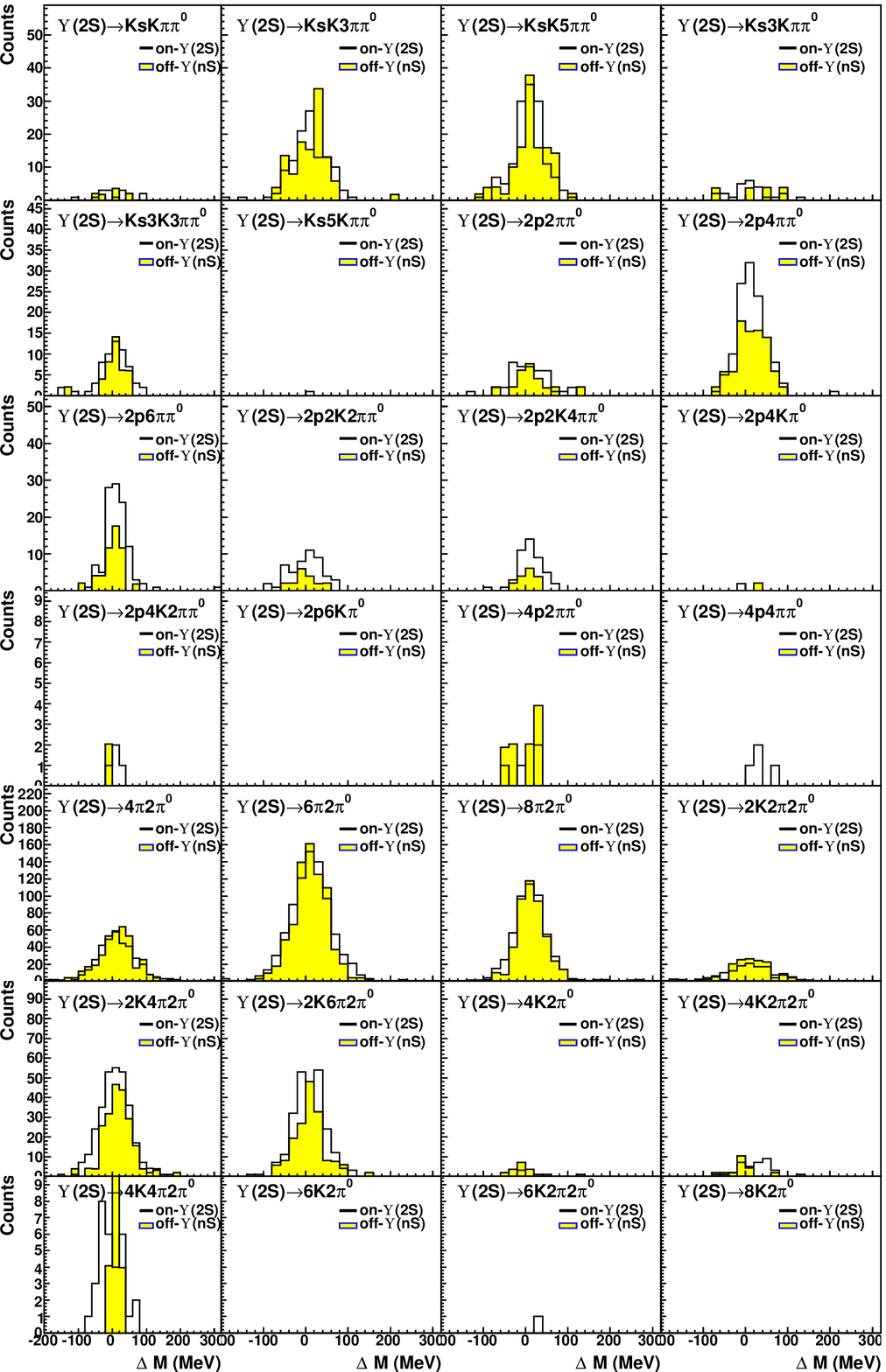}

\end{center}

\flushleft

FIG. B1(c), cont'd: Mass distributions $\Delta M\equiv M(\Upsilon(2S))-M(\mathrm{hadrons})$ for individual decay modes in on--resonance $\Upsilon(2S)$ data and the sum of off--resonance $\Upsilon(1S)$ and $\Upsilon(2S)$ data.
The off--resonance data has been scaled by the luminosity ratio $\mathcal{L}_\mathrm{on}/\mathcal{L}_\mathrm{off} = 2.03$.
The ordinate scale is counts/20~MeV.

\end{figure*}

\begin{figure*}[!p]
\begin{center}

\includegraphics[width=5.in]{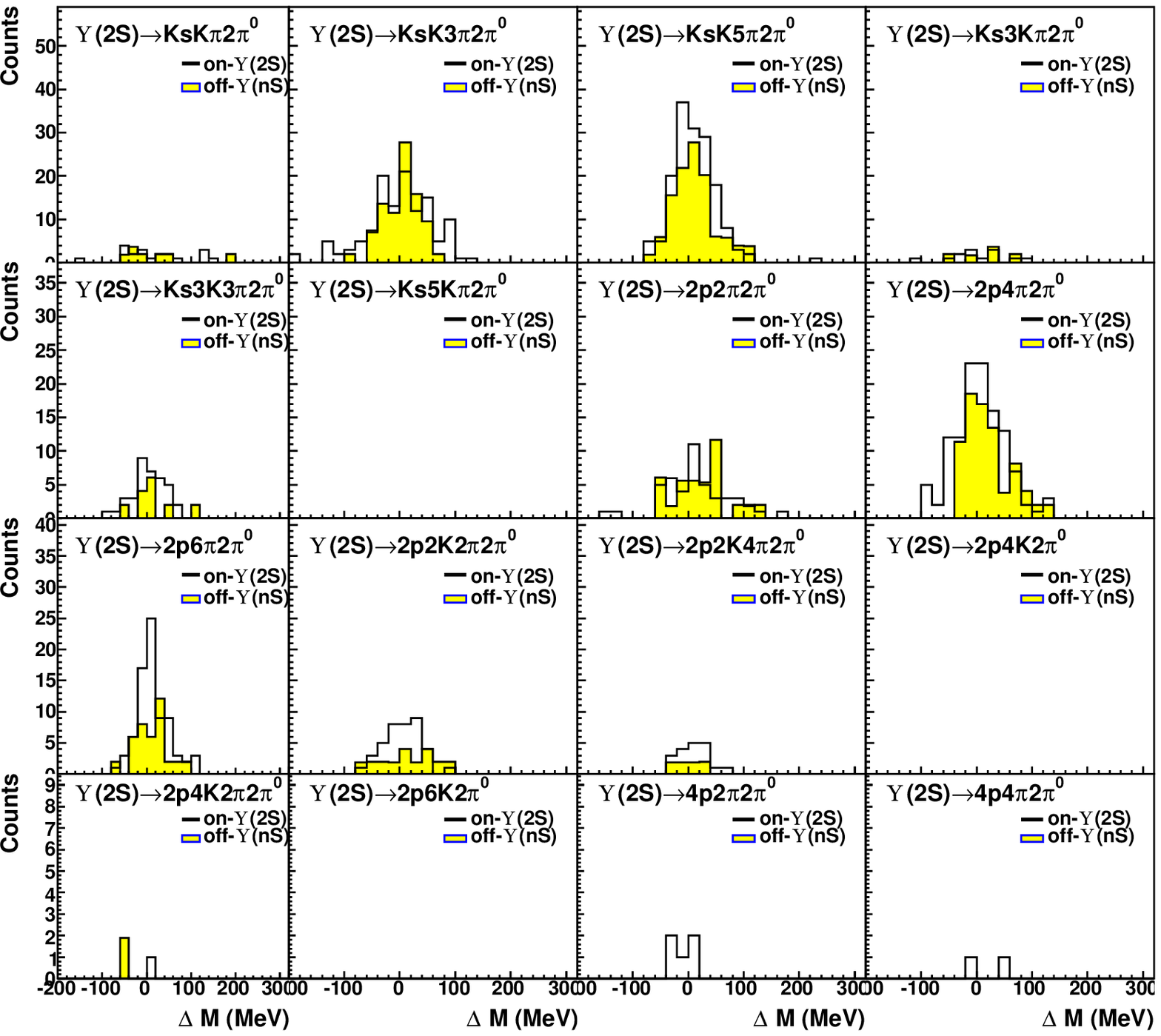}
\end{center}

\vspace*{-3.2in}

\flushleft

FIG. B1(d), cont'd: Mass distributions $\Delta M\equiv M(\Upsilon(2S))-M(\mathrm{hadrons})$ for individual decay modes in on--resonance $\Upsilon(2S)$ data and the sum of off--resonance $\Upsilon(1S)$ and $\Upsilon(2S)$ data.
The off--resonance data has been scaled by the luminosity ratio $\mathcal{L}_\mathrm{on}/\mathcal{L}_\mathrm{off} = 2.03$.
The ordinate scale is counts/20~MeV.

\end{figure*}

\addtocounter{figure}{1}


\end{document}